\documentclass[preprint,floatfix] {revtex4} 
\newcommand{\rvec}{\mathrm {\mathbf {r}}} 
\newcommand{\pvec}{\mathrm {\mathbf {p}}} 
\usepackage{graphicx}
\usepackage{tikz}
\usepackage{subfigure}
\usepackage{xcolor}
\usepackage{amsmath}
\usepackage{enumerate}
\usepackage{tabularx}
\usepackage{nicefrac}
\usepackage{enumitem}
\usepackage{color, soul}
\definecolor{darkblue}{rgb}{0,0,0.5}
\setulcolor{darkblue}

\begin{document}

\title{Shell-confined atom and plasma: incidental degeneracy, metallic character and information entropy}

\author{Neetik Mukherjee}
\altaffiliation{Email: pchem.neetik@gmail.com.}

\author{Amlan K.~Roy}
\altaffiliation{Corresponding author. Email: akroy@iiserkol.ac.in, akroy6k@gmail.com.}
\affiliation{Department of Chemical Sciences\\
Indian Institute of Science Education and Research (IISER) Kolkata, 
Mohanpur-741246, Nadia, WB, India}

\begin{abstract}
Shell confined atom can serve as a generalized model to explain both \emph{free} and \emph{confined} condition. In this scenario, 
an atom is trapped inside two concentric spheres of inner $(R_{a})$ and outer $(R_{b})$ radius. The choice 
of $R_{a}, R_{b}$ renders four different quantum mechanical systems. In hydrogenic atom, they are termed
as (a) free hydrogen atom (FHA) (b) confined hydrogen atom (CHA) (c) shell-confined hydrogen atom (SCHA) (d) left-confined hydrogen 
atom (LCHA). By placing $R_{a}, R_{b}$ at the location of radial nodes of respective \emph{free} $n,\ell$ states, a new kind of 
degeneracy may arise. At a given $n$ of FHA, there exists $\frac{n(n+1)(n+2)}{6}$ number of iso-energic states with energy $-\frac{Z^{2}}{2n^{2}}$. 
Furthermore, within a given $n$, the individual contribution of each of these four potentials has also been enumerated. 
This incidental degeneracy concept is further explored and analyzed in certain well-known \emph{plasma} (Debye and exponential cosine screened) 
systems. Multipole oscillator strength, $f^{(k)}$, and polarizability, $\alpha^{(k)}$, are evaluated for (a)-(d) in some low-lying states $(k=1-4)$. 
In excited states, \emph{negative} polarizability is also observed. In this context, metallic behavior of H-like systems in SCHA is discussed 
and demonstrated. Additionally analytical closed-form expression of $f^{(k)}$ and $\alpha^{(k)}$ are reported for $1s,2s,2p,3d,4f,5g$ 
	states of FHA. Finally, Shannon entropy and Onicescu {\color{red}information} energies are investigated in ground state in SCHA and LCHA in both position and 
momentum spaces. Much of the results are reported here for first time.  

 


\end{abstract}
\maketitle
\section{Introduction}
The discovery and development of modern scientific techniques has triggered an invasive interest on confined quantum systems. 
Particularly, in such environment, the rearrangement of atomic orbitals and increase in coordination number may lead to some 
fascinating, exceptional changes in physical and chemical characteristics \cite{grochala07}, such as, room temperature 
superconductivity \cite{snider20}, metallic behavior in ground state of H-like atoms \cite{sen02} etc. These confined systems have 
profound applications in condensed matter physics, high energy physics, astrophysics and nanotechnology {\color{red}\cite{sen14,aquil21}}. The idea
of quantum confinement has been exploited in construction of \emph{artificial atom} or a \emph{quantum dot} 
{\color{red}\cite{jean20}}. Such systems typically
consist of a group of electrons confined within a potential well. Another important example is the encapsulation of an atom or 
molecule in fullerene cage or zeolite cavity \cite{raggi14, raggi16, carbon18}. 

Atomic polarization plays a key role in explaining a number of processes in physics and chemistry. For example, multipole polarizability of an 
atom reflects quantitative distortion in electronic charge distribution due to presence of an external electromagnetic field. A host 
of macroscopic properties like 
refractive index, dielectric constant can be estimated via dipole polarizability \cite{mitroy10}. The latter plays an important role in 
the determination of physico-chemical properties, like optical response, as well as atomic and molecular interactions \cite{tiihonen17}.

Originally, the confinement model was proposed to understand changes in static dipole polarizability of H atom due to the 
influence of effective pressure acting on a given surface \cite{michel37}. In this fundamental system, H atom was trapped inside an 
impenetrable spherical cavity. The results of the designed model was utilized to gain knowledge of the core of planets like jupiter and 
saturn \cite{sommerfeld38,guillot99}. Of late, this concept was extended to a number of other physical, chemical and biological systems. 
A considerable amount of theoretical work has been published, covering a 
large variety of confining potentials \cite{grochala07,sen14} resulting in a vast literature. A confined H atom (CHA) in a spherical 
enclosure \cite{garza98,loughlin02,burrows06,aquino07,baye08,roy15,centeno17,coll17} represents a prototypical system, whose Schr\"odinger 
equation (SE) can be solved \emph{exactly} \cite{burrows06,montgomery07} in terms of \emph{Kummer confluent Hypergeometric} function.  
Hydrogen atom under the influence of several \emph{penetrable} as well as \emph{impenetrable} cavity was explored with great enthusiasm, 
giving rise to several interesting attractive properties--both from chemical and physical point of view. They offer some unique 
phenomena, especially the rearrangement of atomic orbitals, \emph{simultaneous, incidental} and \emph{inter-dimensional} 
degeneracy \cite{montgomery07} etc. Recently a new virial-like theorem is also formulated for such confinement situation \cite{mukherjee19}. 
Moreover, various properties like hyper-fine splitting constant, dipole shielding factor, nuclear magnetic screening constant, static, and 
dynamic polarizability, information entropy, Compton profiles {\color{red}\cite{mukherjee18,mukherjee18a,jiao17,mukherjee17,mukherjee20,sabin09,sen12}}
were examined for CHA. {\color{red}Further, the information theoretic measures are investigated for H-like atoms in Debye plasmas \cite{zan17}. 
A recent study reports the influence of external electric field on total Shannon entropy ($S_t$) \cite{salazar21}. Benchmark results for R\'enyi, 
Tsallis entropies and Onicescu {\color{red}information} energy ($E^O$) for ground state of Helium atom was studied using Hylleraas method \cite{ou19}.  
The static multipole polarizabilities are estimated for H-like atoms in Hulth\'en potential under both \emph{confined} and \emph{free} conditions 
\cite{he21} and H-atom in ring-shaped potentials \cite{yadav21}. Moreover, generalized pseudospectral (GPS) method was qeen used to explore several 
spectroscopic properties 
like fine structure, hyperfine splitting in confined environment \cite{zhu20}. Photoionization in H-atoms under fullerene cage is also reported for 
low-lying $s$-states \cite{lin13}.} However, an in-depth analysis of multipole oscillator strength and polarizability are
yet to be done for H atom trapped inside a cage, which remains one of the objectives of this work. 

A shell confinement model provides a new, unique boundary condition \cite{sen02,sen05,burrows08}. An appropriate choice of inner ($R_{a}$) 
and outer ($R_{b}$) radius of the shell can describe all possible radial boundary conditions so far reported in literature. For instance, 
when $R_{a}=0, R_{b}=r_{c}$ ($r_{c}$ is a real finite number) it reduces to CHA. On the other hand, choosing $R_{a}=0, R_{b}=\infty$, 
a \emph{free} H atom (FHA) is achieved. When both $R_{a}, R_{b}$ are non-zero and finite, it is termed as \emph{shell confined H-like 
atom} (SCHA). Whereas, a \emph{finite} $R_{a}$ and \emph{infinite} $R_{b}$ indicates \emph{left confined H-like atom} (LCHA). All these four systems, 
in general, are referred as a \emph{generalized confined H atom} (GCHA). The nodal characteristics of orbitals of FHA have played a 
significant role in conceptual development of degeneracy in GCHA. Previously, an attempt was made to solve SE of SCHA \emph{exactly} \cite{burrows08}, 
with limited success. Later, an accurate 
numerical strategy \cite{sen81,sen02} was prescribed to emphasize the occurrence of incidental degeneracy in SCHA \cite{sen05}. This new 
degeneracy can also account for the presence of incidental and simultaneous degeneracy in a CHA. 
The Kirkwood \cite{wood32} and Buckingham \cite{buck37} polarizability were evaluated 
\cite{sen05}. Sternheimer perturbation-numerical method \cite{stern54} was employed to calculate dipole polarizability in ground 
state \cite{sen05}. The Buckingham results are in good agreement with polarizability obtained via perturbation numerical 
procedure \cite{sen02}. The higher value of dipole polarizability in SCHA indicates metallic behavior of H atom in ground state \cite{sen02}.
Eigenvalues and eigenfunctions of $D$-dimensional SCHA has been examined lately \cite{pupyshev19}. 

In practice, a prototypical example of shell confinement is encapsulation of an atom or molecule under fullerene cage and
zeolite cavity \cite{nascimento11} or inside \emph{metal organic fremework} \cite{efros16,fei21}. Such an environment enhances 
stability and activity of \emph{noble}-metal catalysts by inhibiting the sintering effect \cite{peng18,rao18,kumar19,lai21,fei21}, amplifies 
photoluminescence in nano crystals by reducing non-radiative Auger processes \cite{efros16,fan20} and removes defects in polymer crystals 
\cite{shi13,khadilkar18} etc. Apart from these examples, shell confinement has potential applications in pollution control
\cite{qin20,zhang20}, therapeutics \cite{hastings19} and energy storage \cite{kumar18,wei19,wang20}.

In spite of having such versatile characteristics, shell confinement model has been studied only sparingly. As a consequence, 
literature on the topic is rather scarce. In this endeavour, our primary objective is to explore SCHA systematically, mainly through energy 
and other characteristic properties. Towards this goal, we consider incidental degeneracy, multipole ($2^{k}$-pole) oscillator strength, 
$f^{(k)}$ and polarizability, $\alpha^{(k)}$ ($k=1-4$), as well as certain information measures like $S$ and {\color{red}} $E^O$ in ground and a few 
low-lying states. 
Here $k = 1-4$ represent dipole, quadrupole, octupole and hexadecapole moments successively. In GCHA model, dependence of this 
degeneracy on principal ($n$) and orbital ($\ell$) quantum numbers is analyzed. This helps us to find out the exact number of 
degenerate states (in GCHA) associated with a given FHA energy of the form $-\frac{Z^{2}}{2n^{2}}$. Further, we can also estimate the number 
of such degenerate states that exists in GCHA. The calculation of dipole polarizability will guide us to examine the existence of metallic 
character in excited states. To this end, pilot calculations are performed on ground and lower excited states by invoking GPS method. 
To the best of our knowledge, most of the outcomes are presented here for first time. The article is constructed in following parts: Sec.~II 
provides a brief description about the formalism employed in present work. Section~III offers a detailed discussion of results. Finally, we 
conclude with a few remarks in Sec.~IV.      

\section{Theoretical formalism}
The single-particle time-independent non-relativistic radial SE for a spherically confined system is expressed (atomic unit employed unless 
otherwise stated) as,
\begin{equation}\label{eq:1}
\left[-\frac{1}{2} \ \frac{d^2}{dr^2} + \frac{\ell (\ell+1)} {2r^2} + V_{c}(r)\right] \psi_{n,\ell}(r)=\mathcal{E}_{n,\ell}\ \psi_{n,\ell}(r), 
\end{equation}
where $V_{c}$ represents the desired confined potential given below \cite{sen05,sen02}, 
\begin{equation}\label{eq:2}
\begin{aligned}    
V_{c}(r)=\begin{cases}
=v(r) \ \  \mathrm{for}   \ \ \ R_{a} \le r \le R_{b} \\
=\infty \ \ \ \mathrm{for} \ \ \ 0 \le r \le R_{a} \\
=\infty \ \ \ \mathrm{for} \ \ \ r \ge R_{b}.    
\end{cases}
\end{aligned}
\end{equation}
Here, $v(r)=-\frac{Z}{r}$, signifies the electron-nuclear Coulomb attraction potential ($Z$ refers to nuclear charge). Throughout our work, 
$V_{c}(r)$ will be referred as GCHA. Depending upon the values of 
$R_{a},R_{b}$, four distinct possibilities can be envisaged, as in the following,  

\begin{enumerate}
\item
When $R_{a}=0$, $R_{b}=\infty$, it gives rise to FHA. 
\item
When $R_{a}=0$, $R_{b}= r_{c}$, a finite number, it corresponds to CHA.
\item
When $R_{a} \ne 0$, $R_{b} \ne \infty$, with $R_{a}, R_{b}$ finite, it signifies SCHA.
\item
If $R_{a} \ne 0$, $R_{b}=\infty$, it refers to LCHA.
\end{enumerate}     

In order to calculate the energy, spectroscopic properties and information entropy the GPS method was invoked. {\color{red}
This provides a \emph{non-uniform, optimal} spatial discretization maintaining, high accuracy at both small and large 
distances. In contrast to the standard finite-difference methods, reasonably smaller number of grid point suffices, as it facilitates a
denser mesh at small $r$, while a coarser mesh at large $r$. Further, by applying a \emph{symmetrization} technique and a 
\emph{non-linear} mapping procedure, a \emph{symmetric} eigenvalue equation is achieved. It is computationally orders of magnitude faster
than finite-difference or finite element methods. Thus in essence, it combines the simplicity of direct finite-difference or finite element
method, with the fast convergence of finite basis set approaches.}
Over the time, it has been successfully used to estimate various bound-state properties of several central potentials including energy and other properties in CHA and confined 
many-electron atom \cite{roy2014,roy2014a,roy2014b,mukherjee18,mukherjee19,mukherjee20, majumdar2020, majumdar2021}, 

\subsection{Multipole polarizability}
By definition, the static multipole polarizability can be conveniently written as,  
\begin{equation}\label{eq:3}
\alpha^{(k)}_{i}=\alpha^{(k)}_{i}(\mathrm{bound})+\alpha^{k}_{i}(\mathrm{continuum}).
\end{equation} 
Conventionally $\alpha^{(k)}_{i}$ is expressed in terms of compact sum-over states form \cite{das12}. However it can also be directly 
estimated by employing the standard perturbation theory framework \cite{dalgarno62}. In the first procedure, Eq.~(\ref{eq:4}) modifies to {\color{red}\cite{zhu20a}},
\begin{equation}\label{eq:4}
\begin{aligned}
\alpha^{(k)}_{i} & = \sum_{n}\frac{f^{(k)}_{ni}}{(\mathcal{E}_{n}-\mathcal{E}_{i})^{2}}
-c\int\frac{|\langle R_{i}|r^{k}Y_{kq}(\rvec)|R_{\epsilon p}\rangle|^{2}}{(\mathcal{E}_{\epsilon p}-\mathcal{E}_{i})} \ \mathrm{d}\epsilon, \\
\alpha^{(k)}_{i}(\mathrm{bound}) & = \sum_{n}\frac{f^{(k)}_{ni}}{(\Delta \mathcal{E}_{ni}){2}}, \ \ \ \ \ 
\alpha^{k}_{i}(\mathrm{continuum}) = c\int\frac{|\langle R_{i}|r^{k}Y_{kq}(\rvec)|R_{\epsilon p}\rangle|^{2}}{(\mathcal{E}_{\epsilon p}-\mathcal{E}_{i})} \  \mathrm{d}\epsilon.
\end{aligned}
\end{equation}  
In Eq.~(\ref{eq:4}), the summation and integral terms signify bound and continuum contributions respectively, $f^{(k)}_{ni}$ represents 
multipole oscillator strength ($k$ is a positive integer), $c$ is a \emph{real} constant depending only on $\ell$ quantum number. {\color{red}$q$ is 
an integer.} Here, $f^{(k)}_{ni}$
measures the mean probability of transition between an initial ($i$) to final ($n$) state, which is normally expressed as, 
\begin{equation}\label{eq:5}
f^{(k)}_{ni}=\frac{8\pi}{(2k+1)}\Delta\mathcal{E}_{ni}|\langle r^{k} Y_{kq}(\rvec)\rangle|^{2}. 
\end{equation}    
Designating the initial and final states as $|n \ell m\rangle$ and $|n^{\prime}\ell^{\prime}m^{\prime}\rangle$, one can easily derive,  
\begin{equation}\label{eq:6}
f^{(k)}_{ni}=\frac{8\pi}{(2k+1)} \ \Delta\mathcal{E}_{ni} \ \frac{1}{2\ell+1}\sum_{m}\sum_{m^{\prime}} |\langle n^{\prime}\ell^{\prime}m^{\prime}|r^{k}Y_{kq}(\rvec)|n \ell m \rangle|^{2}.
\end{equation}
The application of Wigner-Eckart theorem and sum rule for \emph{3j} symbol further leads to,  
\begin{equation}\label{eq:7}
f^{(k)}_{ni}=2 \ \frac{(2\ell^{\prime}+1)}{(2k+1)} \ \Delta\mathcal{E}_{ni} \ |\langle r^{k}\rangle^{n^{\prime}\ell^{\prime}}_{n \ell}|^{2} \
\left\{\begin{array}{c} \ell^{\prime} \ \ k \ \ \ell\\ 0 \ \ 0 \ \ 0 \end{array}\right\}^{2}. 
\end{equation}
The transition matrix element is then given by following radial integral,
\begin{equation}\label{eq:8}
\langle r^{k} \rangle = \int_{0}^{\infty} R_{n^{\prime} \ell^{\prime}}(r) r^{k} R_{n \ell} (r) r^{2} \mathrm{d}r.
\end{equation}
Note that, $f^{(k)}_{ni}$ depends on $n, \ell$, but independent of magnetic quantum number $m$. In this article, we compute $f^{(k)}$, 
$\alpha^{(k)}$, having $k=1-4$, for states with $\ell=0-4$. It is necessary to point out that, there exists a multipole 
oscillator strength sum rule as follows, 
\begin{equation}\label{eq:9}
S^{(k)}=\sum_{m}f^{(k)}=k\langle \psi_{i}|r^{(2k-2)}|\psi_{i}\rangle, 
\end{equation}
where the summation includes all the bound and continuum states.

\subsection{Information entropy}
{\color{red}Information entropic measures are functionals of density and they quantify density in several complimentary ways. 
They have potential applications in atomic avoided crossing, electron correlation effect, quantum entanglement, orbital-free
density functional theory etc \cite{mukherjee18,mukherjee20}. $S$ is the arithmetic mean of uncertainty, and expressed as an 
expectation value of logarithmic density. $S_{\rvec}$ measures the uncertainty in localization of a particle in $r$ space. A lower $S_{\rvec}$ 
indicates higher accuracy in predicting the localization. Similarly, $S_{\pvec}$ measures the uncertainty
in predicting the momentum of a particle. $S_{\rvec}$ and $S_{\pvec}$ are expressed as,}
\begin{equation}\label{eq:10}
\begin{aligned} 
S_{\rvec} & =  -\int_{{\mathcal{R}}^3} \rho(\rvec) \ \ln [\rho(\rvec)] \ \mathrm{d} \rvec   = 
2\pi \left(S_{r}+S_{(\theta,\phi)}\right), \\   
S_{\pvec} & =  -\int_{{\mathcal{R}}^3} \Pi(\pvec) \ \ln [\Pi(\pvec)] \ \mathrm{d} \pvec   = 
2\pi \left(S_{p}+S_{(\theta, \phi)}\right). 
\end{aligned} 
\end{equation}
Here $\rho(\rvec), \Pi(\pvec)$ signify $r$- and $p$-space densities, both normalized to unity. 
{\color{red}Arguably, $S_{\rvec}$ and $S_{\pvec}$ provide the most appropriate uncertainty relation \cite{bbi75}. $S_{\rvec}$ and $S_{\pvec}$
are the logarithmic functionals of density. As a consequence, the total Shannon entropy is expressed as $S_{\rvec}+S_{\pvec}$,}  
\begin{equation}\label{eq:10a}
\begin{aligned} 
S_{t} & = S_{\rvec}+S_{\pvec} & =2\pi \left[S_{r}+S_{p}+2S_{(\theta, \phi)}\right] \ \ \geq 3(1+\ln \pi). 
\end{aligned} 
\end{equation}

The quantities $S_r, S_p$ and $S_{\theta}$ are defined as \cite{bbi75},   
\begin{equation}\label{eq:11}
\begin{aligned} 
S_{r} & =  -\int_0^\infty \rho(r) \ \ln [\rho(r)] r^2 \mathrm{d}r, \ \ \ \ \   
S_{p}  =  -\int_{0}^\infty \Pi(p) \ln [\Pi(p)]  \ p^2 \mathrm{d}p, \\
\rho(r) & = |\psi_{n,l}(r)|^{2},  \ \ \ \ \ \ \ \    \Pi(p) = |\psi_{n,l}(p)|^{2}, \\
S_{(\theta, \phi)} & =   -\int_0^\pi \chi(\theta) \ \ln [\chi(\theta)] \sin \theta \mathrm{d} \theta, \ \ \ \ \ \
\chi(\theta)   =  |\Theta(\theta)|^2.  \\   
\end{aligned} 
\end{equation}
Another important measure studied in this work is $E^O$, referring to the 2nd order entropic moment \cite{sen12}. {\color{red}It is the 
expectation value of density. It portrays exactly opposite behavior to $S$. It is also termed as dis-equilibrium, as it measures
the deviation of a distribution from equilibrium \cite{shiner99}.} In $r$ and $p$ space, the respective quantities are defined as,  
\begin{equation}\label{eq:12}
E^{O}_{r}= \int_0^\infty [\rho(r)]^{2} r^2 \mathrm{d}r, \ \   
E^{O}_{p}= \int_{0}^\infty [\Pi(p)]^{2} p^2 \mathrm{d}p, \ \   
E^{O}_{\theta, \phi}= \int_0^\pi [\chi(\theta)]^{2} \sin \theta \mathrm{d}\theta, \ \
E^{O}_{t}=E^{O}_{r}E^{O}_{p}\left[E^{O}_{\theta, \phi}\right]^{2}. 
\end{equation}
where, $E^{O}_{t}$ is the total Onicescu 
{\color{red}information} energy. Accurate $r$-space wave functions are obtained by applying GPS method.  
The corresponding $p$-space wave function is generated by Fourier transforming the $r$-space counterpart. This is 
accomplished quite efficiently by following a procedure adopted in \cite{mukherjee18}. 

\section{Result and Discussion}
In \cite{sen05}, all three (CHA, SCHA, LCHA) models were mentioned under the general SCHA heading. However, since they have quite different energy 
characteristics, we discuss them separately here. The demonstrative results are presented for H-atom ($Z=1$) only. However, similar outcome 
can also be extracted for $Z \ne 1$ cases. Thus, at first, we shall analyze the salient features of incidental degeneracy achieved by placing the 
boundary at respective nodal positions of FHA. Then we present $f^{(k)}(Z)$ and $\alpha^{(k)} (Z)$ ($k=1-4$) for selected low-lying 
states in these four potentials. Further, in the realm of Herzfeld criterion of metallic behavior, we have computed $\alpha^{(1)}(Z)$ for 
$1s, 2s, 2p, 3d, 4f, 5g$ states. As a bonus, analytical closed-form expressions of $f^{(k)}(Z)$ and $\alpha^{(k)}(Z)$ are derived for all 
these six states. Finally, we consider $S_{\rvec}, S_{\pvec}, S$ as well as $E^{O}_{\rvec}, E^{O}_{\pvec}, E$ in ground state involving these four 
potentials. It is noteworthy to mention that, in case of degeneracy, radial boundaries are chosen specifically at the nodes of FHA to illustrate their
role. Whereas for other properties ($f^{(k)}, \alpha^{(k)}$, information entropy), no such factor was taken into consideration. Thus they are selected 
so as to illustrate the essential features related to an individual property.

\begingroup           
\squeezetable
\begin{table}
\caption{Incidental degeneracy in GCHA associated with $n=4$ of FHA. See text for details.}
\centering
\begin{ruledtabular}
\begin{tabular}{l|llllllll}
Serial & No. of node & State & $R_{a}$  & $R_{b}$ & Energy & FHA   & $\alpha^{(1)}$ & $S_{\rvec}$\\
\hline
$a$  & 0   &  $1s$ & 0      &  1.87164450  &  $-$0.03125000 & $4s$    & 0.27404101      & 2.22927677 \\
$b$  & 0   &  $1s$ & 1.87164450 &  6.6108150  &  $-$0.03125000 & $4s$ & 169.7527968     & 6.55734580 \\
$c$  & 0   &  $1s$ & 6.6108150 &  15.51755  &  $-$0.03125000 & $4s$   & 8609.5280939    & 9.15073565 \\
$d$  & 0   &  $1s$ & 15.51755 &  100     &  $-$0.03125000 & $4s$      & 322925.0793     & 12.14484806 \\
$e$  & 1   &  $2s$ & 0      &  6.6108150  &  $-$0.03125000 & $4s$     & $-$128.96450306 & 6.35223141 \\
$f$  & 1   &  $2s$ & 1.87164450 &  15.51755  &  $-$0.03125000 & $4s$  & 2265.40684074   & 9.10240049 \\
$g$  & 1   &  $2s$ & 6.6108150 &  100     &  $-$0.03125000 & $4s$     & 173868.83409    & 12.14272018 \\
$h$  & 2   &  $3s$ & 0      &  15.51755  &  $-$0.03125000 & $4s$      & 129.09470914    & 9.01151759 \\
$i$  & 2   &  $3s$ & 1.87164450 &  100     &  $-$0.03125000 & $4s$    & 80408.86663     & 12.09689745 \\
$j$   & 3   &  $4s$ & 0      &  100     &  $-$0.03125000 & $4s$        & 4992.00000000   & 12.07490387 \\
$k$   & 0   &  $2p$ & 0      &  10-2$\sqrt{5}$  &  $-$0.03125000 & $4p$    & 13.56524044     & 5.42877070 \\
$l$   & 0   &  $2p$ & 10-2$\sqrt{5}$  & 10+2$\sqrt{5}$ & $-$0.03125000 & $4p$ & 1711.37938497   & 8.51079967 \\
$m$   & 0   &  $2p$ & 10-2$\sqrt{5}$ &  125  &  $-$0.03125000 & $4p$      & 84939.073612    & 11.63160747 \\
$n$   & 1   &  $3p$ & 0      & 10+2$\sqrt{5}$   &  $-$0.03125000 & $4p$   & $-$977.65896463 & 8.32046807 \\
$o$   & 1   &  $3p$ & 10-2$\sqrt{5}$ & 105  &  $-$0.03125000 & $4p$        & 40632.47423     & 11.61219021 \\
$p$   & 2   &  $4p$ & 0      & 125   &  $-$0.03125000 & $4p$           & 5107.1999999    & 11.53386387 \\
$q$   & 0   &  $3d$ & 0      &  12   &  $-$0.03125000 & $4d$           & 203.03802379    & 7.82189131 \\
$r$   & 0   &  $3d$ & 12     & 120   &  $-$0.03125000 & $4d$           & 18.72296856     & 11.37785020 \\
$s$   & 1   &  $4d$ & 0      & 125   &  $-$0.03125000 & $4d$           & 5760.000000     & 11.26209911 \\
$t$   & 2   &  $4f$ & 0      & 130   &  $-$0.03125000 & $4f$           & 6720.000000     & 10.86085521 \\	
\end{tabular}
\end{ruledtabular}
\end{table}  
\endgroup  

\subsection{Incidental Degeneracy}
{\color{red}Following \cite{sen05}, it may so happen that, the energy of a given confined state becomes equal to that of an unconfined state 
(here it is $-\frac{Z}{2n^{2}}$), when the radius of confinement is suitably chosen at the location of radial nodes in latter state. 
Such a phenomenon is termed as \emph{incidental degeneracy}. It is shown in Eq.~(\ref{eq:2}) that, \emph{shell confined} condition renders 
four different systems. This degeneracy may provide a connection amongst them.} 

\subsubsection{H-like ion}
It is known that, if $R_a, R_b$ of a GCHA coincide with certain specific radial nodes of ($n,\ell$) state of FHA, then there exists $(n'-\ell-1)$ number of 
nodes in between them. Furthermore, energy of such a ($n', \ell$) GCHA state becomes degenerate to that of a FHA state
At first, we wish to determine the \emph{number of degenerate states} associated with a given FHA energy $-\frac{Z^{2}}{2n^{2}}$. 
It is worthwhile mentioning that, in this part we shall only discuss the states that arises due to placing the boundary at nodal points of FHA.  
For demonstrative purpose, we present all the 10 states belonging to $n \leq 4$ of GCHA in Table~I. The corresponding boundaries are chosen 
from radial nodes of $4s, 4p, 4d$ states of FHA. Thus as can be seen, there exists a total of 20 degenerate states in GCHA, all having the 
same energy of $-0.031250$ a.u., corresponding to $n=4$ of FHA. Out of that, the number of $s, p, d,f$ are 10, 6, 3, 1 respectively. It is also 
recognized that, there are $6 (a,e,h,k,n,q), 4 (b,c,f,l), 6 (d,g,i,m,o,r), 4 (j,p,s,t)$ states belonging to CHA, SCHA, LCHA and FHA. The last 
two columns also tabulate the respective $\alpha^{(1)}$ ($\Delta \ell=1$) and $S_{r}$. One can see that, in SCHA $\alpha^{(1)}$ possesses higher  
value compared to CHA and FHA counterparts. 

\begingroup                    
\squeezetable
\begin{table}
\caption{Incidental degeneracy in WCP and ECSCP, for $\lambda_{1},\lambda_2=0.01$ a.u. See text for details.}
\centering
\begin{ruledtabular}
\begin{tabular}{l|llllllll}
Serial & No. of node & State & $R_{a}$  & $R_{b}$ & Energy & Free state & $\alpha^{(1)}$ & $S_{\rvec}$\\
\hline
\multicolumn{8}{c}{WCP} \\
\hline
$2a$ &  0   &   $1s$   &  0  &  2.000390  & $-$0.115293282 & $2s$  & 0.34278641   & 2.39716825 \\
$2b$ &  0   &   $1s$   &  2.000390 &  100 & $-$0.115293282 & $2s$  & 932.14066065 & 8.21415856 \\
$2c$ &  1   &   $2s$   &  0    &  100  & $-$0.115293282 & $2s$ & 120.5848668  & 8.11443243 \\
\hline
$3a$ & 0   &  $1s$   & 0   &  1.902698  &   $-$0.046198857 &  $3s$ & 0.28975431    & 2.27113719 \\
$3b$ & 0   &  $1s$   & 1.902698 & 7.108762 & $-$0.046198857 & $3s$ & 208.23885866  & 6.75134003 \\
$3c$ & 0   &  $1s$   & 7.108762 & 150      & $-$0.046198857 & $3s$ & 28579.047941  & 10.53960365 \\
$3d$ & 1   &  $2s$   & 0        & 7.108762 & $-$0.046198857 & $3s$ & $-$207.14861496  & 6.56892996 \\
$3e$ & 1   &  $2s$   & 1.9026980 & 150     & $-$0.046198857 & $3s$ & 12624.6506    & 10.48406780 \\
$3f$ & 2   &  $3s$   & 0         & 150     & $-$0.046198857 & $3s$ & 1033.70055187 & 10.44196440 \\
\hline
$4a$  &  0   &  $1s$ & 0         & 1.8729343 & $-$0.022356120 & $4s$  & 0.27468622    & 2.23104252 \\
$4b$  &  0   &  $1s$ & 1.8729343 & 6.6268050 & $-$0.022356120 & $4s$  & 170.96931921  & 6.56397087 \\
$4c$  &  0   &  $1s$ & 6.6258050 & 15.6046240 & $-$0.022356120 & $4s$ & 8757.380222   & 9.16720124 \\
$4d$  &  0   &  $1s$ & 15.6046240 & 150     &  $-$0.022356120 & $4s$  & 337593.5180   & 12.18838925 \\
$4e$  &  1   &  $2s$ &  0         & 6.6268050 & $-$0.022356120 & $4s$ & $-$131.07497378 & 6.35963966 \\
$4f$  &  1   &  $2s$ & 1.8729343  & 15.6046240 & $-$0.022356120 & $4s$ & 2309.54693312  & 9.11891006 \\
$4g$  &  1   &  $2s$ & 6.6268050  & 150.0      & $-$0.022356120 & $4s$  & 182824.8322 & 12.18544076 \\
$4h$  &  2   &  $3s$ &  0         & 15.6046240 & $-$0.022356120 & $4s$ & 131.30125522 & 9.02899807 \\
$4i$  &  2   &  $3s$ &  1.8729343 & 150 & $-$0.022356120 & $4s$ & 85268.9345 & 12.14058812 \\
$4j$  &  3   &  $4s$ &  0         & 150  & $-$0.022356120 & $4s$ & 5294.641728 & 12.11924015 \\  
\hline
\multicolumn{8}{c}{ECSCP} \\
\hline
$2a$ & 0   &   $1s$   &  0  &  2.0000208  & $-$0.115013458 & $2s$  & 0.34257005   & 2.39669558   \\
$2b$ & 0   &   $1s$   &  2.0000208 &  100 & $-$0.115013458 & $2s$  & 928.75702501 & 8.21114396   \\
$2c$ & 1   &   $2s$   &  0 &  100 & $-$0.115013458 & $2s$  & 120.07106878    & 8.11127833   \\
\hline
$3a$ & 0   &  $1s$   & 0         &  1.90200530  & $-$0.045619079 & $3s$ & 0.28939390   & 2.27020331 \\
$3b$ & 0   &  $1s$   & 1.90200530  & 7.0994429   & $-$0.045619079 & $3s$ & 207.43129440 & 6.74776731 \\
$3c$ & 0   &  $1s$   & 7.0994429  & 150        & $-$0.045619079 & $3s$ & 28215.90613  & 10.52774639 \\
$3d$ & 1   &  $2s$   & 0         & 7.0994429   & $-$0.045619079 & $3s$ & $-$205.22594788  & 6.56497789 \\
$3e$ & 1   &  $2s$   & 1.902000530 & 150        & $-$0.045619079 & $3s$ & 12430.91176  & 10.47209351 \\
$3f$ & 2   &  $3s$   & 0 & 150        & $-$0.045619079 & $3s$ & 1017.71439735  & 10.42967937 \\
\hline
$4a$ & 0   &  $1s$ & 0          & 1.87185526   & $-$0.021437465 & $4s$  & 0.27414583   & 2.22956395 \\
$4b$ & 0   &  $1s$ & 1.87185526 & 6.61372660   & $-$0.021437465 & $4s$  & 169.96865645 & 6.55854163 \\
$4c$ & 0   &  $1s$ & 6.61372660 & 15.5362170  & $-$0.021437465 & $4s$ & 8639.641193 & 9.15424609 \\
$4d$ & 0   &  $1s$ & 15.5362170 & 150        & $-$0.021437465 & $4s$  & 327409.3061 & 12.15969133 \\
$4e$ & 1   &  $2s$ &  0         & 6.61372660  & $-$0.021437465 & $4s$ & $-$129.33805232 & 6.35357332 \\
$4f$ & 1   &  $2s$ & 1.87185526 & 15.5362170 & $-$0.021437465 & $4s$ & 2274.9433235 & 9.10593888 \\
$4g$ & 1   &  $2s$ & 6.61372660  & 150.0      & $-$0.021437465 & $4s$  & 176835.0570 & 12.15740646 \\
$4h$ & 2   &  $3s$ &  0         & 15.6046240 & $-$0.021437465 & $4s$ & 129.58544973 & 9.02879432 \\
$4i$ & 2   &  $3s$ &  1.87185526 & 150 & $-$0.021437465 & $4s$ & 82114.5084 & 12.11197447 \\ 
$4j$ & 3  &  $4s$ &  0 & 150 & $-$0.021437465 & $4s$ & 5112.60426 & 12.09022342 \\
\end{tabular}
\end{ruledtabular}
\end{table}  
\endgroup  

It is well known that, in FHA, energies of all the $\ell$ states (0 to $n-1$) within a given $n$ are same. Now, from an observation of the results in 
Table~I, it can be extracted that, there appears $\frac{(n-\ell)(n-\ell+1)}{2}$ number of iso-energic states in GCHA. Thus, for a given 
$n$-state of FHA, the total number of degenerate GCHA states work out to be, 
\begin{equation}\label{eq:14}
\begin{aligned}
 & =  \frac{1}{2} \sum^{(n-1)}_{\ell=0} (n-\ell)(n-\ell+1)  
  =  \frac{1}{2} \sum^{(n-1)}_{\ell=0} (n-\ell)^{2} + \sum^{n-1}_{\ell=0}(n-\ell) \\
 & =  \frac{n(n+1)}{2} +\frac{n(n+1)(2n+1)}{12} \\
 & =  \frac{n(n+1)(n+2)}{6}.
\end{aligned}
\end{equation}
It suggests that, this number \emph{does not} depend on $\ell$. Now, we can determine the contribution of each 
of these four categories in this degeneracy, as follows.

\begin{enumerate}[label=(\Roman*)]
\item 
\textbf{FHA:} For a particular $n$, there exists $n$ number of degenerate states. 
\item
\textbf{CHA:} In this case, an $\ell$-orbital contributes $(n-\ell-1)$ number of degenerate states. Thus, the total number of degenerate 
states are then given by, 
\begin{equation}\label{eq:15}
\begin{aligned}
& \sum^{(n-1)}_{\ell=0} (n-\ell-1)
& = \frac{n(n-1)}{2}. 
\end{aligned}
\end{equation}
\item

\textbf{SCHA:} In $s$ orbital, first SCHA state occurs with energy equal to $n=3$ of FHA \cite{sen05}. Similarly, for $p$ orbital, 
it has energy equal to $n=4$ of FHA. So, for a given $\ell$, 1st degenerate SCHA state appears at $n=(\ell+3)$ having energy 
$-\frac{z^{2}}{2(\ell+3)^{2}}$. For a given $n$, such states can be achieved up to: $\ell=(\ell_{max}-2)=(n-3)$. Therefore, at a fixed $n$, 
a given $\ell$ state contributes as $\frac{(n-\ell-2)(n-\ell-1)}{2}$, giving the total number as,  
\begin{equation}\label{eq:16}
\begin{aligned} 
&   \sum^{(n-3)}_{\ell=0}\frac{(n-\ell-2)(n-\ell-1)}{2}   
& = \frac{n(n-1)(n-2)}{6}
\end{aligned}         
\end{equation}

\item
\textbf{LCHA:} Similar to CHA, here also a particular $\ell$ orbital will contribute $(n-\ell-1)$ degenerate states, giving the same 
total as in CHA, namely, $\sum^{(n-1)}_{\ell=0} (n-\ell-1) =  \frac{n(n-1)}{2}$. 
\end{enumerate}  

Next, we estimate the individual contribution of all four systems in above degeneracy. On the basis of above discussion and 
following Table~I, one can find certain characteristics. To facilitate this, we use $n, \ell$ to denote principal 
and orbital quantum number of FHA, whereas $n_{k}, \ell_{j}$ signify the same for other three systems ($k, j$ are integers).    

\begin{figure}                         
\begin{minipage}[c]{0.50\textwidth}\centering
\includegraphics[scale=0.8]{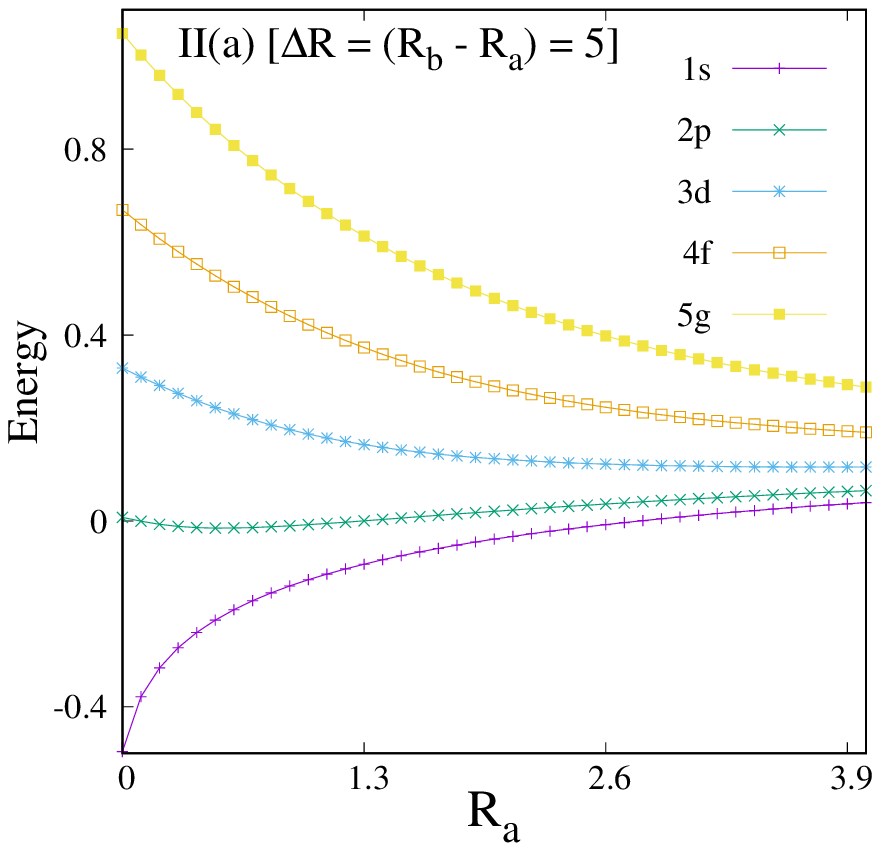}
\end{minipage}%
\begin{minipage}[c]{0.50\textwidth}\centering
\includegraphics[scale=0.8]{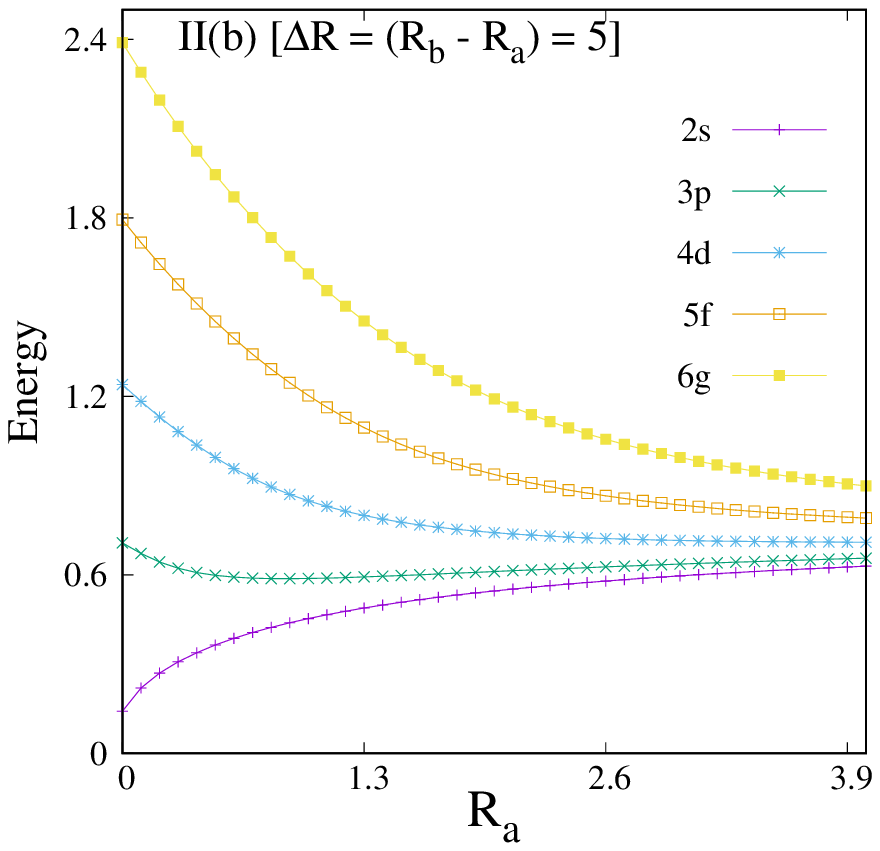}
\end{minipage}%
\vspace{2mm}
\begin{minipage}[c]{0.50\textwidth}\centering
\includegraphics[scale=0.8]{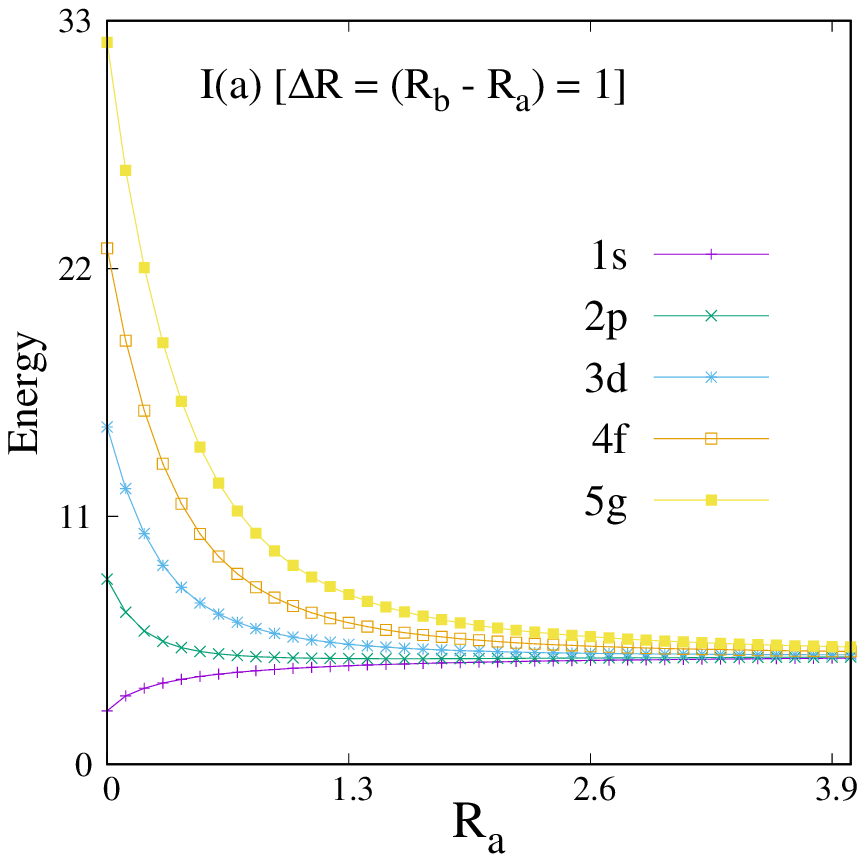}
\end{minipage}%
\begin{minipage}[c]{0.50\textwidth}\centering
\includegraphics[scale=0.8]{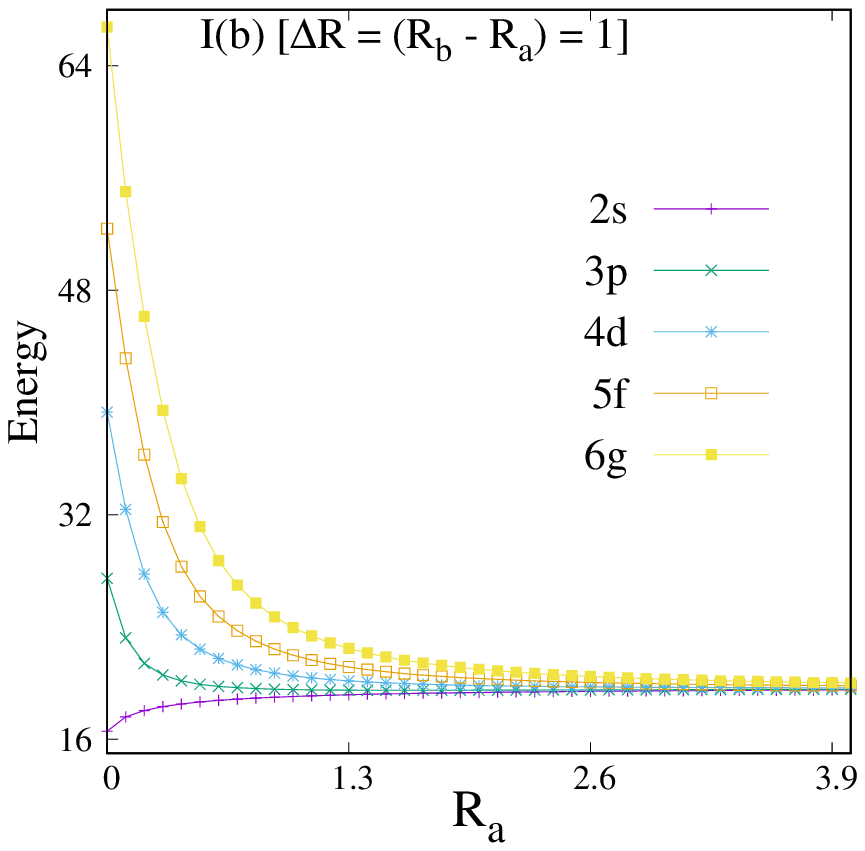}
\end{minipage}%
\caption{Energy as function of $R_{a}$ (in a.u) in SCHA, for (I) $\Delta R =(R_{b}-R_{a})=1$ (II) $\Delta R =(R_{b}-R_{a})=5$.
Labels (a), (b), refer to circular and single-node states. See text for details.}
\end{figure}

\begin{enumerate}[label=(\roman*)]
\item
Corresponding to the $n$th state of FHA, there are $\frac{n(n+1)(n+2)}{6}$ number of degenerate GCHA states, each having 
the same energy $-\frac{Z^{2}}{2n^{2}}$. 
\item
Each $\ell$-state belonging to a certain $n$, contributes $\frac{(n-\ell)(n-\ell+1)}{2}$ number of GCHA states. 
\item
Number of incidental degenerate state increases with $n$. However, at a fixed $n$, the number of such states reduce with rise in 
$\ell$.
\item 
The first occurrence of degenerate SCHA state, takes place at $n=3$. 
\item
At a fixed $\ell$, $n_{1} < n$. It suggests that, $n_{1}$ takes values from $(\ell+1)$ to $n-1$. 
If we choose $n=4$, then $\mathcal{E}_{4}=-0.03125$. Therefore, for $\ell=0$, $n_{1}=1,2,3$; for $\ell=1$, 
$n_{1}=2,3$; and when $\ell=2$, $n_{1}=3$.  
\item
Two arbitrary states $(n_{1},\ell_{1})$, $(n_{2},\ell_{2})$ are degenerate when $n_{1} < n, \ell_{1} < n$ and 
$n_{2} < n, \ell_{2} < n$. Note that, they may belong to any of the systems in GCHA, except FHA. 
\end{enumerate}

\begingroup              
\squeezetable
\begin{table}
\caption{$f^{(1)}$ values for $1s,2s,2p$ states in CHA, SCHA and LCHA. See text for details.}
\centering
\begin{ruledtabular}
\begin{tabular}{ll|ll|ll|llll}
$R_{a}$  & $R_{b}$ & $1s \rightarrow 2p$  & $1s \rightarrow 3p$ & $2s \rightarrow 2p$ & $2s \rightarrow 3p$ & 
$2p \rightarrow 1s$  & $2p \rightarrow 2s$ & $2p \rightarrow 3d$ & $2p \rightarrow 4d$ \\
\hline	
0    & 1  & 0.98455839  & 0.00772592   & $-$0.60825789 & 1.56032656 & $-$0.32818613  & 0.20275263 & 1.08482483 & 0.01857681  \\
0.1  & 1  & 0.89910222  & 0.09117798   & $-$0.54875155 & 1.26759877 & $-$0.29970074  & 0.18291718 & 1.07826147 & 0.02579334  \\
0.2  & 1  & 0.81158829  & 0.17664462   & $-$0.46434643 & 1.01215223 & $-$0.27052943  & 0.15478214 & 1.04448022 & 0.05966553  \\
0.5  & 1  & 0.69746119  & 0.28964404   & $-$0.35084205 & 0.73447901 & $-$0.23248706  & 0.11694735 & 0.92955029 & 0.17316032  \\
0.8  & 1  & 0.66991458  & 0.31699958   & $-$0.32345071 & 0.67370461 & $-$0.22330486  & 0.10781690 & 0.89321899 & 0.20918396  \\
\hline                                                                                                                     
0    & 2  & 0.99105877  & 0.00000217   & $-$0.61189926 & 1.57832558 & $-$0.33035292  & 0.20396642 & 1.09062730 & 0.01308847 \\
0.1  & 2  & 0.96414685  & 0.02875100   & $-$0.60923016 & 1.46958646 & $-$0.32138228  & 0.20307672 & 1.08969744 & 0.01424386 \\
0.5  & 2  & 0.78114997  & 0.20643653   & $-$0.43425912 & 0.93184418 & $-$0.26038332  & 0.14475304 & 1.02245081 & 0.08147398 \\
1    & 2  & 0.69749557  & 0.28958819   & $-$0.35086313 & 0.73452648 & $-$0.23249852  & 0.11695438 & 0.92959980 & 0.17310450 \\
1.2  & 2  & 0.68355578  & 0.30344579   & $-$0.33700278 & 0.70354791 & $-$0.22785193  & 0.11233426 & 0.91134321 & 0.19119914 \\
1.5  & 2  & 0.67205870  & 0.31486957   & $-$0.32557999 & 0.67836486 & $-$0.22401957  & 0.10852666 & 0.89607618 & 0.20634732 \\
1.8  & 2  & 0.66739177  & 0.31950490   & $-$0.32094504 & 0.66823564 & $-$0.22246392  & 0.10698168 & 0.88985569 & 0.21252327 \\
\hline                                                                                                                     
0    & 5  & 0.84879929  & 0.10827497   & $-$0.45637469 & 1.42333674 & $-$0.28293310  & 0.15212490 & 1.10303346 & 0.00102259 \\
1    & 5  & 0.82132944  & 0.16452278   & $-$0.47393100 & 1.02743235 & $-$0.27377648  & 0.15797700 & 1.05560747 & 0.04906997 \\
2    & 5  & 0.72015051  & 0.20643653   & $-$0.37324792 & 0.78555850 & $-$0.24005017  & 0.12441597 & 0.95825952 & 0.14467096 \\
2.5  & 5  & 0.69759377  & 0.28958819   & $-$0.35089758 & 0.73466209 & $-$0.23253126  & 0.11696586 & 0.92974209 & 0.17292899 \\
3    & 5  & 0.68357365  & 0.30344579   & $-$0.33700060 & 0.70357230 & $-$0.22785788  & 0.11233353 & 0.91136832 & 0.19116227 \\
4    & 5  & 0.66991478  & 0.31486957   & $-$0.32344989 & 0.67370488 & $-$0.22330493  & 0.10781663 & 0.89321926 & 0.20918302 \\
4.5  & 5  & 0.66739177  & 0.31950490   & $-$0.32094500 & 0.66823564 & $-$0.22246392  & 0.10698167 & 0.88985569 & 0.21252323 \\
\hline                                                                                                                     
0   & 10 & 0.49203980  & 0.25817376  & $-$0.07781857 & 0.96754959   & $-$0.16401327  & 0.02593952 & 1.07006404 & 0.02801599 \\
0.5 & 10 & 0.97045316  & 0.01437069  & $-$0.62098640 & 1.56331874   & $-$0.32348439  & 0.20699547 & 1.08623978 & 0.01685871 \\
1   & 10 & 0.94767400  & 0.02402487  & $-$0.58992446 & 1.36743381   & $-$0.31589133  & 0.19664149 & 1.10659524 & 0.00041483 \\
2   & 10 & 0.82883015  & 0.14973954  & $-$0.47587757 & 1.03956806   & $-$0.27627672  & 0.15862586 & 1.06686692 & 0.03701308 \\
3   & 10 & 0.75955649  & 0.22459445  & $-$0.41041279 & 0.87499038   & $-$0.25318550  & 0.13680426 & 1.00438637 & 0.09839538 \\
5   & 10 & 0.69774060  & 0.28903455  & $-$0.35085740 & 0.73486495   & $-$0.23258020  & 0.11695247 & 0.92995783 & 0.17260954 \\
7   & 10 & 0.67494769  & 0.31198282  & $-$0.32843445 & 0.68465713   & $-$0.22498256  & 0.10947815 & 0.89992293 & 0.20251950 \\
9.5 & 10 & 0.66683858  & 0.32005426  & $-$0.32039568 & 0.66703853   & $-$0.22227953  & 0.10679856 & 0.88911811 & 0.21325570 \\
\hline
0    & $\infty$ & 0.41619672  & 0.07910156   &    0.00000000 & 0.43486544 & $-$0.13873224  & 0.00000000 & 0.69578470 & 0.12179511 \\
0.1  & $\infty$ & 0.61261825  & 0.08716313   & $-$0.27197479 & 0.76653089 & $-$0.20420608  & 0.09065826 & 0.69643208 & 0.12177348 \\
0.5  & $\infty$ & 0.91228737  & 0.03992888   & $-$0.50501203 & 1.34020061 & $-$0.30409579  & 0.16833734 & 0.73703538 & 0.11966976 \\
1    & $\infty$ & 0.95710827  & 0.00509331   & $-$0.46634518 & 1.42628570 & $-$0.31903609  & 0.15544839 & 0.84505124 & 0.10660158 \\
2    & $\infty$ & 0.91518127  & 0.00537086   & $-$0.38224386 & 1.32047405 & $-$0.30506042  & 0.12741462 & 1.01517164 & 0.05572610 \\
5    & $\infty$ & 0.82683246  & 0.05128960   & $-$0.29816602 & 1.10002016 & $-$0.27561082  & 0.09938867 & 1.08031802 & 0.00001907 \\
7    & $\infty$ & 0.79907239  & 0.07217889   & $-$0.27883906 & 1.02880135 & $-$0.26635746  & 0.09294635 & 1.05975869 & 0.00511577 \\
8    & $\infty$ & 0.78916499  & 0.08038843   & $-$0.27256657 & 1.00309177 & $-$0.26305500  & 0.09085552 & 1.04975557 & 0.00959527 \\
9    & $\infty$ & 0.78094995  & 0.08751295   & $-$0.26761185 & 0.98164753 & $-$0.26031665  & 0.08920395 & 1.04067662 & 0.01429067 \\
10   & $\infty$ & 0.77400929  & 0.09376540   & $-$0.26360162 & 0.96343656 & $-$0.25800310  & 0.08786721 & 1.03254711 & 0.01891643 \\
\end{tabular}
\end{ruledtabular}
\end{table}  
\endgroup          

When both $R_a, R_b$ are finite, non-zero, i.e, it is SCHA, the behavior of particle is deeply influenced by $R_{a}$, $R_{b}$ and $\Delta R =(R_{b}-R_{a})$. 
However, controlling any two parameters would also serve the purpose of the remaining one. It is found that, at a fixed $R_{b}$, energy of a given state 
progresses with $R_{a}$ (smaller $\Delta R$). Conversely, at a given $R_{a}$, a reverse pattern is noticed with rise in $R_{b}$ (larger $\Delta R$). 
It is of interest to monitor the energy pattern for a fixed $\Delta R$, with modulations in both $R_{a}$ and $R_{b}$, which is depicted in 
Fig.~1. The bottom and top panels display energy as a function of $R_{a}$ at two selected $\Delta R$, namely, 1 and 5 in (I) and (II).       
The left and right sides record first five circular ($1s, 2p, 3d, 4f, 5g$) and single-node ($2s, 3p, 4d, 5f, 6g$) states, having labels (a), (b). 
A careful observation reveals that, in either case, energy of $\ell=0$ states ($1s, 2s$) gradually advances with $R_{a}$ (squeezing of box). 
On the contrary, $\ell \ne 0$ states record a decay in the same with growth in $R_{a}$; initially at a quick pace and then slows down until becoming 
flat at sufficiently large $R_a$. However, for $\ell > 0$, there is harmony amongst the states with slight differences in lower $R_a$ region. 

\subsubsection{H-plasma}
Now, an important question arises: is this special degeneracy a unique feature of one-electron Coulombic systems? To examine it, we 
extend the calculations on two familiar plasma models: (i) weakly coupled plasma (WCP) or Debye plasma, governed by a potential, 
$V(r)=-\frac{Z}{r}e^{-\lambda_{1} r}$ (ii) exponential cosine screened Coulomb potential (ECSCP), given by, 
$V(r)=-\frac{Z}{r}e^{-\lambda_{2} r}\cos \lambda_{2} r$. Here, $\lambda_{i}$ is the inverse of Debye radius and represents the interaction 
between electron and ions in a plasma \cite{solyu12}. In particular, $\lambda_{1}=\sqrt{\frac{4\pi e^{2}n_{e}}{k_{b}T}}$ ($n_{e},k_{b},T$ 
signify ion density, Boltzmann constant and plasma temperature respectively), while 
$\lambda_{2}=\frac{k_{q}}{\sqrt{2}}=\sqrt{\frac{n_{e}\omega_{pe}}{\hbar}}$ ($k_{q}$ is the electron plasma wave number connected to plasma 
frequency and number density). Note that, in WCP, classical interactions are considered, while the quantum effect in a plasma can be added 
by invoking a $\cos \lambda r$ term in WCP \cite{jiao21}. These two prototypical systems were studied heavily, offering a vast literature. 
Thus, the influence of screening on energy spectrum \cite{solyu12,paul09,bahar14,bahar16}, photo-ionization cross section 
\cite{jung95,jung96,song03}, electron-impact excitations \cite{lin10,lin11} were investigated with appreciable interest. Generally speaking, 
plasma systems have finite number of bound states, which reduces with enhancement of $\lambda$. 

\begin{figure}                         
\begin{minipage}[c]{0.32\textwidth}\centering
\includegraphics[scale=0.53]{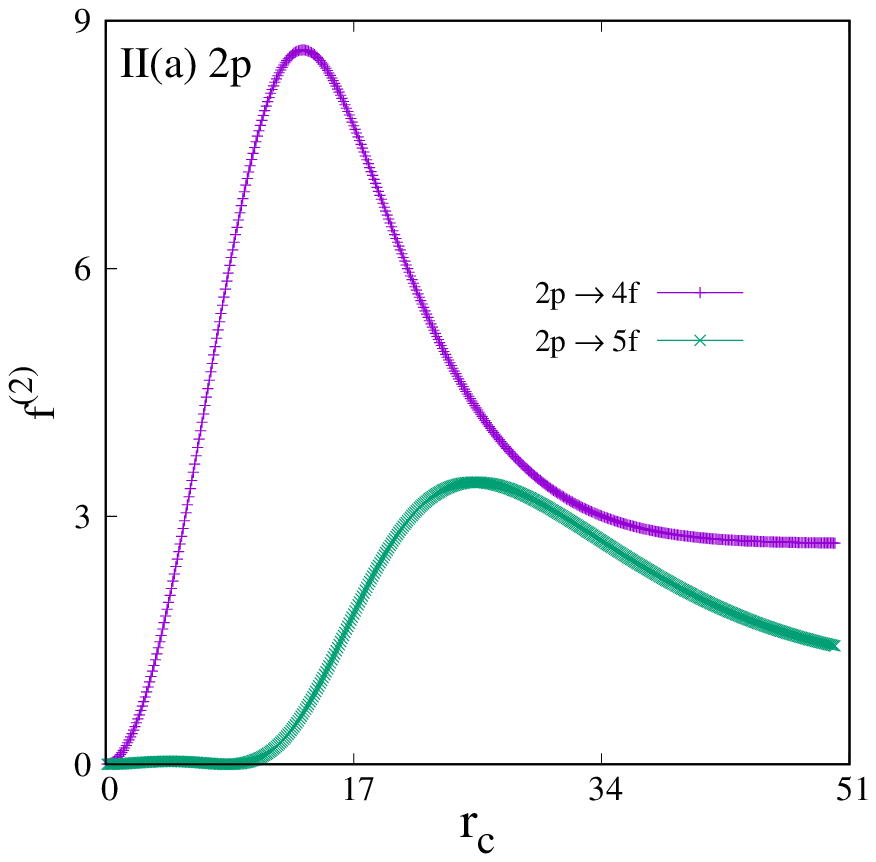}
\end{minipage}%
\begin{minipage}[c]{0.32\textwidth}\centering
\includegraphics[scale=0.53]{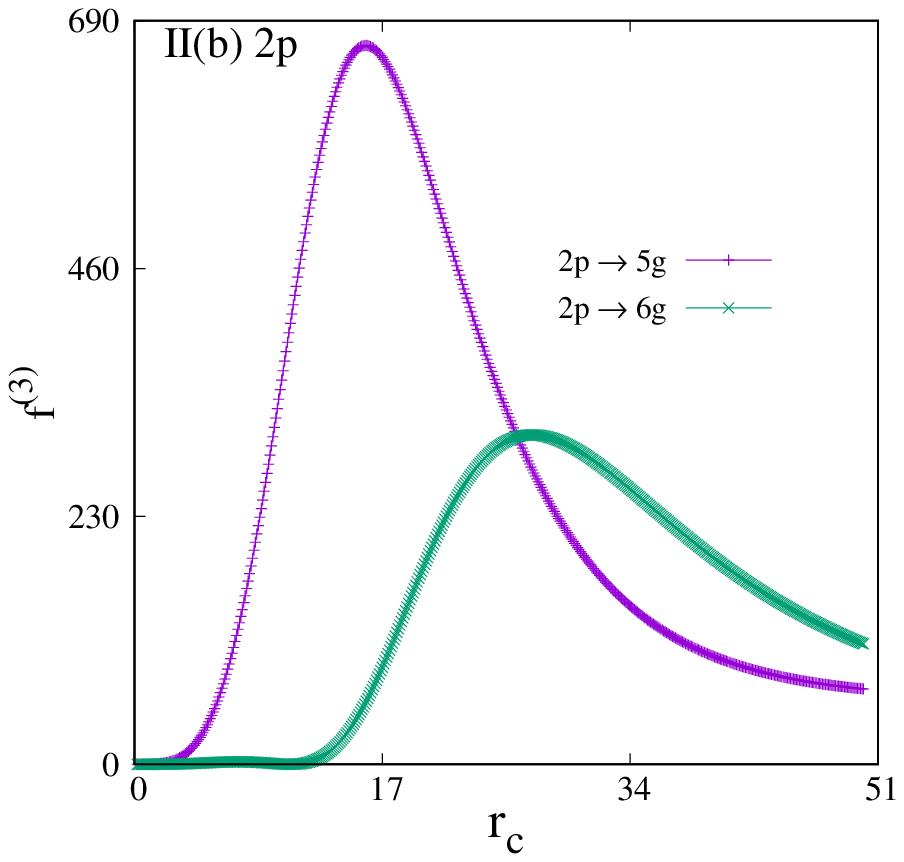}
\end{minipage}%
\begin{minipage}[c]{0.32\textwidth}\centering
\includegraphics[scale=0.53]{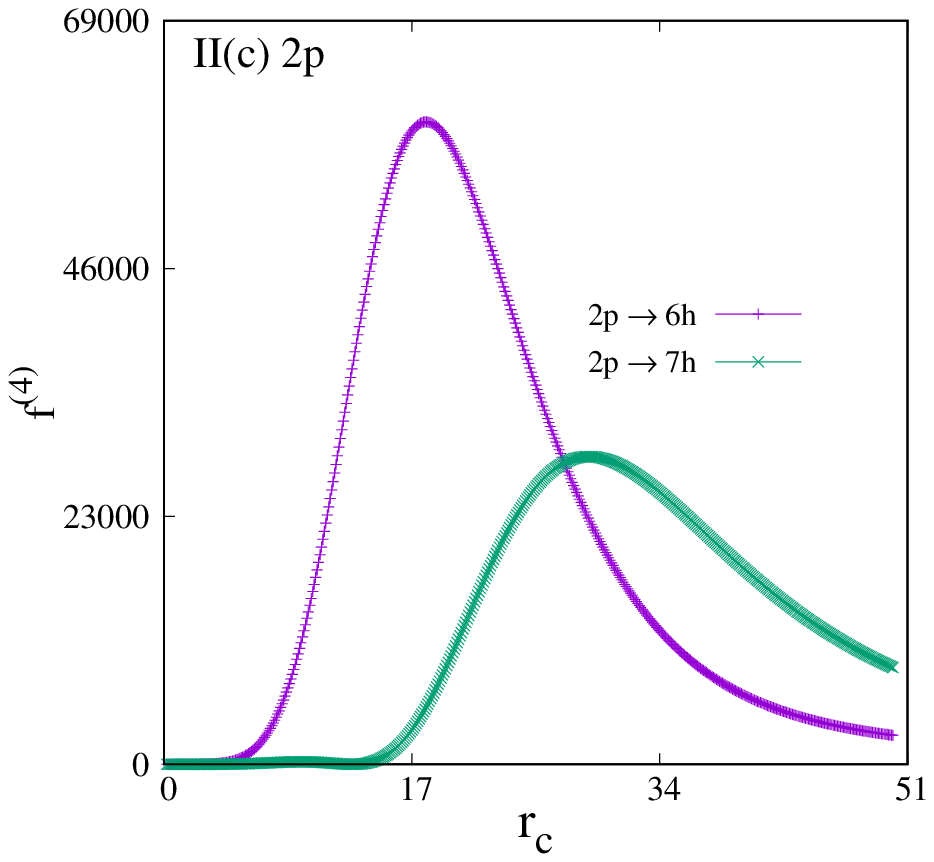}
\end{minipage}%
\vspace{2mm}
\begin{minipage}[c]{0.32\textwidth}\centering
\includegraphics[scale=0.53]{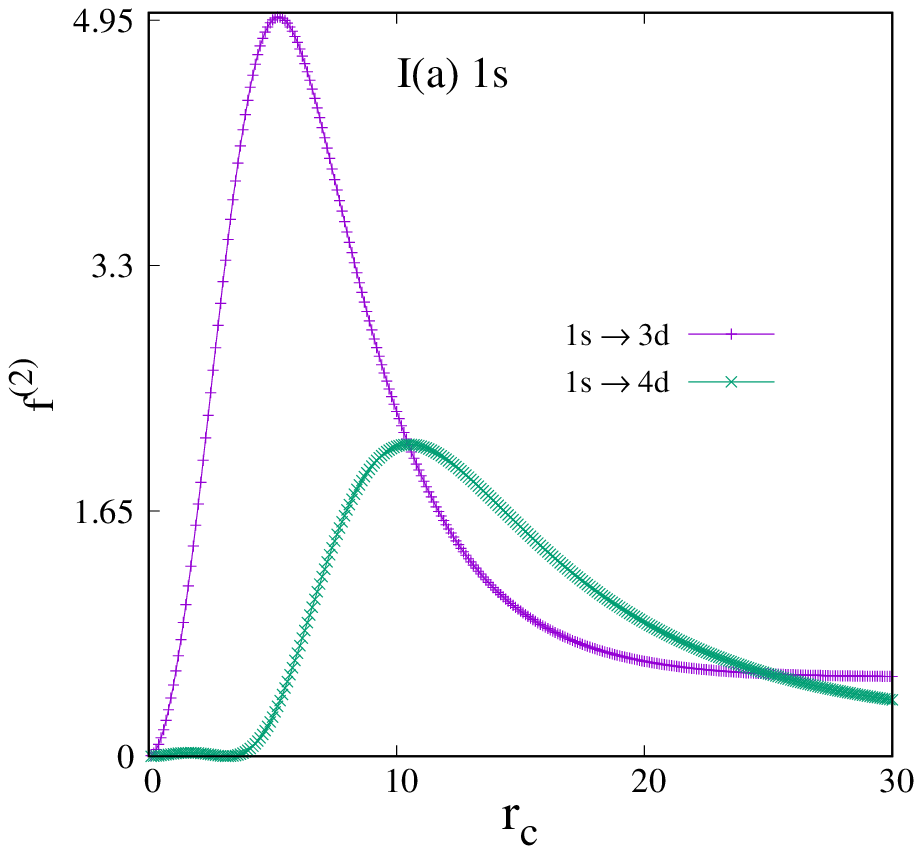}
\end{minipage}%
\begin{minipage}[c]{0.32\textwidth}\centering
\includegraphics[scale=0.53]{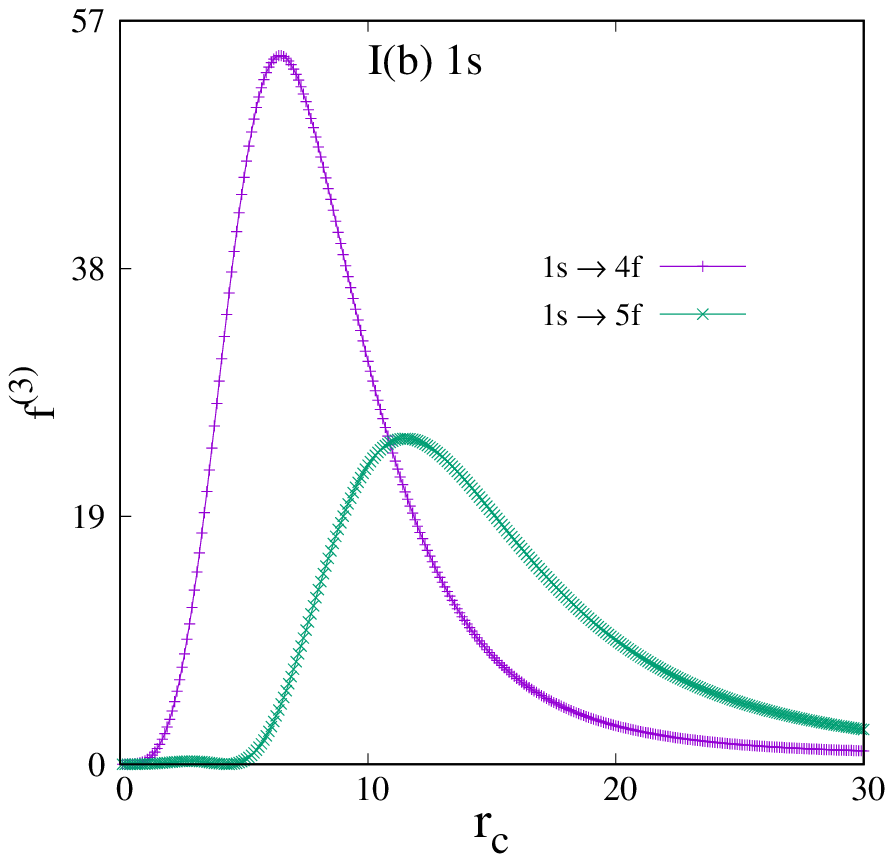}
\end{minipage}%
\begin{minipage}[c]{0.32\textwidth}\centering
\includegraphics[scale=0.53]{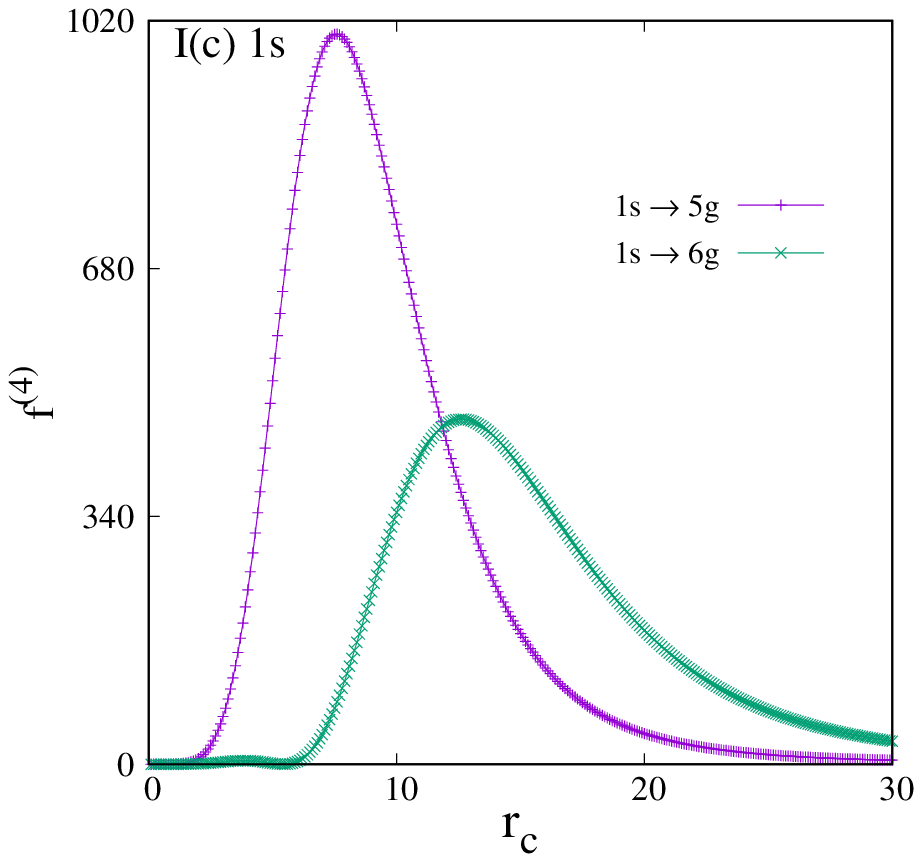}
\end{minipage}%
\caption{Plot of $f^{(k)}$ as function of $r_{c}$ (in a.u), for (I) $1s$ (II) $2p$ states in CHA. First two transitions are shown; 
(a), (b), (c) refer to $k=2,3,4$. See text for details.}
\end{figure}

Let us recall that, in contrast to FHA, the plasma models, WCP and ECSCP are \emph{devoid} of accidental degeneracy. Table~II illustrates the 
incidental degeneracy for these, taking $n=2-4$ in $s$ (or $\ell=0$) states. Like GCHA, here also $R_{a}, R_{b}$ are placed at the nodal 
positions of respective $s$ states in \emph{free} plasmas. In WCP, at $n=2$ (energy $=-0.1152930$ a.u.), a three-fold degeneracy exists with 
one confined ($2a$), one left-confined ($2b$) and one free ($2c$) WCP. This degeneracy in \emph{shell-confined} WCP, 
however, arises at $n=3$ with energy $-0.04619881$ a.u. Thus, at $n=3$, there survives six degenerate states in \emph{generalized confined} WCP, 
namely, confined $(3a, 3d)$, shell-confined $(3b)$, left-confined $(3c, 3e)$ and free $(3f)$ WCP. Further, at $n=4$ (energy $=-0.0223561$ a.u.), 
there are ten degenerate states belonging to confined $(4a, 4e, 4h)$, shell-confined $(4b, 4c, 4f)$, left-confined $(4d, 4g, 4i)$ and free 
$(4j)$ WCP respectively. Moving to ECSCP system, one encounters exactly identical pattern of degeneracy as WCP, with obvious energy 
differences between the two. This amply displays the existence of incidental degeneracy in these two plasma systems, implying that such a 
degeneracy is not necessarily limited to FHA, and may occur in other quantum systems as well. The last two columns represent 
$\alpha^{(1)}$ and $S_{r}$ successively. It may be mentioned here that, WCP and ECSCP were invoked to establish the existence of incidental 
degeneracy in {\color{red}plasma} potentials. We have not gone beyond this point to undertake an elaborate study of incidental degeneracy in 
these two {\color{red}or other} systems and may be explored in future.

\begin{figure}                         
\begin{minipage}[c]{0.32\textwidth}\centering
\includegraphics[scale=0.53]{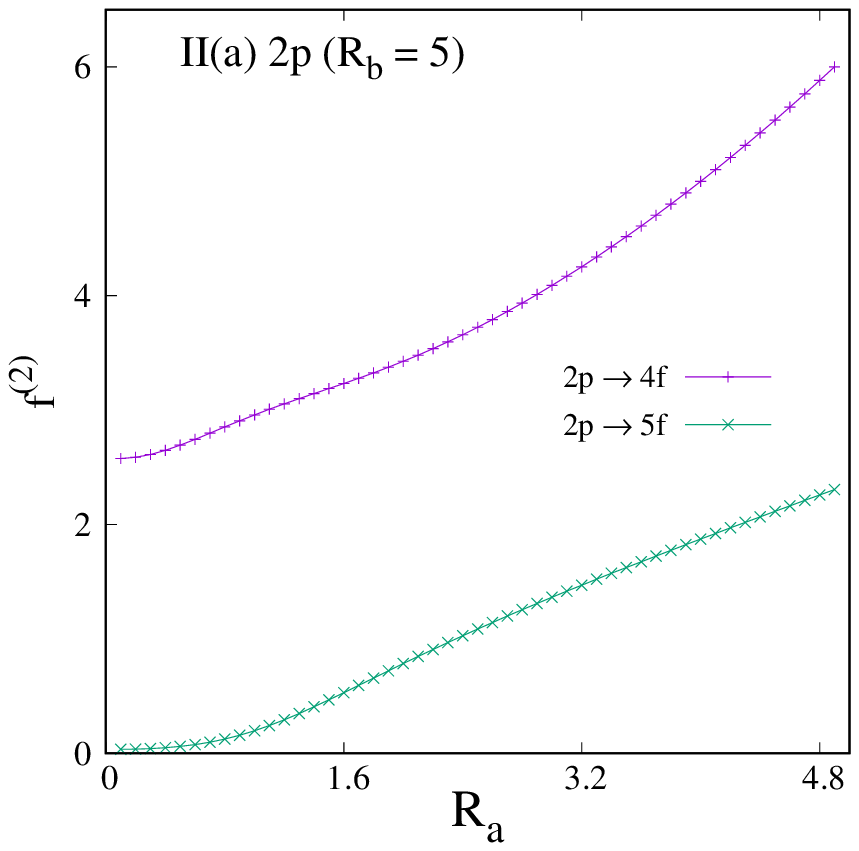}
\end{minipage}%
\begin{minipage}[c]{0.32\textwidth}\centering
\includegraphics[scale=0.53]{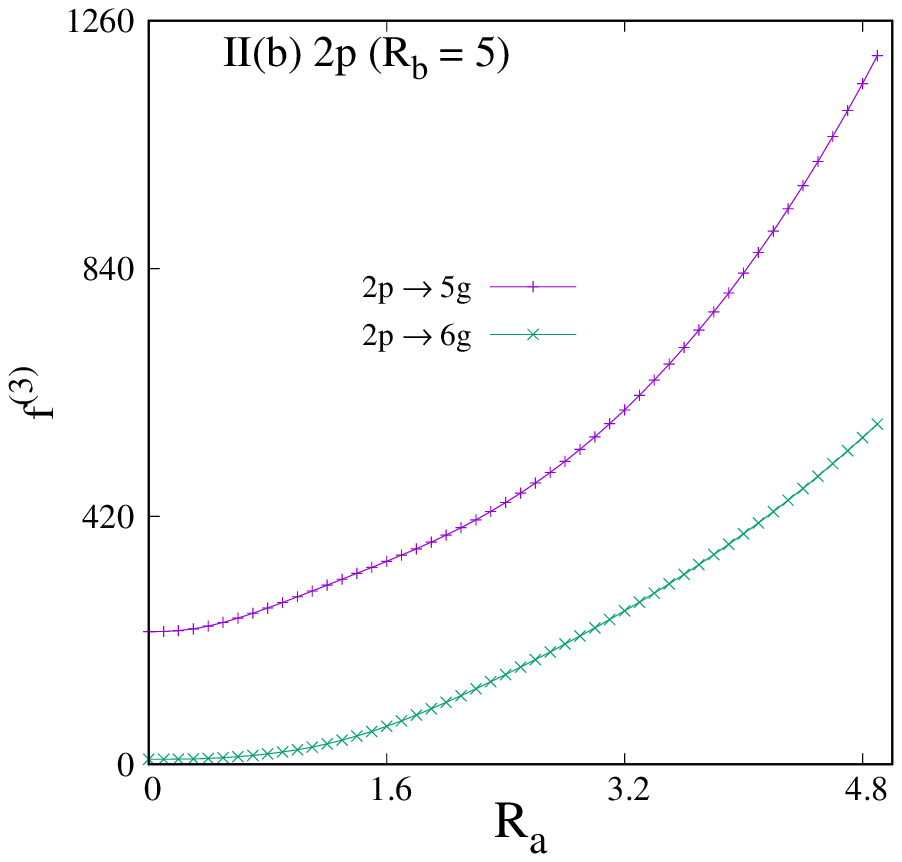}
\end{minipage}%
\begin{minipage}[c]{0.32\textwidth}\centering
\includegraphics[scale=0.53]{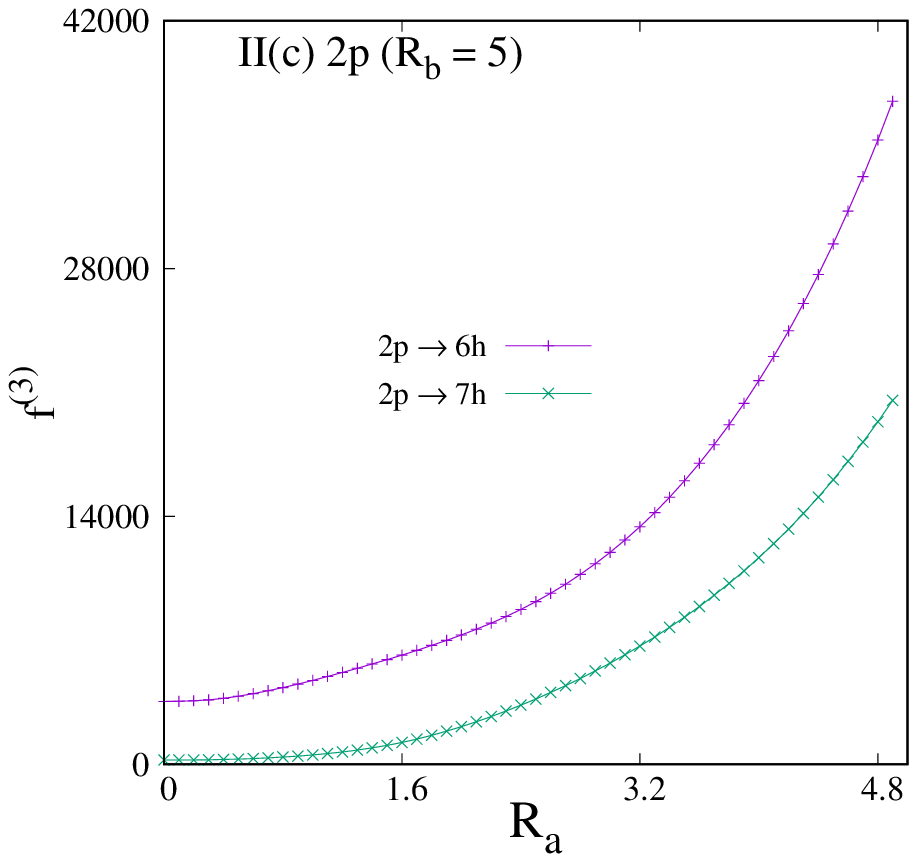}
\end{minipage}%
\vspace{2mm}
\begin{minipage}[c]{0.32\textwidth}\centering
\includegraphics[scale=0.53]{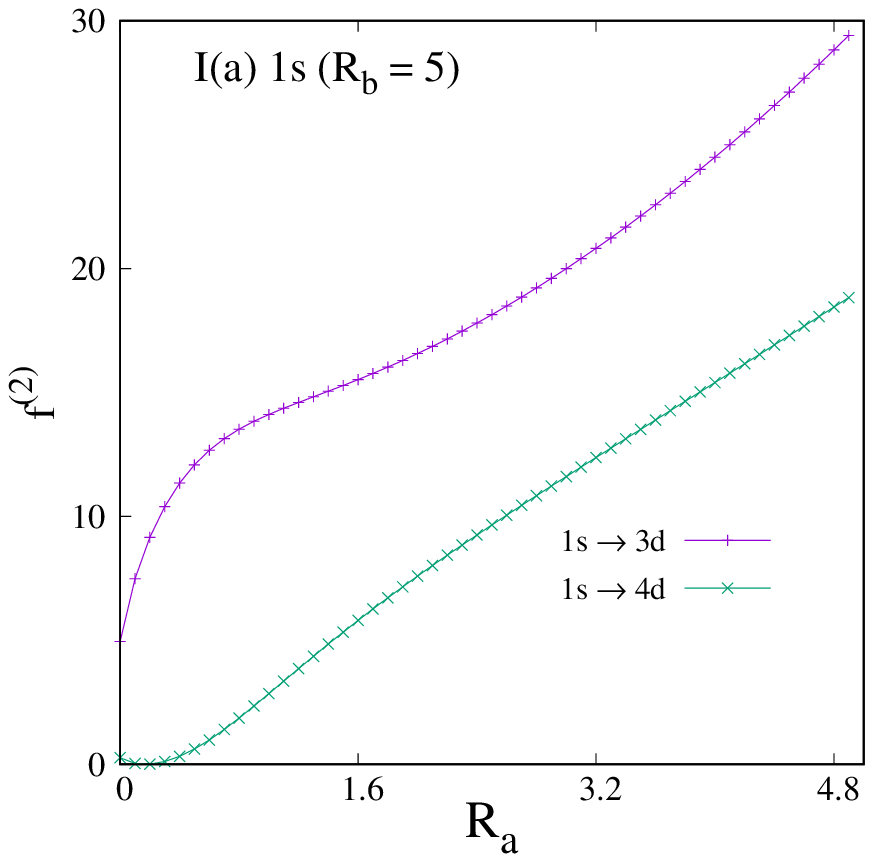}
\end{minipage}%
\begin{minipage}[c]{0.32\textwidth}\centering
\includegraphics[scale=0.53]{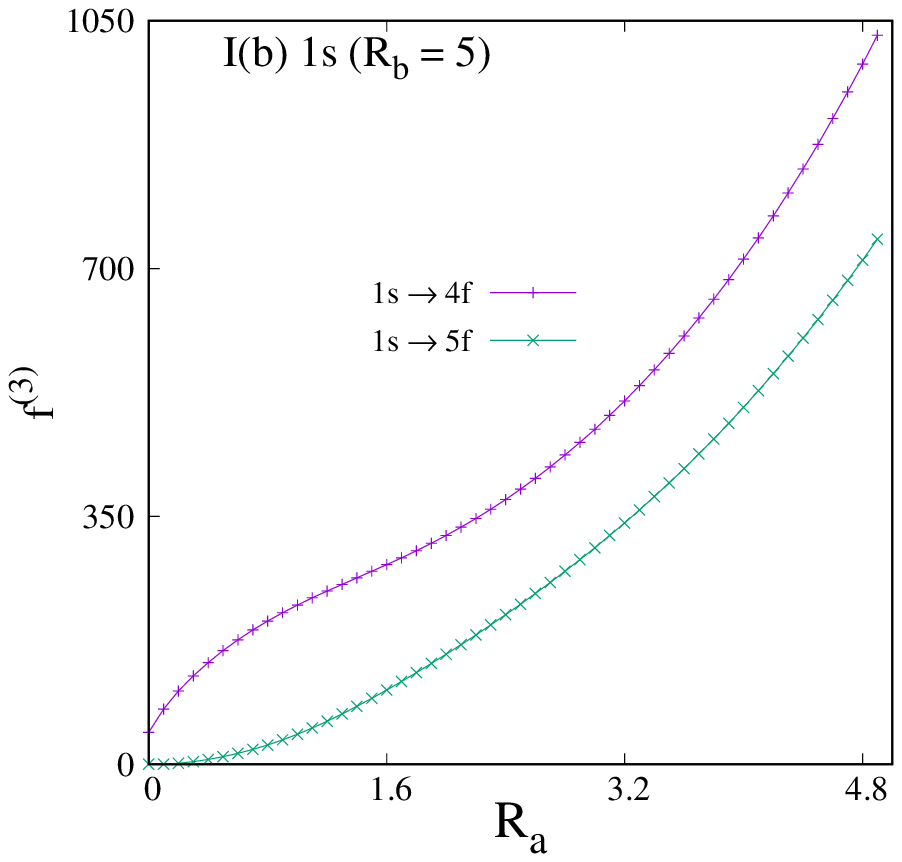}
\end{minipage}%
\begin{minipage}[c]{0.32\textwidth}\centering
\includegraphics[scale=0.53]{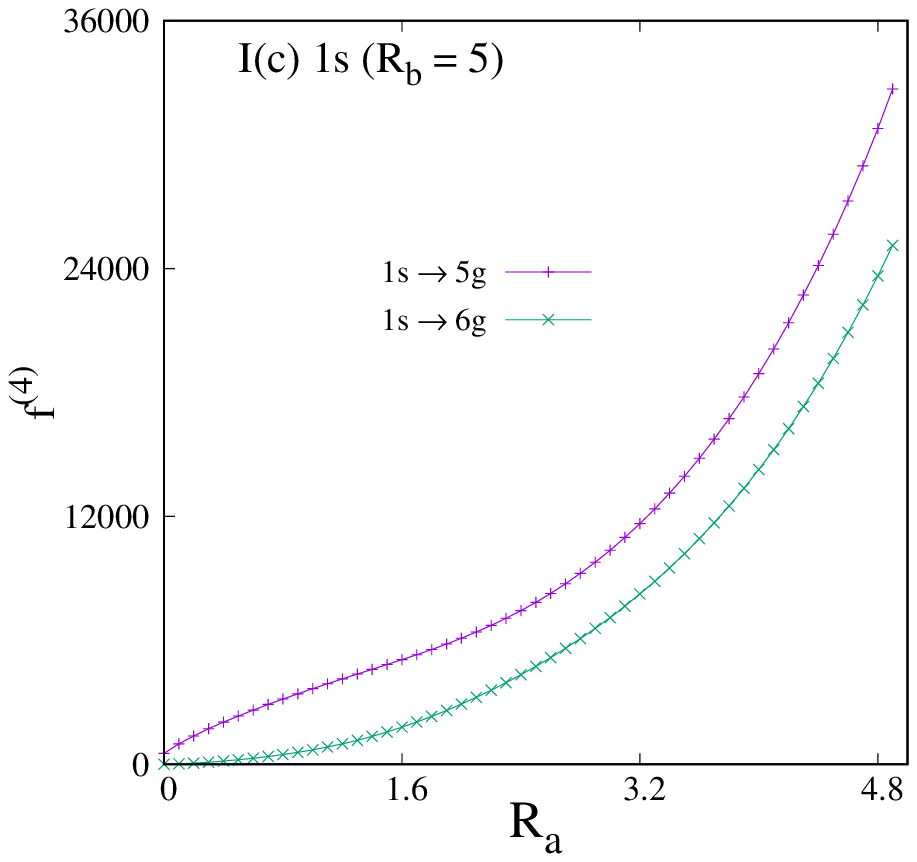}
\end{minipage}%
\caption{Plot of $f^{(k)}$ as function of $R_{a}$ (in a.u), keeping $R_{b}$ fixed at $5$, for (I) $1s$ (II) $2p$ states in SCHA. 
First two transitions are shown; (a), (b), (c) refer to $k=2,3,4$. See text for details.}
\end{figure} 

\subsection{Multipole oscillator strength and polarizability}
Unlike the previous section on energy, here we have split the discussion of $f^{(k)}$ and $\alpha^{(k)} (k=1-4)$ on confined H-like ions and 
its \emph{free} counterpart, in some low-lying states. Except $\alpha^{(1)}$ of $1s$ in SCHA, no such results are reported so far in any 
other GCHA model. Wherever possible, these are compared with available literature. As an offshoot, analytical closed-form expression of $f^{(k)}$ 
and $\alpha^{(k)}$ (considering the bound-state contribution) are {\color{red} presented in Appendix A for $k=1,2,3,4$, in case of FHA.}  

At the outset, we note that, the oscillator strength sum rule, Eq.~(\ref{eq:9}) is verified for all states in CHA, SCHA, LCHA, for $k = 1-4$. By 
definition, $f^{(k)}$ determines the probability of transition from an initial to a final state. For absorption/emission, it is $(+)$ve/$(-)$ve. 

\begin{figure}                         
\begin{minipage}[c]{0.32\textwidth}\centering
\includegraphics[scale=0.53]{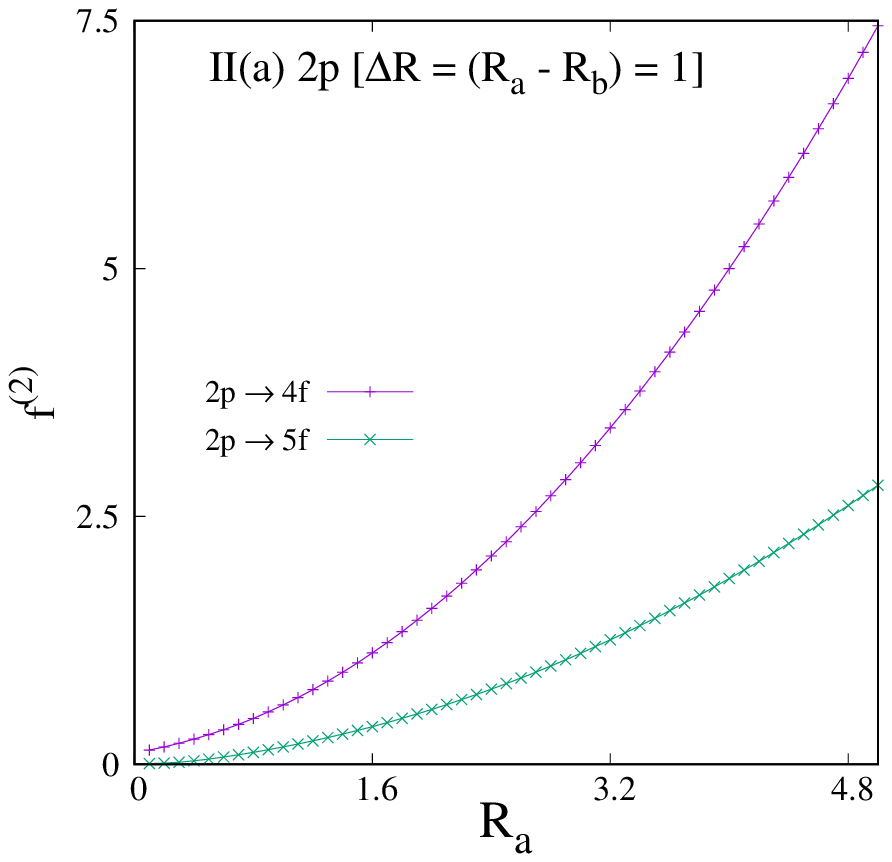}
\end{minipage}%
\begin{minipage}[c]{0.32\textwidth}\centering
\includegraphics[scale=0.53]{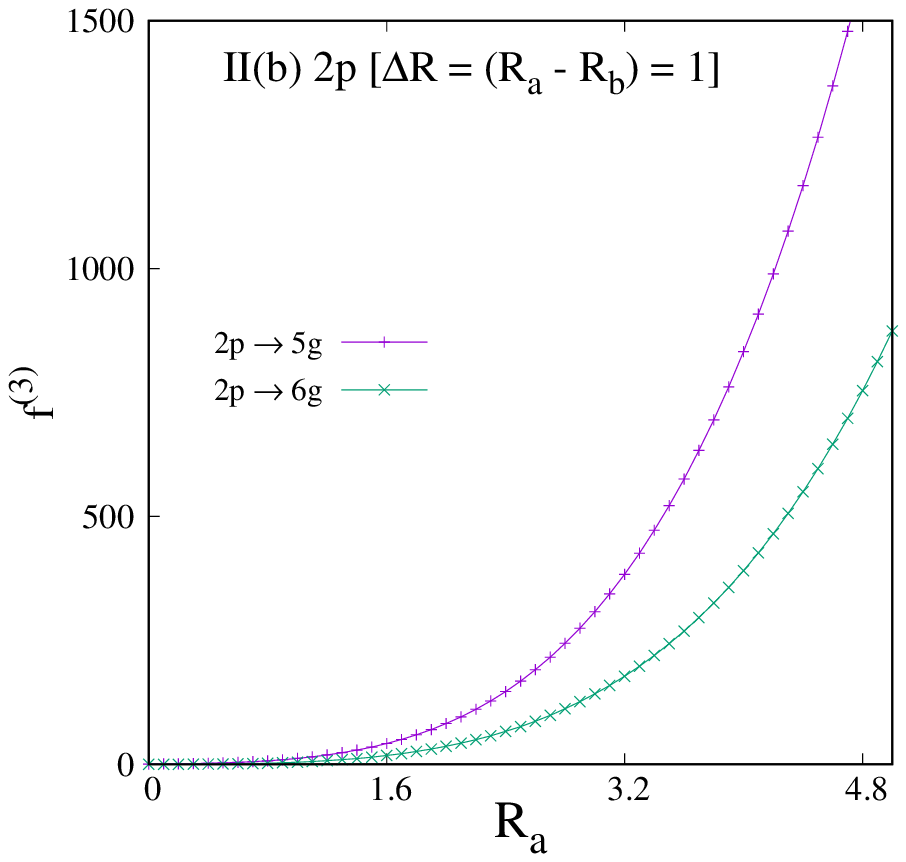}
\end{minipage}%
\begin{minipage}[c]{0.32\textwidth}\centering
\includegraphics[scale=0.53]{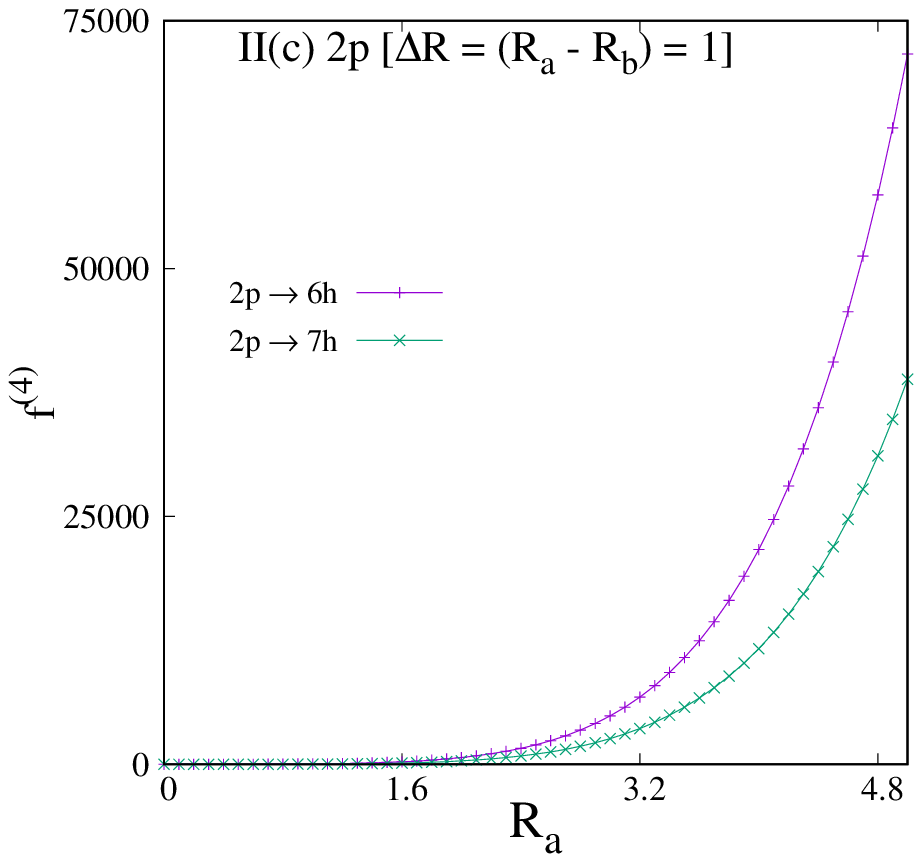}
\end{minipage}%
\vspace{2mm}
\begin{minipage}[c]{0.32\textwidth}\centering
\includegraphics[scale=0.53]{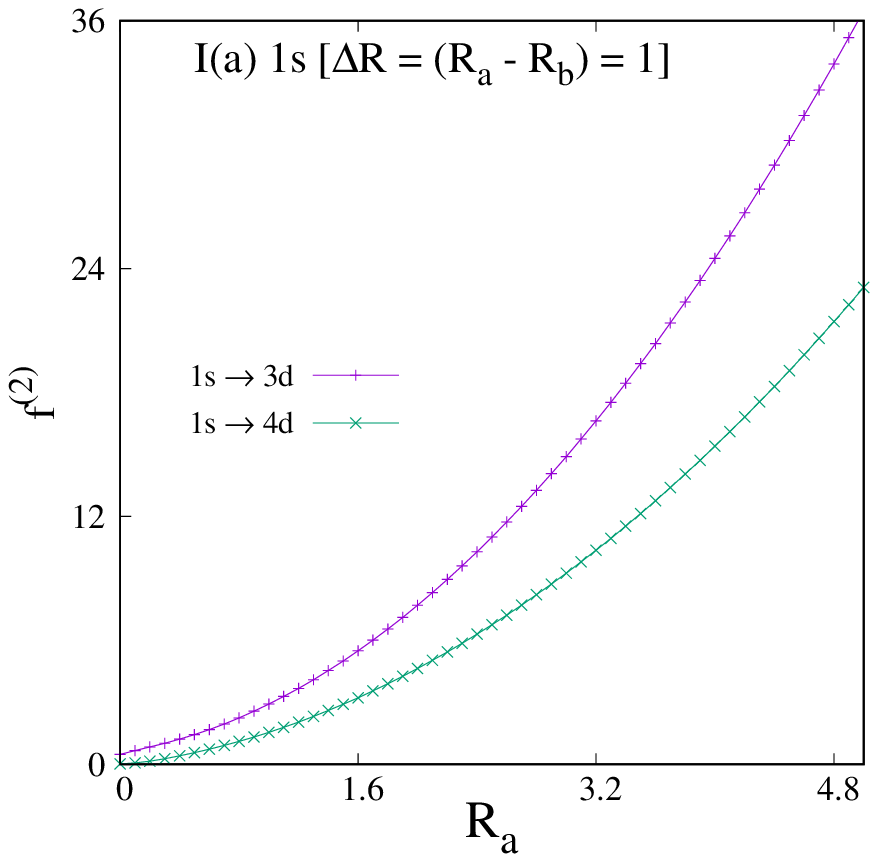}
\end{minipage}%
\begin{minipage}[c]{0.32\textwidth}\centering
\includegraphics[scale=0.53]{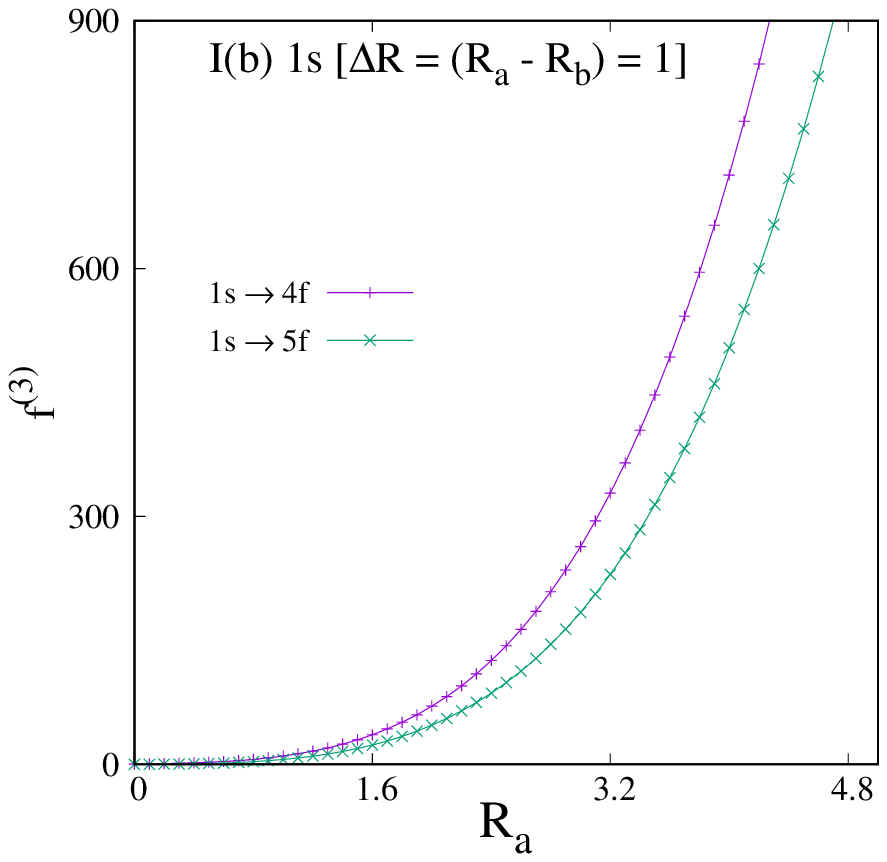}
\end{minipage}%
\begin{minipage}[c]{0.32\textwidth}\centering
\includegraphics[scale=0.53]{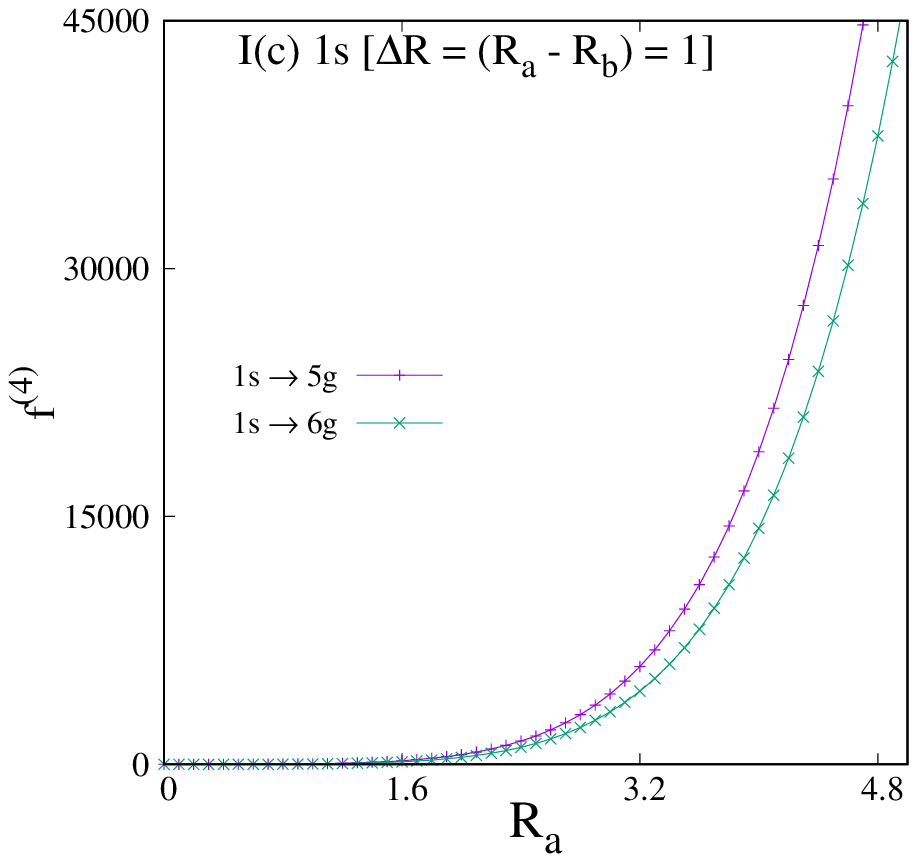}
\end{minipage}%
\caption{Plot of $f^{(k)}$ (in a.u.) as function of $R_{a}$, for $\Delta R =(R_{b}-R_{a})=1$ in (I) $1s$ (II) $2p$ states 
in SCHA. First two transitions are shown; (a), (b), (c) refer to $k=2,3,4$. See text for details.}
\end{figure} 

The selection rule for $f^{(1)}$ is $\Delta \ell= \pm 1$. Note that, from an $s$ state, transition can take place only to a $p$ state. 
However, from $p$, transition can happen to both $s$ and $d$ states. Table~III imprints calculated $f^{(1)}$ for $1s, 2s, 2p$ for 
$n, \ell \rightarrow n', (\ell+1)$ ($n=1,2; n'=2,3,4$) transitions. In this context, SCHA results are offered for four $R_{b}$, namely, 
$1,2,5,10$; for each $R_b$, $R_{a}$ lies in between $0-R_{b}$. The bottom part shows results for LCHA, having ten separate $R_{a}$ 
(including 0, leading to the special case of FHA), for $R_{b}= \infty$. Further, one recovers a CHA situation when $R_{a}=0$, while 
$r_{c}=R_{b}=1,2,5,10$. It is noticed that, $f^{(1)}_{1s \rightarrow 2p}$ in CHA increases with $r_{c}$ to attain a maximum, and then falls 
down to merge to FHA. In SCHA, for $R_{b} \leq 5$, it decreases with rise in $R_{a}$; but at $R_{b}=10$, it slowly reaches a maximum 
before finally declining. On the contrary, in LCHA, it grows with $R_{a}$ to attain a maximum, and then decays down. 
The behavior of $f^{(1)}_{1s \rightarrow 3p}$ is, however, somehow different from $f^{(1)}_{1s \rightarrow 2p}$, e.g., a reverse trend is 
recorded in case of CHA; however, at $r_{c} \rightarrow \infty$, eventually it converges to FHA. In SCHA, the pattern generated for a given 
$R_{b}$ for various $R_a$, generally differs with change in $R_b$. However, in LCHA it travels through a maximum, then a minimum and again 
rises. In case of $2s \rightarrow 2p$ transition, $f^{(1)}$ is always $(-)$ve, which implies that, except FHA (where they are degenerate), 
the former has higher energy than latter. As usual, in CHA $f^{(1)}_{2s \rightarrow 2p}$ approaches the FHA limit for $R_a \to 0, R_b \to \infty$. 
In SCHA, at $R_{b}=1$ (and 2), it enhances with $R_a$, but for $R_{b}=5$ (and 10), it reaches a minimum and then advances. A similar pattern is 
also noticed in LCHA for $R_b=5$ (and 10). In case of $2s \rightarrow 3p$ transition, $f^{(1)}$ in CHA and LCHA imprint resembling nature, 
i.e., decays after attaining a maximum. But the trend in SCHA differs from $R_{b} \leq 5$ (falls as $R_{a}$ progresses). At $R_{b}=10$, a 
reverse trend is recorded. In contrast to $1s$ and $2s$, the behavior of $f^{(1)}$ in $2p$ is not straightforward. 
Nevertheless, a few comments can be made: (i) at $r_{c} \rightarrow \infty$ limit, CHA results converge to FHA (ii) $f^{(1)}$ in SCHA 
(at $R_{b}=10$) and LCHA display analogous character. Though in respective cases this pattern alters. (iii) $f^{(1)}_{2p \rightarrow 3d}$ and 
$f^{(1)}_{2p \rightarrow 4d}$ show opposite features.           

\begin{figure}                         
\begin{minipage}[c]{0.32\textwidth}\centering
\includegraphics[scale=0.53]{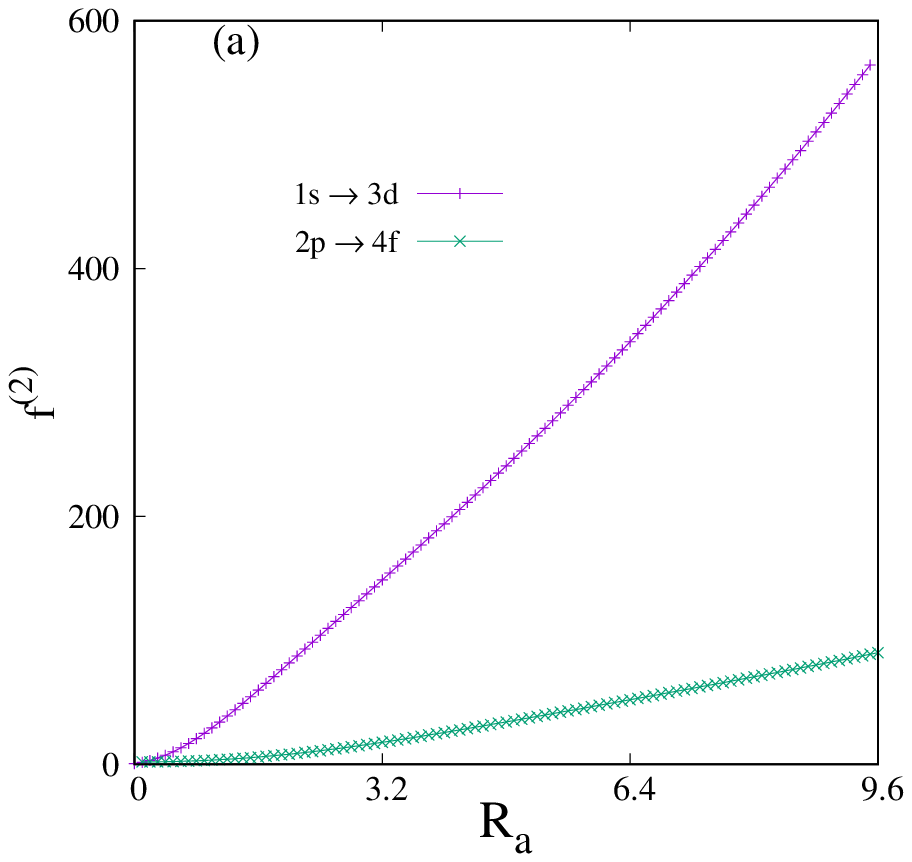}
\end{minipage}%
\begin{minipage}[c]{0.32\textwidth}\centering
\includegraphics[scale=0.53]{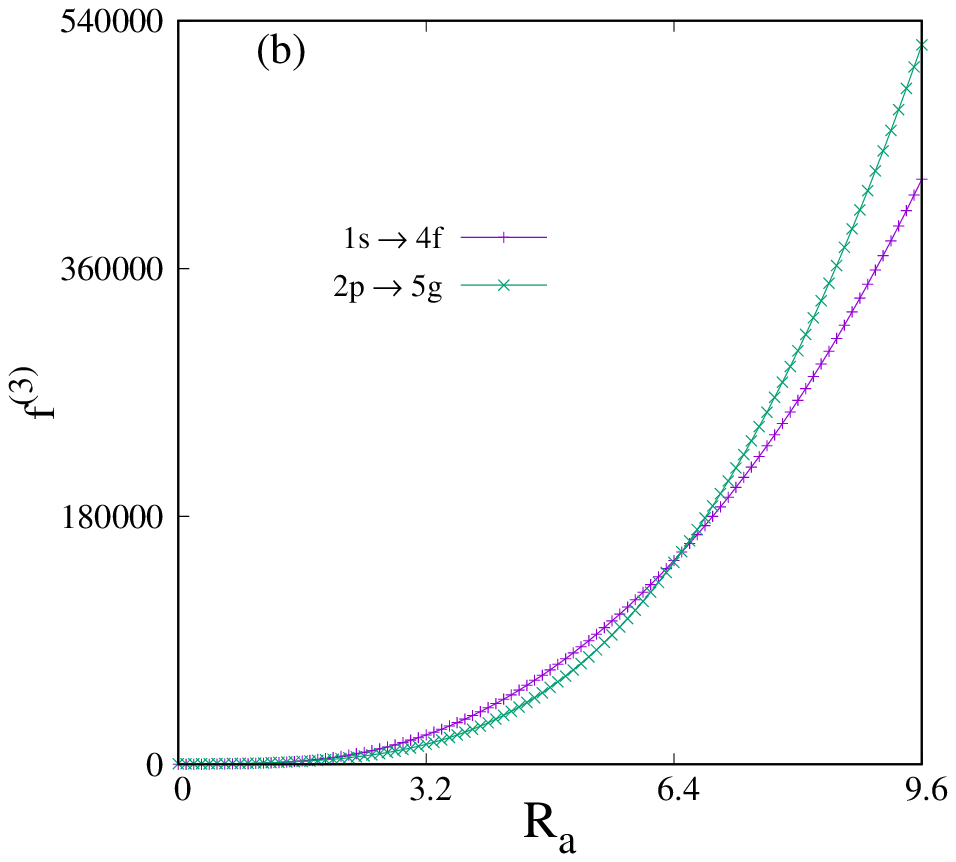}
\end{minipage}%
\begin{minipage}[c]{0.32\textwidth}\centering
\includegraphics[scale=0.53]{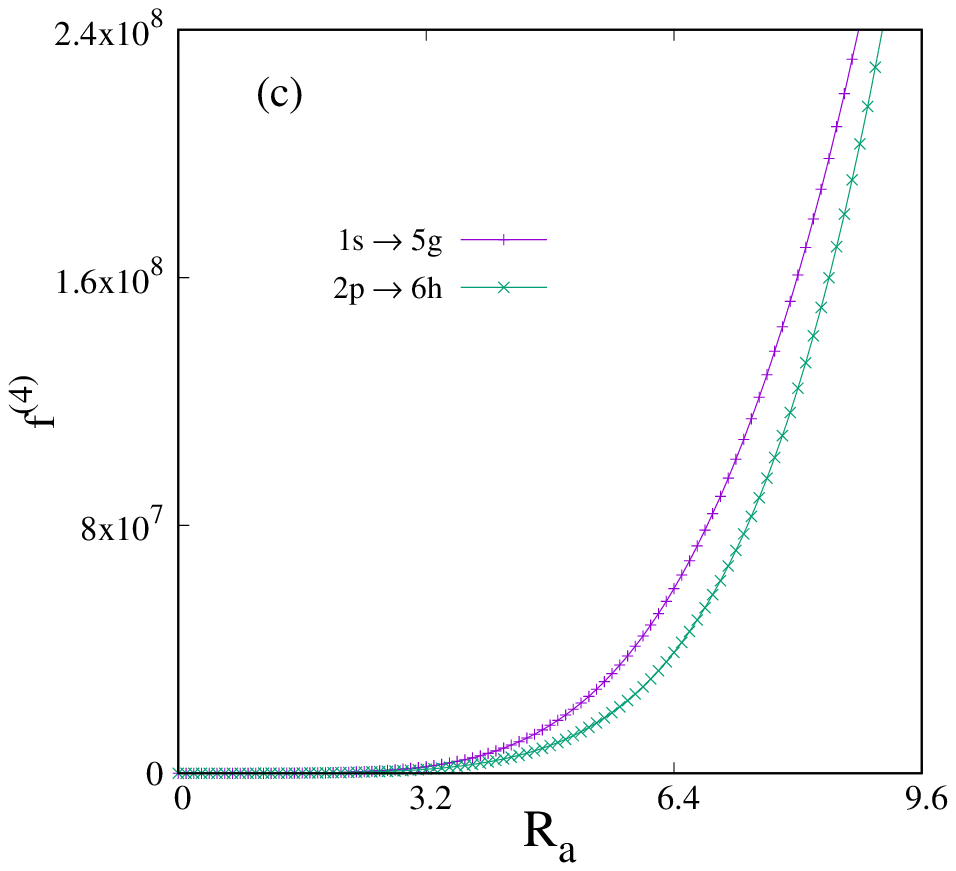}
\end{minipage}%
\caption{Plot of $f^{(k)}$ (in a.u.) as function of $R_{a}$ for $1s$ and $2p$ states in LCHA. Panels (a), (b), (c) 
refer to $k=2,3,4$. See text for details.}
\end{figure} 

Now, the focus is on higher order $f^{(k)}$, for which results are depicted graphically in case of $k=2-4$, related to \emph{quadrupole, 
octupole and hexadecapole} transitions. Corresponding selection rules are: $\Delta \ell = 0, \pm 2$, $\Delta \ell = \pm 1, \pm 3$ and 
$\Delta \ell =0 \pm 2, \pm 4$ respectively. In Fig.~2, bottom (I) and top (II) rows represent transitions from $1s$ and $2p$ states, 
for the respective maximum $\Delta \ell$ values, having $k=2,3,4$, in panels designated as (a), (b), (c). For each $k$ first two transitions 
from these two states are exhibited in terms of $f^{(k)}$ versus $r_{c}$ in CHA. In case of $1s \rightarrow (3d,4f,5g)$ and 
$2p \rightarrow (4f,5g,6h)$ transitions, respective $f^{(k)}$'s pass through a distinct maximum. But for remaining six transitions, 
\emph{viz.}, $1s \rightarrow (4d,5f,6g)$ and $2p \rightarrow (5f,6g,7h)$, there appears a shallow maximum followed by a prominent one.  
Figure~3 now exhibits variation of $f^{(2)}, f^{(3)}, f^{(4)}$ in left, middle and right panels, as function of $R_{a}$ in SCHA keeping 
$R_{b}$ stationary at $5$. The same two states generating same transitions of previous table are considered. In this instance, 
$f^{(k)}$ always advances with $R_{a}$. Figure~4 plots $f^{(k)}$ against $R_{a}$, keeping $\Delta R =(R_{b}-R_{a})$ fixed at 1, in SCHA. The 
presentation strategy is similar to that in Figs.~2 and 3. In all occasions, $f^{(k)}$'s progress with $R_{a}$. Likewise, Fig.~5 depicts 
$f^{(2)}$ for $1s \to 3d, 2p \to 4f$; $f^{(3)}$ for $1s \to 4f, 2p \to 5g$; and $f^{(3)}$ for $1s \to 5g, 2p \to 6h$, transitions in LCHA, 
in panels (a)-(c). One finds that, like the previous figure, here also $f^{(k)}$'s grow with $R_{a}$; the y-axis dramatically increases, 
as $k$ goes from 2 to 4. 

\begingroup     
\squeezetable
\begin{table}
\caption{$\alpha^{(1)}$ values for $1s,2s,3s,4s,2p,3d$ states in GCHA. See text for details.}
\centering
\begin{ruledtabular}
\begin{tabular}{ll|l|llll|ll}
$R_{a}$  & $R_{b}$ & $V$ & $\alpha^{(1)}_{1s}$ & $\alpha^{(1)}_{2s}$ & $\alpha^{(1)}_{3s}$ & $\alpha^{(1)}_{4s}$ & $\alpha^{(1)}_{2p}$ &
$\alpha^{(1)}_{3d}$  \\
\hline
0    & 1 & 1     & 0.02879202$^{\dag}$  & 0.00441401     & 0.00188747 & 0.00105362  & 0.01715126 & 0.00894345  \\
0.1  & 1 & 0.999 & 0.04759422$^{\dag}$  & 0.02720296     & 0.02357306 & 0.02219162  & 0.01004334 & 0.00815930  \\
0.2  & 1 & 0.992 & 0.07284697$^{\dag}$  & 0.05487311     & 0.05124972 & 0.04988689  & 0.00583028 & 0.00565908  \\
0.5  & 1 & 0.875 & 0.20188347$^{\dag}$  & 0.19133640     & 0.18923265 & 0.18848057  & 0.00083203 & 0.00083480  \\
0.8  & 1 & 0.488 & 0.43528355           & 0.43282926     & 0.43236928 & 0.43220786  & 0.00002109 & 0.00002109  \\
\hline
0    & 2 & 8     & 0.34255811$^{\ddag}$ & $-$0.0168850   & $-$0.00436115 & $-$0.00163080 & 0.30828166 & 0.15185462 \\
0.1  & 2 & 7.999 & 0.51258523$^{\ddag}$ & 0.22332074     & 0.19042816 & 0.17549582  & 0.21885819 & 0.149192555 \\
0.5  & 2 & 7.875 & 1.37743250$^{\ddag}$ & 1.14984366     & 1.09461376 & 1.07278640  & 0.07037524 & 0.07079627 \\
1    & 2 & 7     & 3.22129727$^{\ddag}$ & 3.0635182      & 3.02914678 & 3.01659244  & 0.01330690 & 0.01335487 \\
1.2  & 2 & 6.272 & 4.25881264           & 4.1404792      & 4.11614071 & 4.10740597  & 0.00542262 & 0.00542936 \\
1.5  & 2 & 4.625 & 6.20198032           & 6.1449042      & 6.13390400 & 6.13002002  & 0.00082430 & 0.00082441 \\
1.8  & 2 & 2.168 & 8.67861281           & 8.6676719      & 8.66563354 & 8.66491917  & 0.00002106 & 0.00002108 \\
\hline
0    & 5 & 125     & 3.42245422         & $-$21.10657309 & $-$5.69164044 & $-$2.65407833 & 18.08924616 & 7.21196971 \\
1    & 5 & 124     & 38.3097689         & 34.92837261    & 32.75186815   & 31.71727294   & 3.57857074  & 3.72434520 \\
2    & 5 & 117     & 90.3835855         & 85.34545419    & 83.48828979   & 82.73474712   & 1.08353807  & 1.09660163 \\
2.5  & 5 & 109.375 & 124.798441         & 119.92301897   & 118.49342276  & 117.94158361  & 0.51897086  & 0.52129425 \\
3    & 5 & 98      & 165.908991         & 161.85822945   & 160.86291102  & 160.49304359  & 0.21171092  & 0.21200753 \\
4    & 5 & 61      & 272.010054         & 270.53059396   & 270.23821589  & 270.13447254  & 0.01317982  & 0.01318049 \\
4.5  & 5 & 33.875  & 339.006228         & 338.58155287   & 338.50168038  & 338.47363247  & 0.0008232   & 0.00082310 \\
\hline
0    & 10 & 1000    & 4.49681419$^{\S}$  & 37.23973625   & $-$376.86905909 & $-$143.24860953 & 793.3231266 & 171.8366872 \\
0.5  & 10 & 999.875 & 57.8605712  & 87.17950256          & 107.31693132    & 110.54059768    & 107.3856268 & 152.9035064 \\
1    & 10 & 999     & 163.642519  & 277.6753968          & 254.03479328    & 238.63416165    & 80.69486982 & 113.7656686 \\
2    & 10 & 992     & 485.283240$^{\S}$  & 583.9858189   & 540.82756495    & 518.84632115    & 52.316849   & 60.8621082  \\
3    & 10 & 973     & 900.888347$^{\S}$  & 936.7193051   & 895.52440838    & 876.03156272    & 31.4884761  & 33.2715571  \\
5    & 10 & 875     & 1969.35287$^{\S}$  & 1925.755087   & 1900.38211125   & 1889.89454238   & 8.2726984   & 8.32191160  \\
7    & 10 & 657     & 3430.37098$^{\S}$  & 3390.605514   & 3380.46513160   & 3376.69265692   & 1.0688590   & 1.06926410  \\
9.5  & 10 & 142.625 & 6023.026447        & 6021.229170   & 6020.89385098   & 6020.77626011   & 0.0008310   & 0.000836    \\
\hline
0    & $\infty$ & $-$ & 4.50000000$^\pm$ & 120.0000000$^\pm$ & 1012.5000000$^\pm$ & 4992.0000000$^\pm$ & 176.0000000$^\pm$ & 1863.0000000$^\pm$ \\
0.1  & $\infty$ & $-$ & 11.0436170 & $-$832.82391        & $-$19470.0124  & $-$157741.0999 & 551.13773744  & 1862.6259940 \\
0.5  & $\infty$ & $-$ & 62.0058551 & 1191.23431          & 6853.5900      & 20163.4888     & 300.6834286   & 24243.2503  \\
1    & $\infty$ & $-$ & 206.890953 & 3978.47424          & 29050.9501     & 131865.02      & 311.4948323   & 7601.124077 \\
2    & $\infty$ & $-$ & 928.385427 & 13601.4303          & 89661.9116     & 389708.7       & 466.836635    & 3831.84946 \\
5    & $\infty$ & $-$ & 10180.1611 & 90653.1600          & 458038.2       & 1685829.4      & 1425.94246    & 3701.13787 \\
7    & $\infty$ & $-$ & 27077.8811 & 199530.231          & 900453.4       & $-$             & 2448.04066   & 4859.03373 \\
8    & $\infty$ & $-$ & 40427.5127 & 276691.870          & $-$            & $-$             & 3080.4164    & 5613.81181 \\
9    & $\infty$ & $-$ & 57889.5118 & 371559.069          & 1538454.7      & $-$             & 3797.0113    & 6473.34645 \\
10   & $\infty$ & $-$ & 80147.5187 & 486116.385          & 1941232.3      & $-$             & 4600.2812    & 7434.64009 \\
\end{tabular}
\end{ruledtabular}
\begin{tabbing}
$^\dag$Literature results \cite{sen02} for $\alpha^{(1)}_{1s}$ $(R_{b}=1)$ at $R_{a}=$ $0$, $0.1$, $0.2$, $0.5$ are: $0.0284$, $0.0473$, $0.0716$, $0.2000$. \\
$^\ddag$Literature results \cite{sen02} for $\alpha^{(1)}_{1s}$ $(R_{b}=2)$ at $R_{a}=$ $0$, $0.1$, $0.5$, $1.0$ are: $0.3405$, $0.5095$, $1.3588$, $3.2041$. \\
$^\S$Literature results \cite{sen02} for $\alpha^{(1)}_{1s}$ $(R_{b}=10)$ at $R_{a}=$ $0$, $2$, $3$, $5$, $7$ are: $4.4851$, $474.3865$, $880.0750$, $1962.3385$, $3394.0953$. \\
$^\pm$Literature results \cite{baye12} for $\alpha^{(1)}_{1s}$, $\alpha^{(1)}_{2s}$, $\alpha^{(1)}_{3s}$, $\alpha^{(1)}_{4s}$, $\alpha^{(1)}_{2p}$, $\alpha^{(1)}_{3d}$ in FHA are:
$4.5$, $120$, $1012.5$, $4992$, $176$, $1863$. 
\end{tabbing}
\end{table}  
\endgroup  

Now we move to investigate $\alpha^{(1)}$ in GCHA by means of sample calculations on $1s,2s,2p,3s,3d,4s$ states. For $p$ and $d$ states,  
allowed transitions occur to final states having $\ell$ as $(0,2)$ and $(1,3)$ respectively. The collected results in Table~IV include 
contributions from both $\ell$, for same numerical values of $R_a, R_{b}$ of previous table. The third column provides the volume of 
ring ($V=R_{b}^{3}-R_{a}^{3}$) having inner and outer radii $R_{a},R_{b}$. Note that, $(R_{a},R_{b})= (0,1), (0,2), (0,5), (0,10)$ 
represent CHA cases. LCHA results are tabulated in the bottom segment, while its first row correspond to a FHA. Some of these for CHA and 
SCHA were reported in \cite{sen02}, which are duly quoted in footnote. Our calculated $\alpha^{(1)}$ values show excellent agreement 
with these. A careful analysis of this table uncovers several interesting features, some of which are enumerated below: 

\begin{enumerate}
\item 
\textbf{CHA:} In FHA, $\alpha^{(1)}$ is a $(+)$ve quantity. At a given $\ell$, it rises in $n$, while, at a fixed $n$, it progresses with 
$\ell$. However, in CHA, the pattern behavior is not so consistent, recording distinct changes with $r_{c}$, offering both $(+)$ve and $(-)$ve 
values. A straightforward inference is that, with growth in $r_{c}$, $\alpha^{(1)}$ in a given state with an arbitrary $\ell$ progresses 
as $r_{c}$ proceeds towards the respective FHA limit. At $r_{c}=2,5$, $2s,3s,4s$ states offer $(-)$ve polarizability; the same is also found 
for latter two states at $r_c=10$. 

\item
\textbf{SCHA:} In this situation, characteristics of $\alpha^{(1)}$ changes with $\ell$. For $s$-wave states, (at a fixed $R_{b}$) it 
increases with $R_{a}$. A resembling nature in $\alpha^{(1)}$ is also achieved by varying $R_{b}$ keeping $R_{a}$ fixed. These two outcomes 
suggest that, it depends on $(R_{a}, R_{b})$ pair, but not on their difference, $\Delta R$. On the contrary, for $\ell \neq 0$ states, 
at a specific $R_{b}$, it abates with $R_{a}$, but for a given $R_{a}$, it advances with $R_{b}$. Thus, in this scenario, $\alpha^{(1)}$ is 
controlled by all three quantities, $R_{a},R_{b}, \Delta R$.    

\item
\textbf{LCHA:} In $\ell=0$ states, it grows with $R_{a}$, but a \emph{zigzag} pattern is seen for $2p$ and $3d$.
\end{enumerate}
   
It is observed from Table~IV that, for s-waves, at a fixed $R_{b}$, $\alpha^{(1)}$ progresses with $R_{a}$. However, after some 
characteristic $R_a$, it prevails over volume, given in third column. According to \emph{Herzfeld criterion} \cite{harzfeld27,sen02} insulator 
$\rightarrow$ metal conversion occurs under the condition, 
\begin{equation}\label{eq:18}
\frac{4 \pi}{3}V \le \left(\frac{4 \pi}{3}\right) \alpha^{(1)},  \ \ \ \ \ V = \left(R_{b}^{3}-R_{a}^{3}\right) \le \alpha^{(1)}. 
\end{equation}
Now, applying the above criterion, one can easily discern that, in SCHA metallic character can be observed in all the $s$ states. This feature  
was reported before \cite{sen02} in the \emph{ground state} of SCHA. This work, however, shows that, it can be extended to excited states as well 
having $\ell =0$. The threshold $R_{a}$ at which $\alpha^{(1)}$ surpasses $V$ (symbolized as $R_{m}$), is produced in Table~V, for four $s$ states, 
at ten selected $R_{b}$ ($1,2,3,4,5,6,7,8,9,10$). For a given $\alpha^{(1)}$, $R_{m}$ tends to assume larger values with growth of $R_{b}$. Moreover, 
with an enhancement of $R_{b}$, the metallic zone ($R_{b}-R_{m}$) extends. The lone reference results are quoted in the footnote, which shows 
decent qualitative agreement. 

Now we present a cross section of results on $2^{k}$-pole polarizabilities ($k=2,3,4$) of $1s$ state in CHA, SCHA and LCHA graphically in Fig.~6. 
The left (a), middle (b) and right (c) panels represent these for quadrupole, octupole and hexadecapole polarizabilities. The bottom row (I) for CHA 
indicates, for all $k$, $\alpha^{(k)}$ sharply increases with $r_c$, before finally merging to the respective FHA limit. Parallel results for SCHA
against $R_a$ are depicted in panels marked (II), while keeping $R_{b}$ fixed at $5$--reflecting a steady monotonic growth. Similarly panels III(a), 
III(b), III(c) offer SCHA results varying both $R_{a},R_{b}$, maintaining a constant $\Delta R=1$. However here also the monotonic increasing trend 
is maintained for all $k$, as in previous SCHA situation. At last, the top-most panels IV(a), IV(b) and IV(c) show the respective plots in case of 
LCHA, bearing a close resemblance to SCHA scenario. 

\begingroup        
\squeezetable
\begin{table}
\caption{Estimated $R_{a}=R_{m}$ at ten $R_{b}$, for $1s,2s,3s,4s$ states in SCHA. See text for details.}
\centering
\begin{ruledtabular}
\begin{tabular}{l|llll}
  &   \multicolumn{4}{c}{$R_{a}=R_{m}$}        \\
\hline 
$R_{b}$ & $1s$ & $2s$ & $3s$ & $4s$  \\
\hline
1    & 0.81776350            & 0.81844574  & 0.818572206  & 0.818616480 \\
2    & 1.380547989$^{\dag}$  & 1.386963908 & 1.388196667  & 1.388631630 \\
3    & 1.78705451$^{\dag}$  & 1.806803958 & 1.810946733  & 1.812435703 \\
4    & 2.091846379$^{\dag}$  & 2.130115046 & 2.139377180  & 2.1427985178 \\
5    & 2.329536517$^{\dag}$  & 2.385985164 & 2.40275065   & 2.4091529423 \\
6    & 2.524110980$^{\dag}$  & 2.591789087 & 2.618502823  & 2.629078449  \\
7    & 2.692788378$^{\dag}$  & 2.758505970 & 2.7976385904 & 2.813701635  \\
8    & 2.847829929$^{\dag}$  & 2.893365851 & 2.9474474821 & 2.9704167040 \\
9    & 2.997651615$^{\dag}$  & 3.00123877  & 3.0729537329 & 3.104343080  \\
10   & 3.147719215$^{\dag}$  & 3.08540221  & 3.177732377  & 3.219149730  \\
\end{tabular}
\end{ruledtabular}
\begin{tabbing}
$^{\dag}$For $R_{b}= 2,3,4,5,6,7,8,9,10$, literature results \cite{sen02} of $R_m$ are: $1.73,2.08,2.34,2.54,2.71,2.87,3.01,3.15,3.29$ respectively. 
\end{tabbing}
\end{table}  
\endgroup  

\begin{figure}                         
\begin{minipage}[c]{0.32\textwidth}\centering
\includegraphics[scale=0.53]{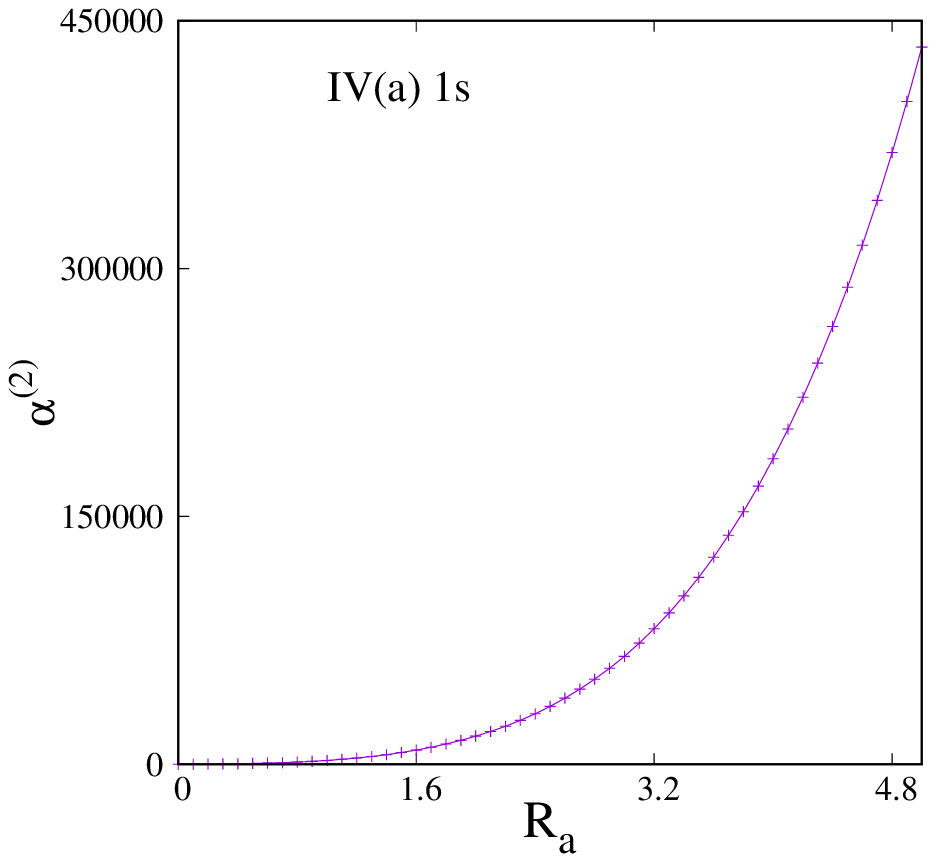}
\end{minipage}%
\begin{minipage}[c]{0.32\textwidth}\centering
\includegraphics[scale=0.53]{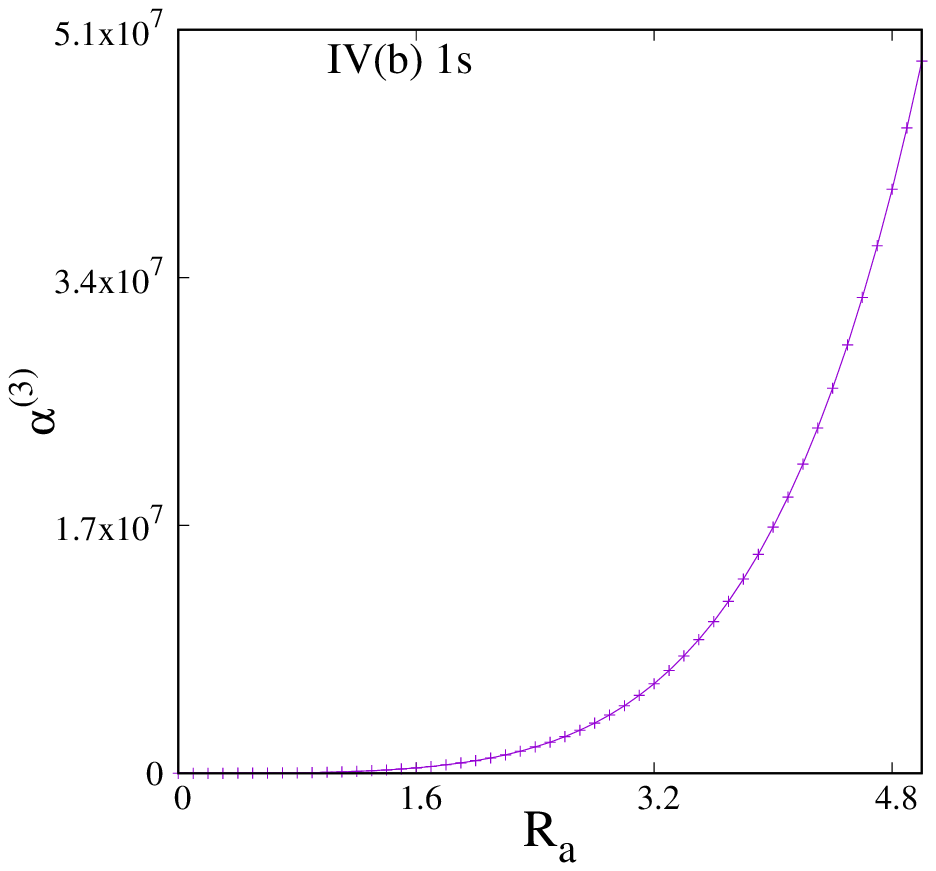}
\end{minipage}%
\begin{minipage}[c]{0.32\textwidth}\centering
\includegraphics[scale=0.53]{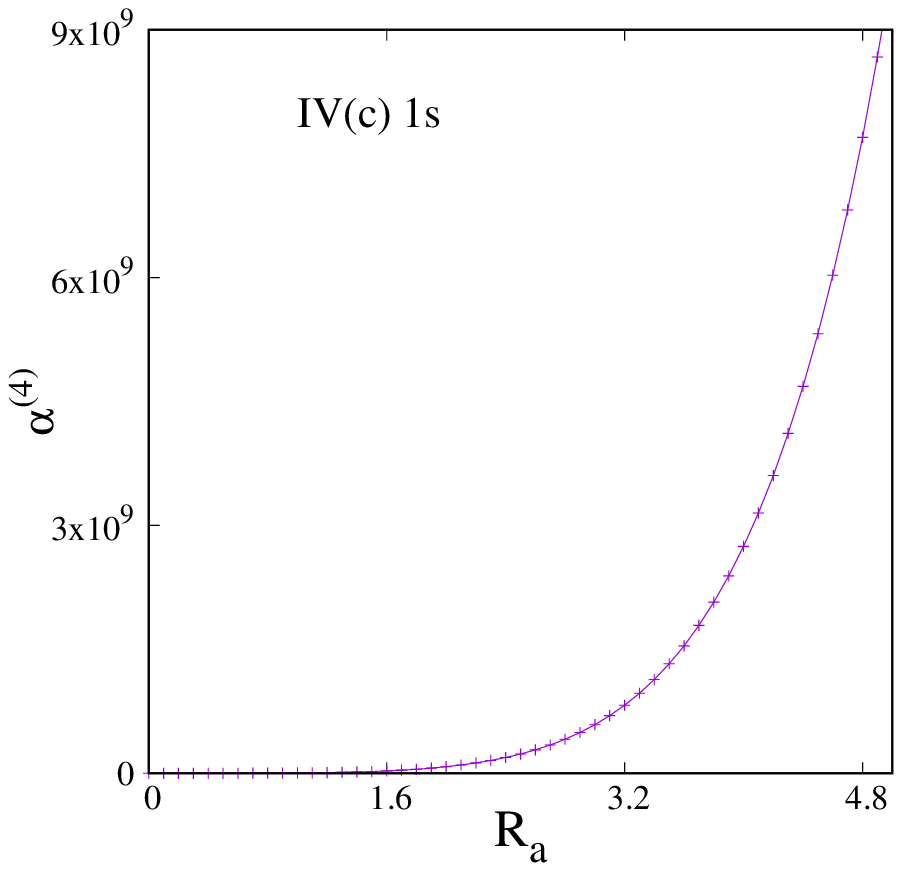}
\end{minipage}%
\vspace{2mm}
\begin{minipage}[c]{0.32\textwidth}\centering
\includegraphics[scale=0.53]{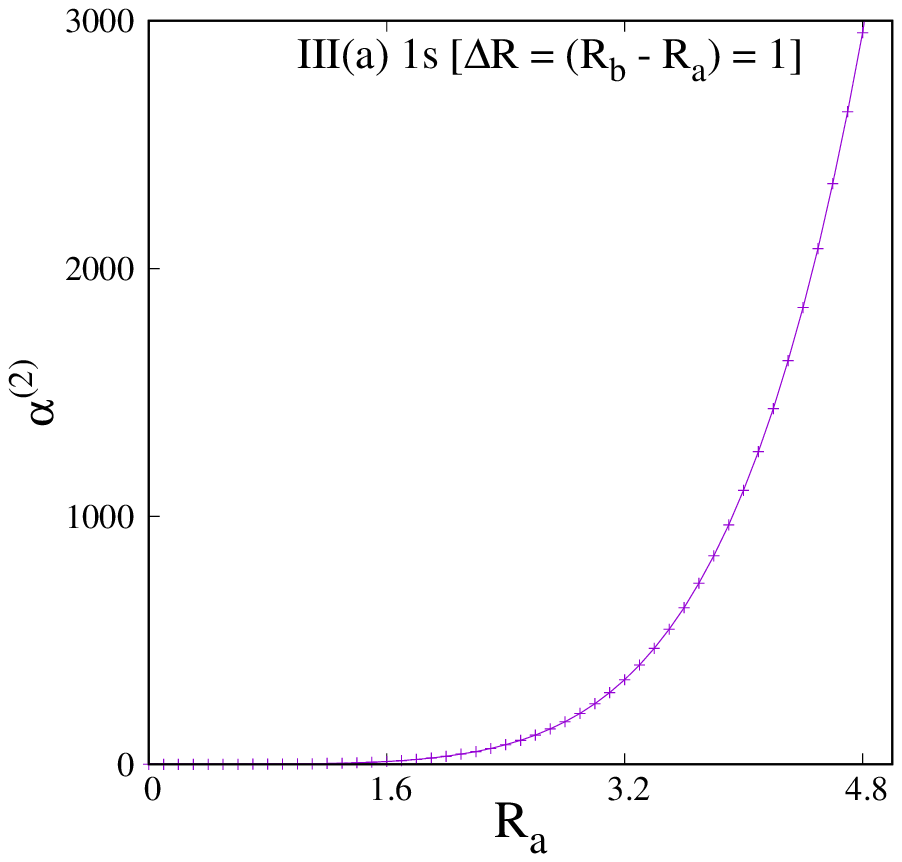}
\end{minipage}%
\begin{minipage}[c]{0.32\textwidth}\centering
\includegraphics[scale=0.53]{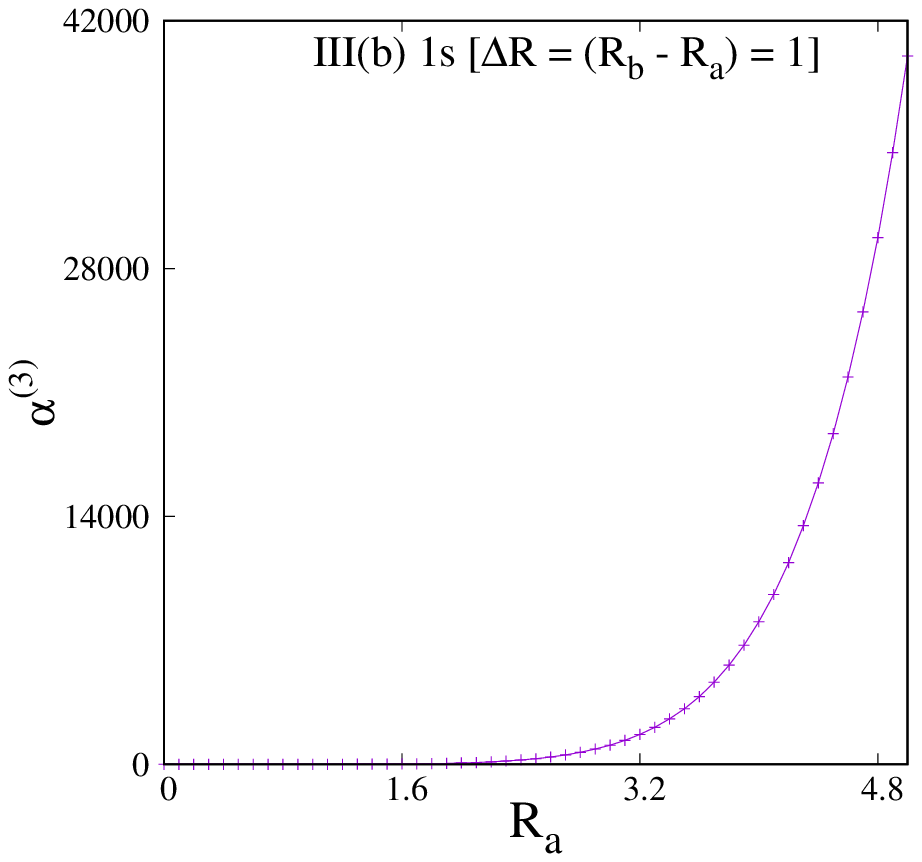}
\end{minipage}%
\begin{minipage}[c]{0.32\textwidth}\centering
\includegraphics[scale=0.53]{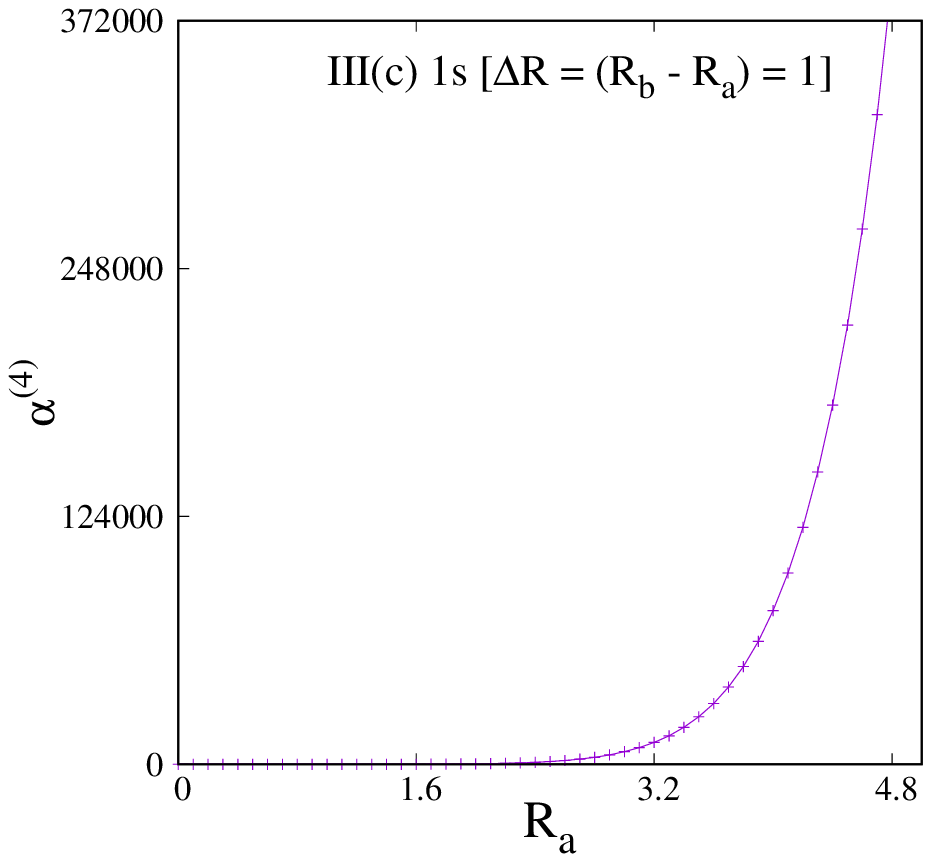}
\end{minipage}%
\vspace{2mm}
\begin{minipage}[c]{0.32\textwidth}\centering
\includegraphics[scale=0.53]{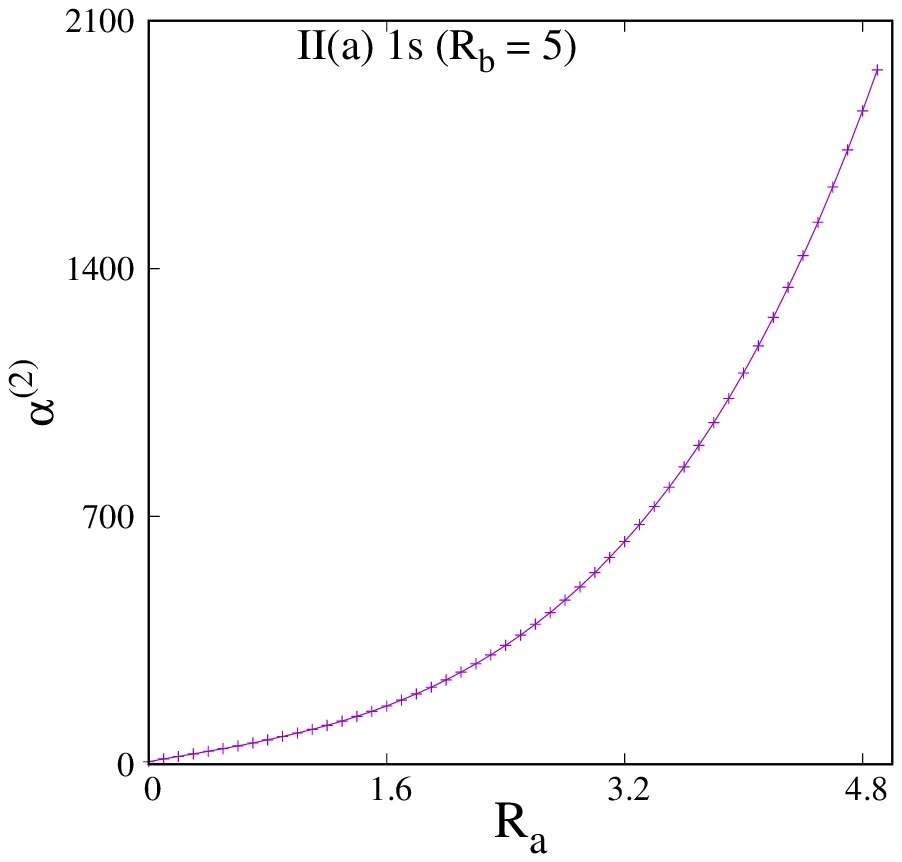}
\end{minipage}%
\begin{minipage}[c]{0.32\textwidth}\centering
\includegraphics[scale=0.53]{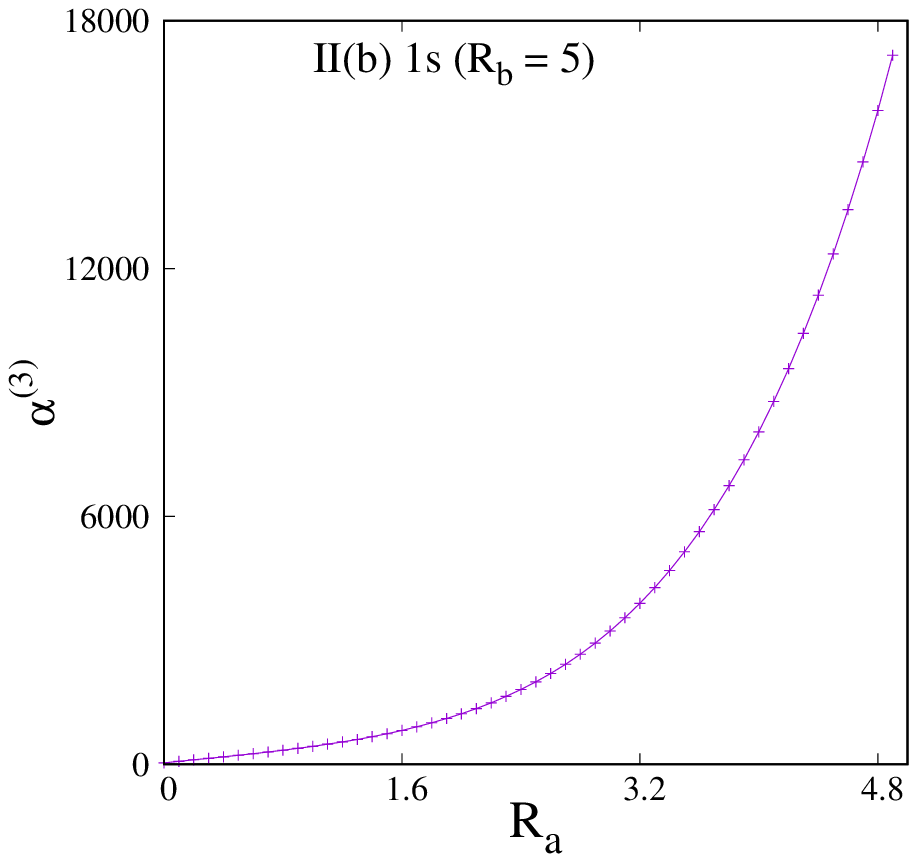}
\end{minipage}%
\begin{minipage}[c]{0.32\textwidth}\centering
\includegraphics[scale=0.53]{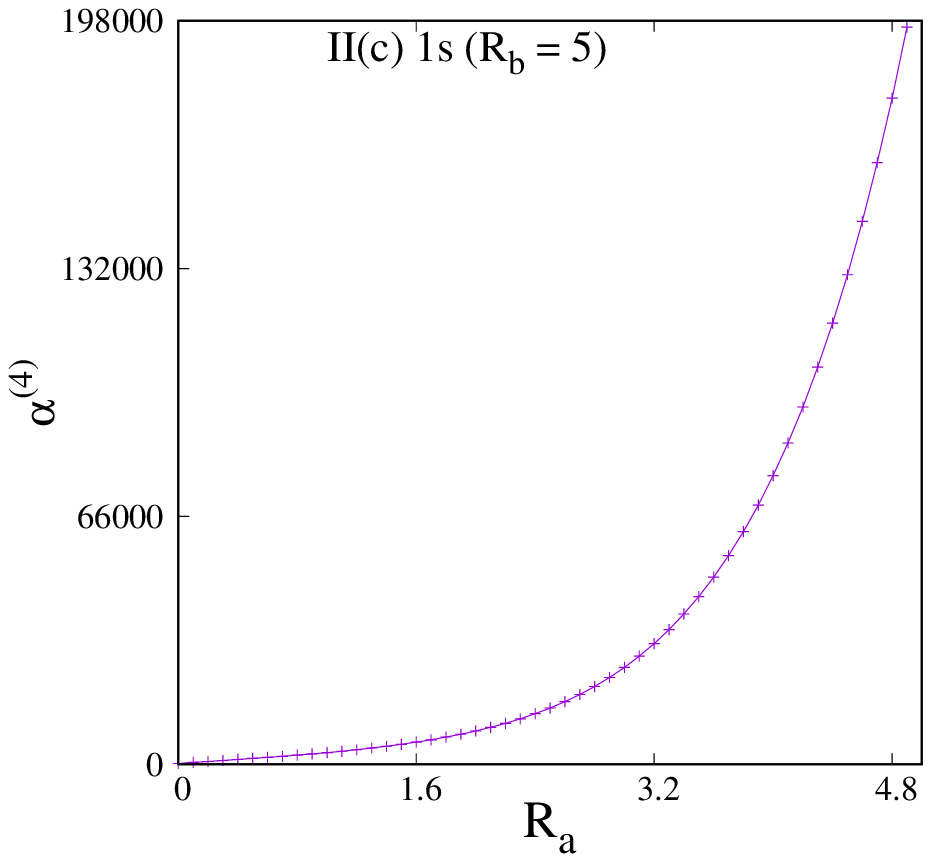}
\end{minipage}%
\vspace{2mm}
\begin{minipage}[c]{0.32\textwidth}\centering
\includegraphics[scale=0.53]{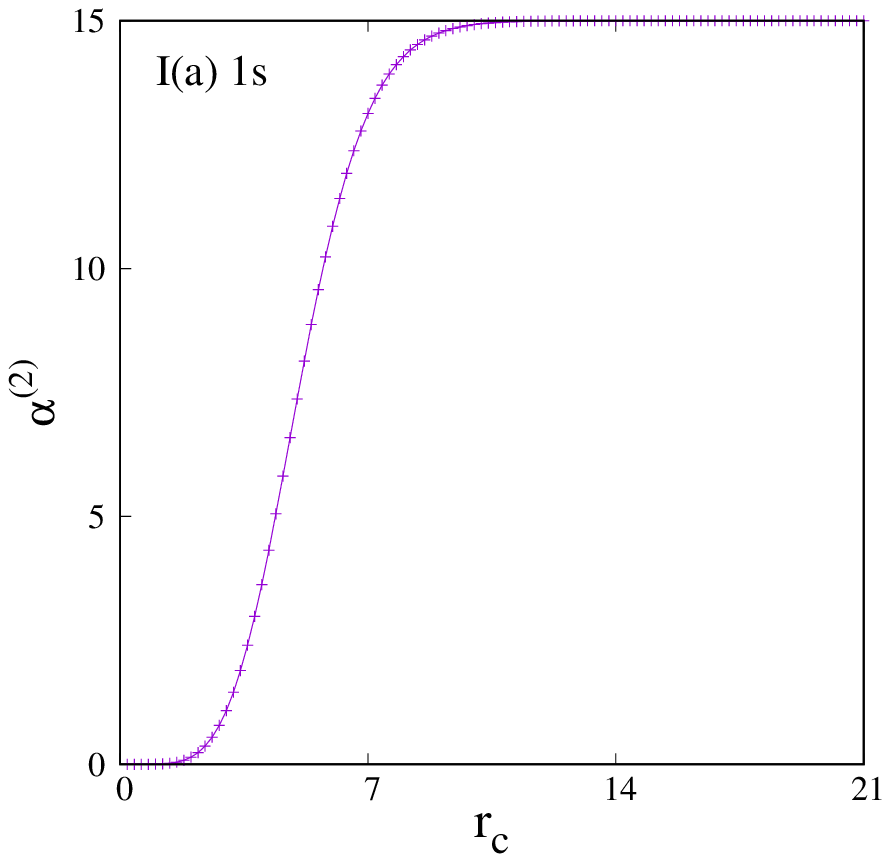}
\end{minipage}%
\begin{minipage}[c]{0.32\textwidth}\centering
\includegraphics[scale=0.53]{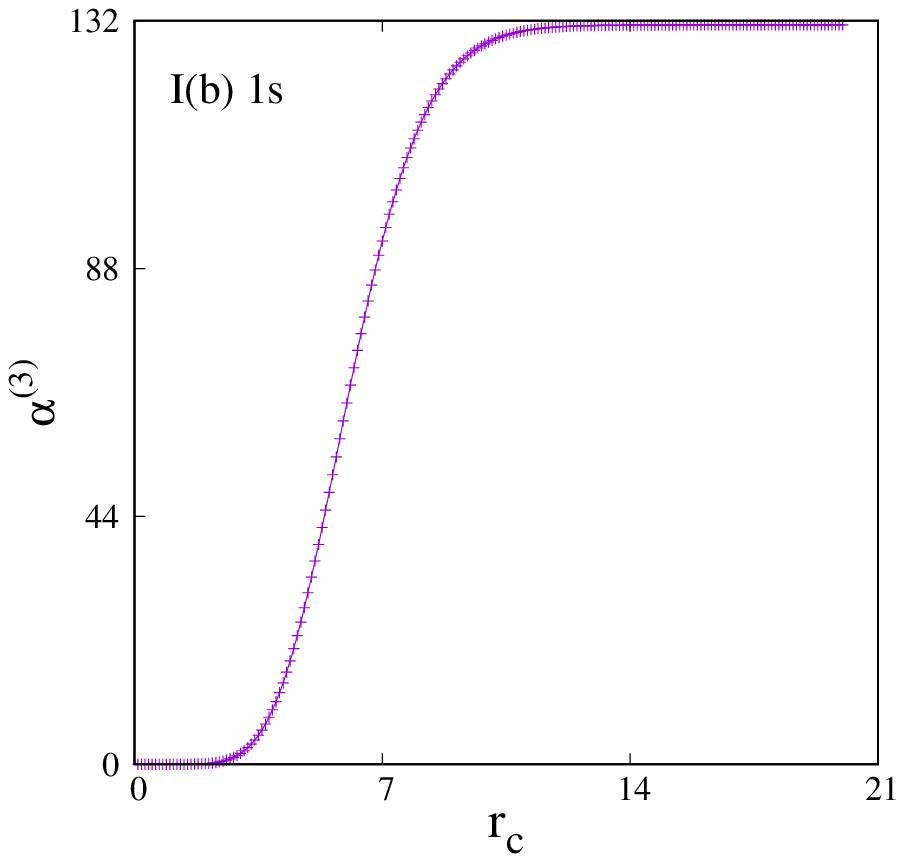}
\end{minipage}%
\begin{minipage}[c]{0.32\textwidth}\centering
\includegraphics[scale=0.53]{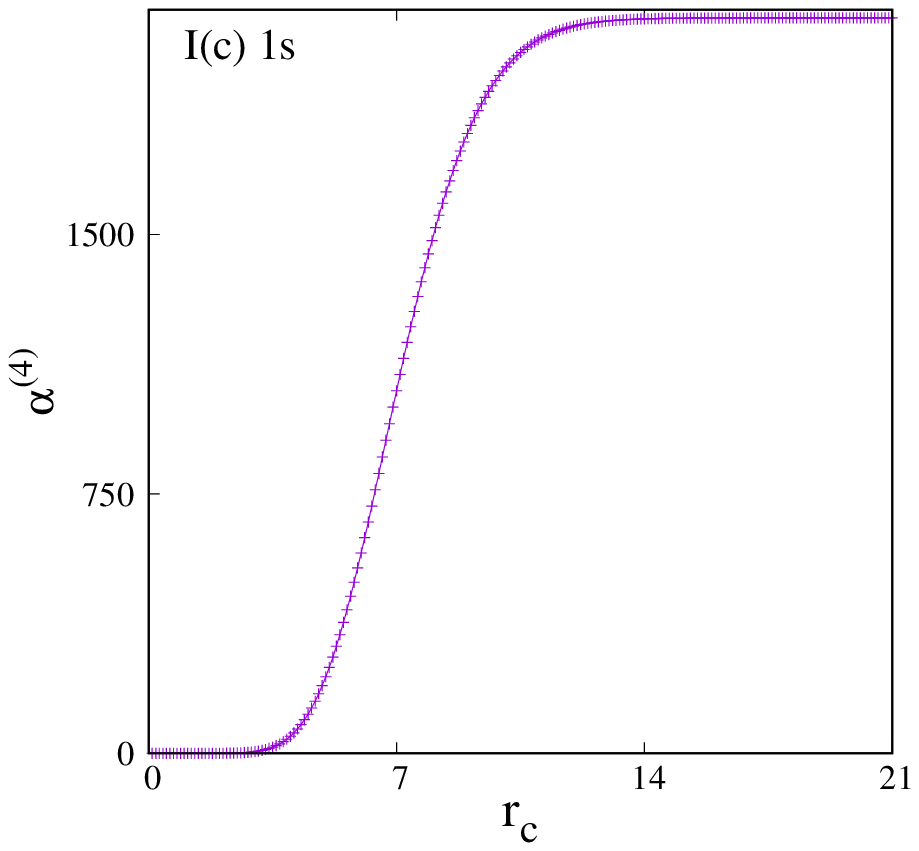}
\end{minipage}%
\caption{Plot of $\alpha^{(k)}$ $(k=2-4)$, in panels (a)-(c) in $1s$ state in GCHA. In CHA (I) they are plotted as function of $r_{c}$ (in a.u). 
In (II) and (IV) they are shown against $R_{a}$ for SCHA and LCHA. In (III), these are given, for SCHA, considering $\Delta R =(R_{b}-R_{a})=1$. 
See text for details.}
\end{figure} 

\subsection{Information entropy}
Now we will present some results on information entropy. A few points are worth noting before that. The net information measure in $r$ and $p$ 
space in a central potential, may be separated into two parts, \emph{viz.}, (i) radial and (ii) angular contributions, as mentioned in 
Eq.~(\ref{eq:10}). The latter remains unchanged in two conjugate spaces in these systems; furthermore, the same is true for different confinement 
situations of GCHA, modelled by various boundary conditions. Moreover, we have opted magnetic quantum number $m$ as $0$, unless stated otherwise.  
In all reported cases, $S_{\rvec}+S_{\pvec}=S_{t}$ satisfy the lower bound: $3(1+\ln \pi)$. 

Representative $S_{\rvec}$ and $E^{O}_{\rvec}$ for $1s$ state in GCHA are tabulated in Table~VI. The SCHA results provided at four $R_{b}$ 
($1,2,5,10$) shows that $S_{\rvec}$ progresses with $R_{a}$ to reach a plateau and then declines. That means, for each $R_{b}$, there is a 
range of $R_{a}$, where $S_{\rvec}$ grows with decrease in $\Delta R$. However, the changes in $S_{\pvec}$ is not so straightforward. 
At $R_{b}=1$, it increases with rise in $R_{a}$, while at $R_{b}=2, 5$ or $10$, it passes through a shallow minimum. Thus, with reduction
in shell radius ($\Delta R$), $S_{\rvec}, S_{\pvec}$ increases and decreases, signifying a gain and loss in uncertainty in $r$ 
and $p$ spaces. This is a reverse trend to what is observed in CHA \cite{mukherjee18}, where a shortening of $r_c$, causes fall and rise in 
$S_{\rvec}$ and $S_{\pvec}$. However, at a fixed $R_{a}$, their sum, $S_{\rvec}+S_{\pvec}=S_{t}$ always advances with growth of 
$R_{b}$. As usual, $E^{O}_{\rvec}, E^{O}_{\pvec}$ portray opposite trend with respect to $S_{\rvec}, S_{\pvec}$. At a fixed $R_{b}$, 
$E^{O}_{\rvec}$ collapses to a flat minimum and $E^{O}_{\pvec}$ rises up to a shallow maximum. Moreover, $E^{O}_{t}$ always declines with 
rise in $R_{a}$. This pattern complements the outcome of $S$. The bottom segment provides results for LCHA. Unlike SCHA, here the 
trend of $S_{\rvec}$ and $S_{\pvec}$ very candid; with growth in $R_{a}$, former enhances, while latter declines. On the contrary, 
$E^{O}_{\rvec}$ reduces and $E^{O}_{\pvec}$ advances with change in $R_{a}$. Further, $S_{t}$ decays to reach a shallow minimum and 
$E^{O}_{t}$ approaches to a flat maximum.   
  
The above results of SCHA and LCHA in ground state drives us to extend the study in $\ell \ne 0$ states. Thus, we present $S_{\rvec}$ 
for first five circular (node-less) states in SCHA and LCHA in Fig.~7. The three rows I, II, III, from bottom to top correspond to 
SCHA, SCHA having a fixed $\Delta R=1$ and LCHA; $\ell=0-4$ states are given labels (a)-(e). Like the previous plots of $f^{(k)}$ and 
$\alpha^{(k)}$, here also, SCHA graphs are presented in two separate forms. At first, $S_{\rvec}$ is plotted against $R_{a}$ 
keeping $R_{b}$ fixed at $5$ in bottom row panels I. In $1s$, a distinct dome-shaped structure appears, while for remaining states, 
the initial shape is partially lost retaining the sharp decay at large $R_a$. Secondly, in middle row panels marked II, $S_{\rvec}$ 
is shown as function of $R_{a}$ keeping $\Delta R$ constant at $1$. In all five states $S_{\rvec}$ firmly progresses with $R_{a}$. 
The top row marked III, displays the respective plot for LCHA. In $1s$ (panel (a)), it grows uninterruptedly. For other states, in 
the beginning, there is a resistance to change. In other words, it remains invariant up to a certain $R_{a}$, and then enhances. 
  
\begingroup         
\squeezetable
\begin{table}
\caption{$S$ and $E^O$ for ground state in GCHA. See text for details.}
\centering
\begin{ruledtabular}
\begin{tabular}{ll|lll|lll}
	$R_{a}$  & $R_{b}$ & $S_{\rvec}$  & $S_{\pvec}$ & $S_{t}=S_{\rvec}+S_{\pvec}$ &
	$E^{O}_{\rvec}$ & $E^{O}_{\pvec}$  & $E^{O}_{t}=E^{O}_{\rvec}E^{O}_{\pvec}$ \\
\hline	
0    & 1  & 0.52903053 & 6.0114 & 6.5404 & 0.84791729  & 0.00364537 & 0.00309098 \\
0.1  & 1  & 0.77666759 & 6.040785 & 6.817452 & 0.56908172  & 0.004180 & 0.002378   \\
0.2  & 1  & 0.89751313 & 6.23932  & 7.13683 & 0.47934610  & 0.004133 & 0.001981   \\
0.5  & 1  & 0.93965484 & 7.35022  & 8.28987 & 0.43687955  & 0.0025784 & 0.0011264   \\
0.8  & 1  & 0.40237874 & 9.9389   & 10.3412 & 0.73903793  & 0.00056417 & 0.00041694   \\
\hline
0    & 2  & 2.39666961 & 4.09171 & 6.48837 & 0.14785297  & 0.0254120 & 0.0027572 \\
0.1  & 2  & 2.66321083 & 3.93985 & 6.60306 & 0.09448635  & 0.030017 & 0.002836 \\
0.5  & 2  & 3.00050972 & 4.30210 & 7.30261 & 0.05789843  & 0.03171950 & 0.00183651 \\
1    & 2  & 3.01785556 & 5.28370 & 8.30155 & 0.05469677  & 0.020779 & 0.001136 \\
1.2  & 2  & 2.93197744 & 5.88332 & 8.81529  & 0.05923939  & 0.014951 & 0.000885 \\
1.5  & 2  & 2.64748816 & 7.20552 & 9.85300  & 0.07834727  & 0.0068029 & 0.0005329 \\
1.8  & 2  & 1.89807736 & 9.88472 & 11.78280 & 0.16543757  & 0.00124238 & 0.0002055 \\
\hline
0    & 5  & 4.01744418 & 2.5243610 & 6.5418051 & 0.04217574 & 0.16706123 & 0.00704593 \\
1    & 5  & 5.63473499 & 1.4621 & 7.0968 & 0.00432817 & 0.490356 & 0.002122 \\
2    & 5  & 5.78652038 & 2.06176 & 7.84828 & 0.00347700 & 0.404385 & 0.001406 \\
2.5  & 5  & 5.76289185 & 2.52188 & 8.28477 & 0.00351784 & 0.321734 & 0.001131 \\
3    & 5  & 5.67961160 & 3.12263 & 8.80224 & 0.00379731 & 0.231656 & 0.000879 \\
4    & 5  & 5.23062327 & 5.0882 & 10.3188 & 0.00591282 & 0.071235 & 0.000421 \\
4.5  & 5  & 4.64694680 & 7.1204 & 11.7673 & 0.01058804 & 0.01925812 & 0.00020390      \\
\hline
0   & 10  & 4.14461075 & 2.42193665 & 6.5665474 & 0.03978966 & 0.20886414 & 0.00831063 \\
0.5 & 10  & 6.22396712 & 0.317019   & 6.540986  & 0.00339560 & 1.3629551  & 0.00462805 \\
1   & 10  & 6.96632717 & $-$0.31664 & 6.64968   & 0.00140630 & 2.47609    & 0.00348    \\
2   & 10  & 7.56814293 & $-$0.5044  & 7.0637    & 0.00065283 & 3.5697296 & 0.0023304 \\
3   & 10  & 7.77837713 & $-$0.3423  & 7.4360    & 0.00049254 & 3.621882  & 0.001783 \\
5   & 10  & 7.83556437 & 0.4567     & 8.2922    & 0.00044356 & 2.561340  & 0.001136 \\
7   & 10  & 7.59776142 & 1.8617     & 9.4594    & 0.00055562 & 1.154784  & 0.000641 \\
9.5 & 10  & 6.08547241 & 7.057      & 13.142    & 0.00251172 & 0.040427  & 0.000101 \\
\hline
0    & $\infty$ &  4.14472988 & 2.42186234  & 6.56659222 & 0.03978873 & 0.20897494 & 0.00831484 \\
0.1  & $\infty$ &  4.90515587 & 1.6122737   & 6.51742957 & 0.01564470 & 0.41905958 & 0.01912784 \\
0.5  & $\infty$ &  6.26312315 & 0.28264261  & 6.54576576 & 0.00333958 & 1.48254647 & 0.00495108 \\
1    & $\infty$ &  7.15077757 & $-$0.521160 & 6.629617   & 0.00126846 & 3.4015178  & 0.00431468 \\
2    & $\infty$ &  8.21112685 & $-$1.436799 & 6.774327   & 0.00040908 & 9.1531967  & 0.0037443  \\
5    & $\infty$ &  9.84346840 & $-$2.789102 & 7.054366   & 0.00007410 & 41.251854  & 0.0030567  \\
7    & $\infty$ & 10.49963587 & $-$3.29565  & 7.20398    & 0.00003758 & 72.517286  & 0.002725  \\
8    & $\infty$ & 10.76761132 & $-$3.4717   & 7.2959     & 0.00002851 & 88.50086   & 0.002523  \\
9    & $\infty$ & 11.00741529 & $-$3.60159  & 7.40582    & 0.00002228 & 102.82136  & 0.002290  \\
10   & $\infty$ & 11.22460206 & $-$3.91190  & 7.31270    & 0.00001782 & 146.1012   & 0.0026   \\
\end{tabular}
\end{ruledtabular}
\end{table}  
\endgroup 

\begin{figure}                         
\begin{minipage}[c]{0.2\textwidth}\centering
\includegraphics[scale=0.33]{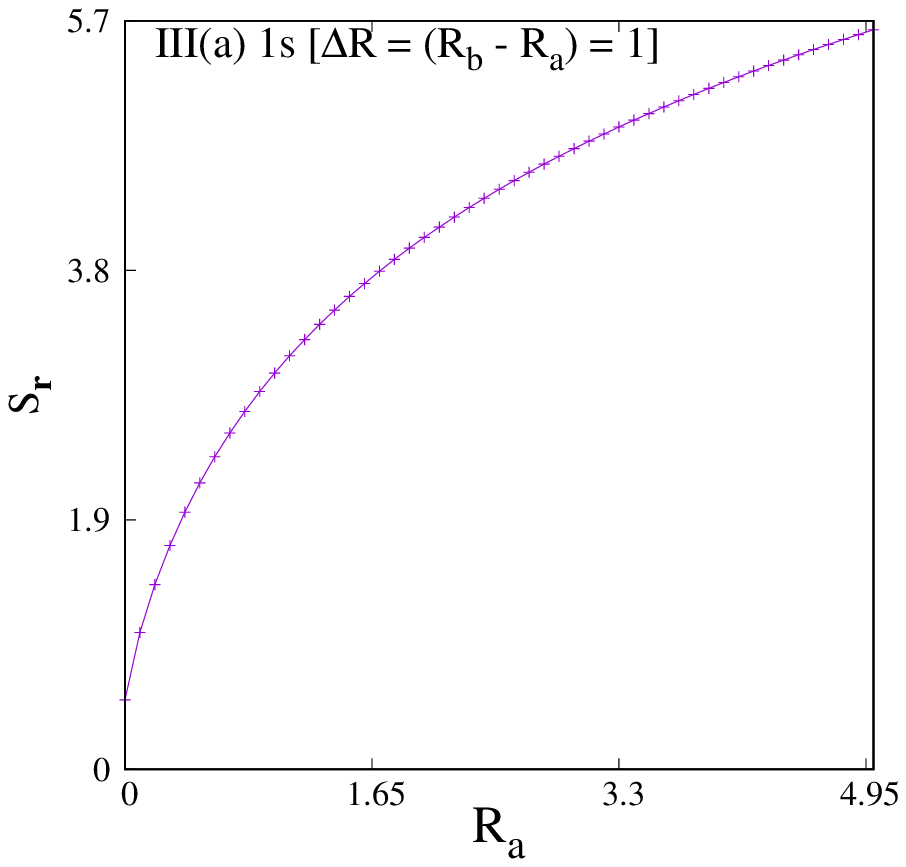}
\end{minipage}%
\begin{minipage}[c]{0.2\textwidth}\centering
\includegraphics[scale=0.33]{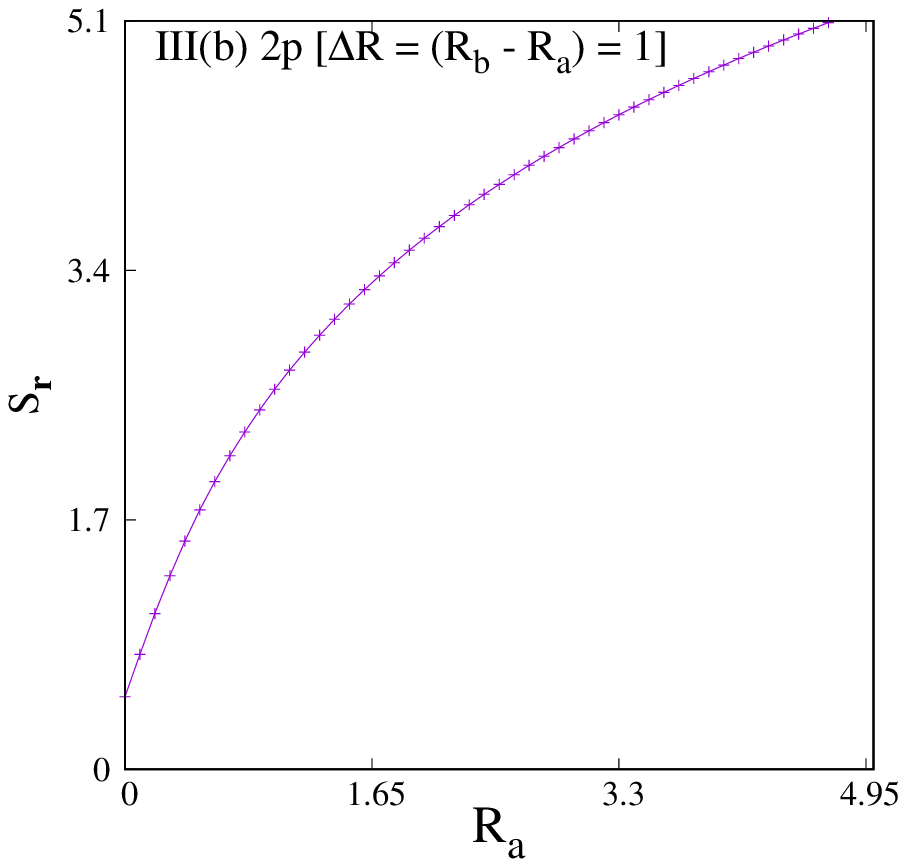}
\end{minipage}%
\begin{minipage}[c]{0.2\textwidth}\centering
\includegraphics[scale=0.33]{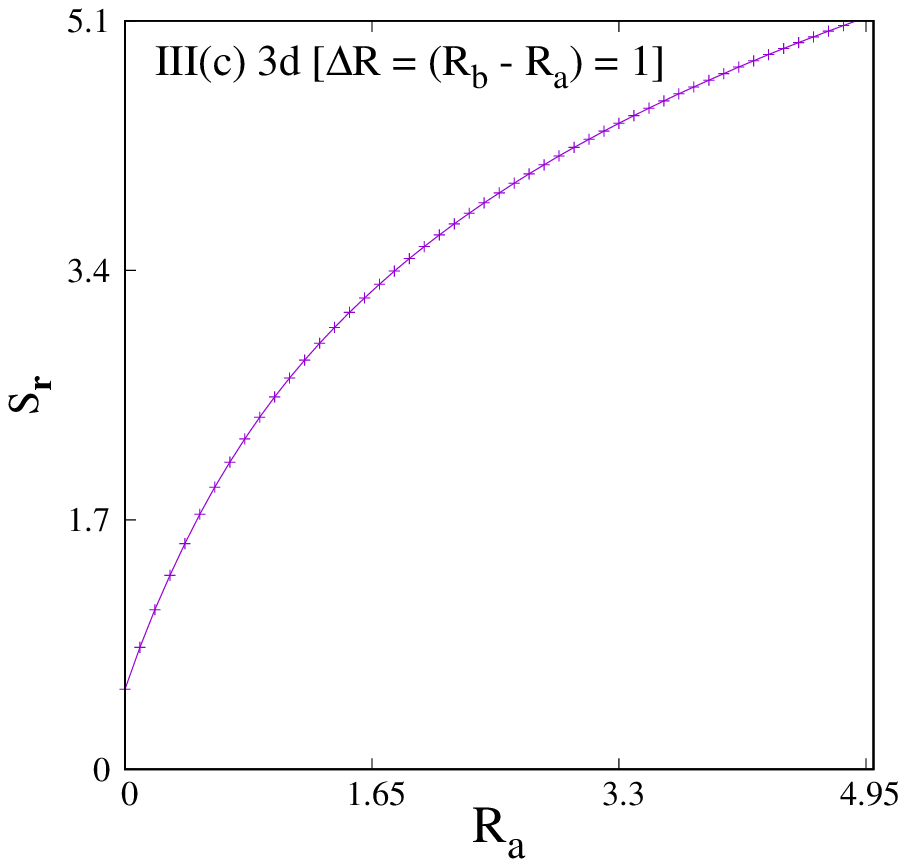}
\end{minipage}%
\begin{minipage}[c]{0.2\textwidth}\centering
\includegraphics[scale=0.33]{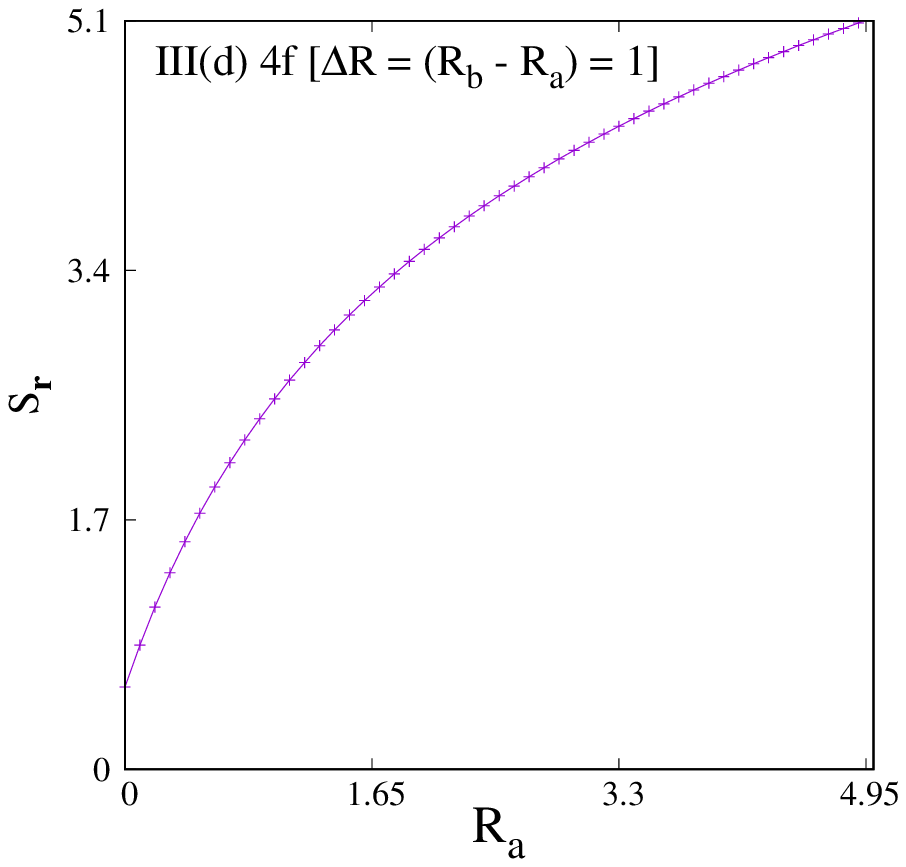}
\end{minipage}%
\begin{minipage}[c]{0.2\textwidth}\centering
\includegraphics[scale=0.33]{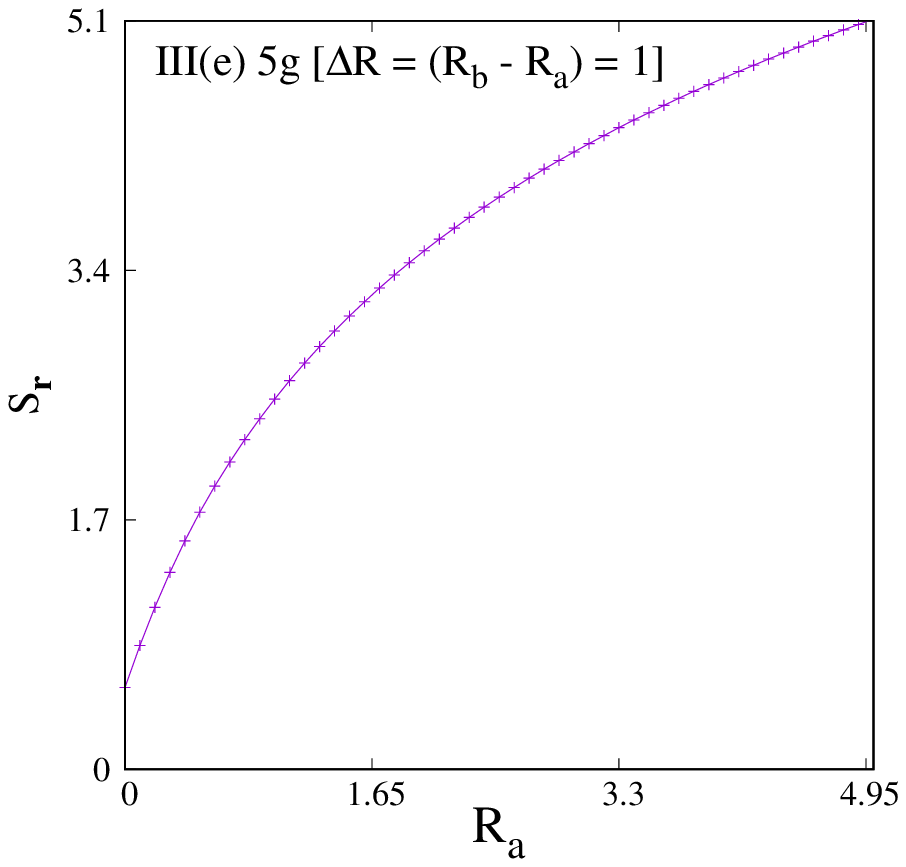}
\end{minipage}%
\vspace{2mm}                        
\begin{minipage}[c]{0.2\textwidth}\centering
\includegraphics[scale=0.33]{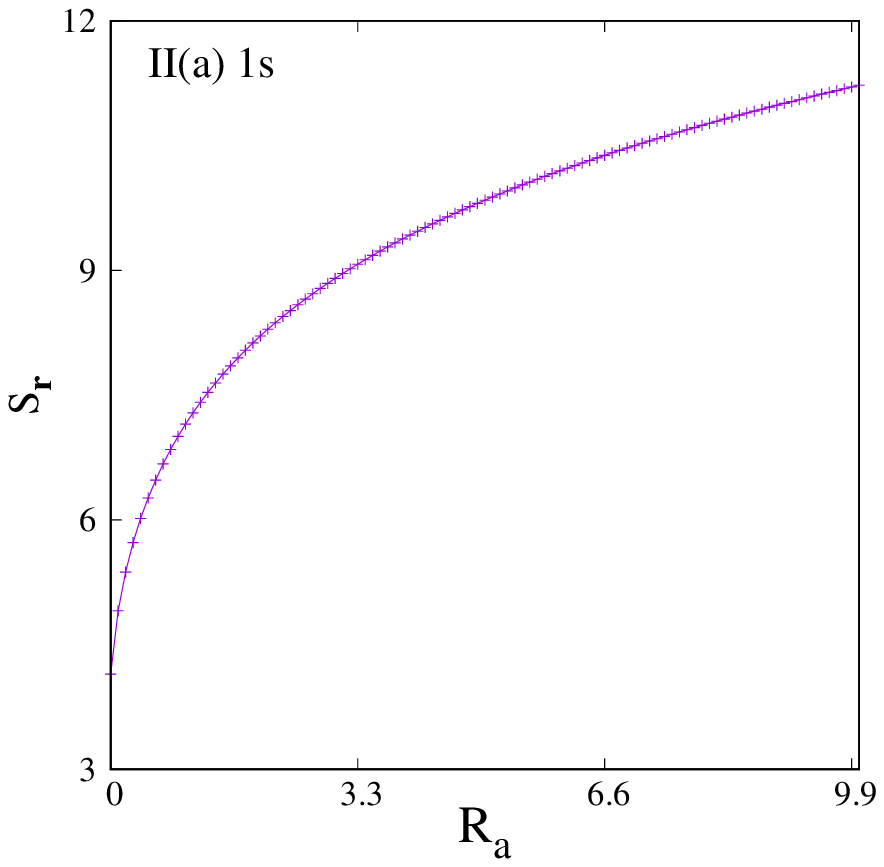}
\end{minipage}%
\begin{minipage}[c]{0.2\textwidth}\centering
\includegraphics[scale=0.33]{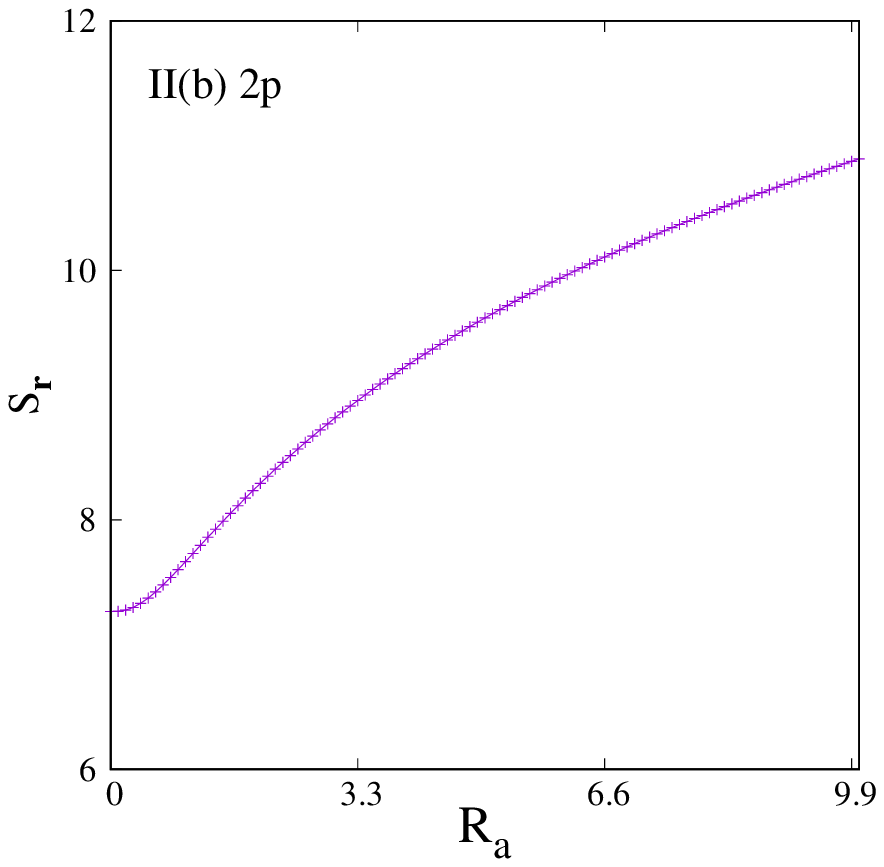}
\end{minipage}%
\begin{minipage}[c]{0.2\textwidth}\centering
\includegraphics[scale=0.33]{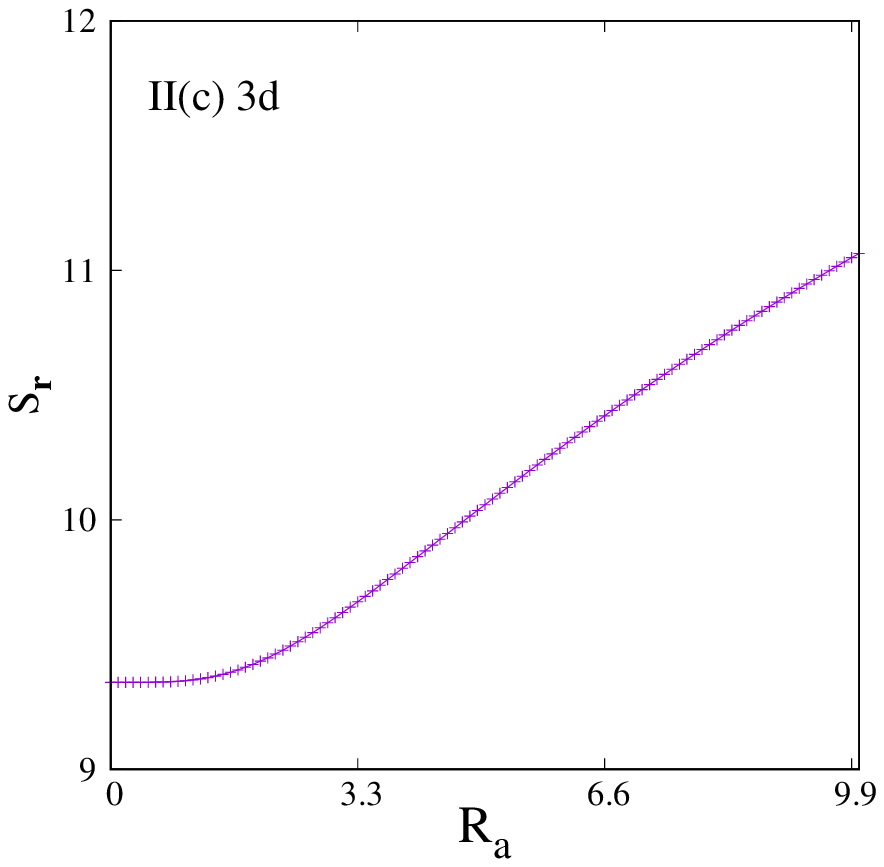}
\end{minipage}%
\begin{minipage}[c]{0.2\textwidth}\centering
\includegraphics[scale=0.33]{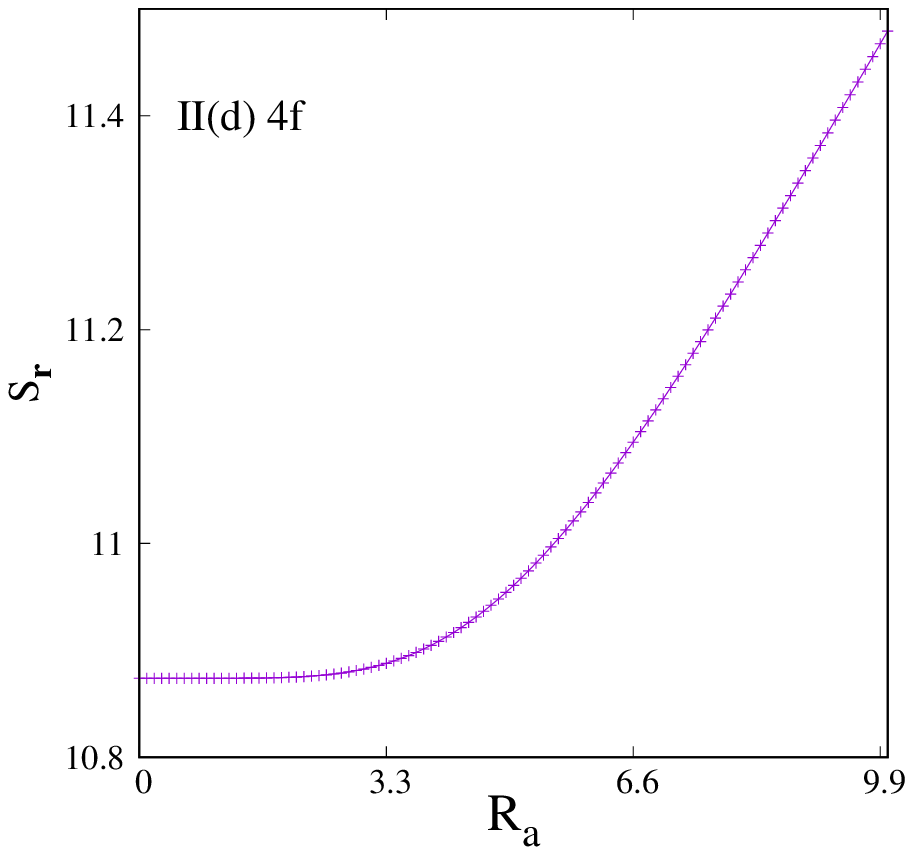}
\end{minipage}%
\begin{minipage}[c]{0.2\textwidth}\centering
\includegraphics[scale=0.33]{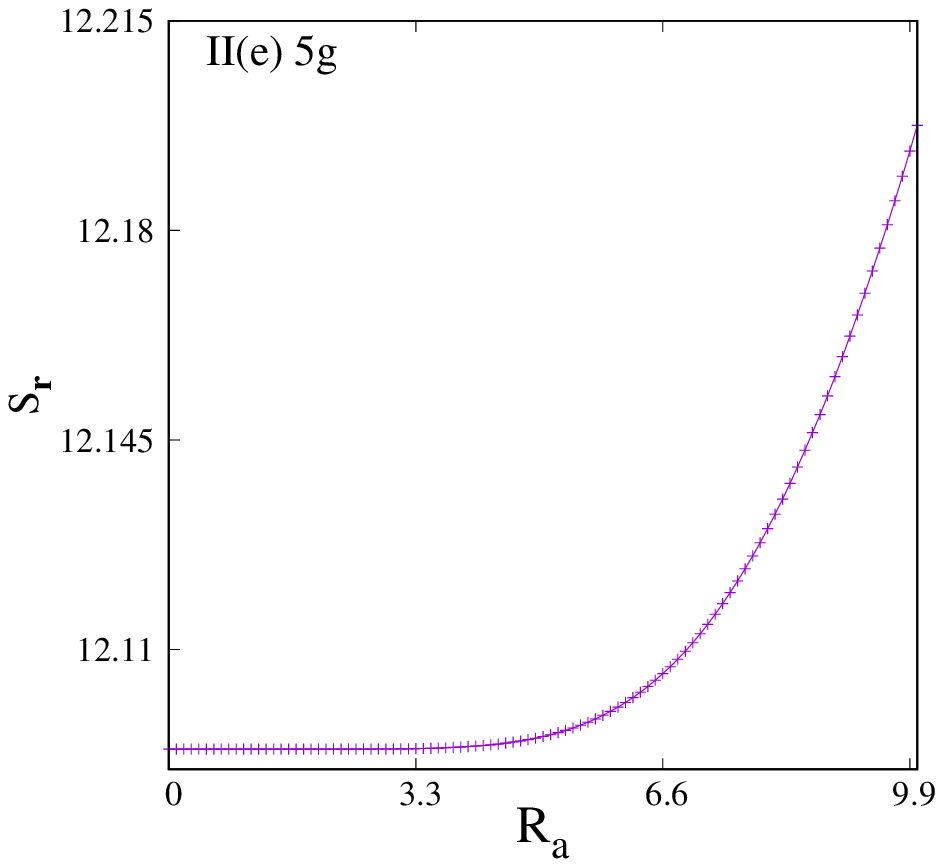}
\end{minipage}%
\vspace{2mm}                         
\begin{minipage}[c]{0.2\textwidth}\centering
\includegraphics[scale=0.33]{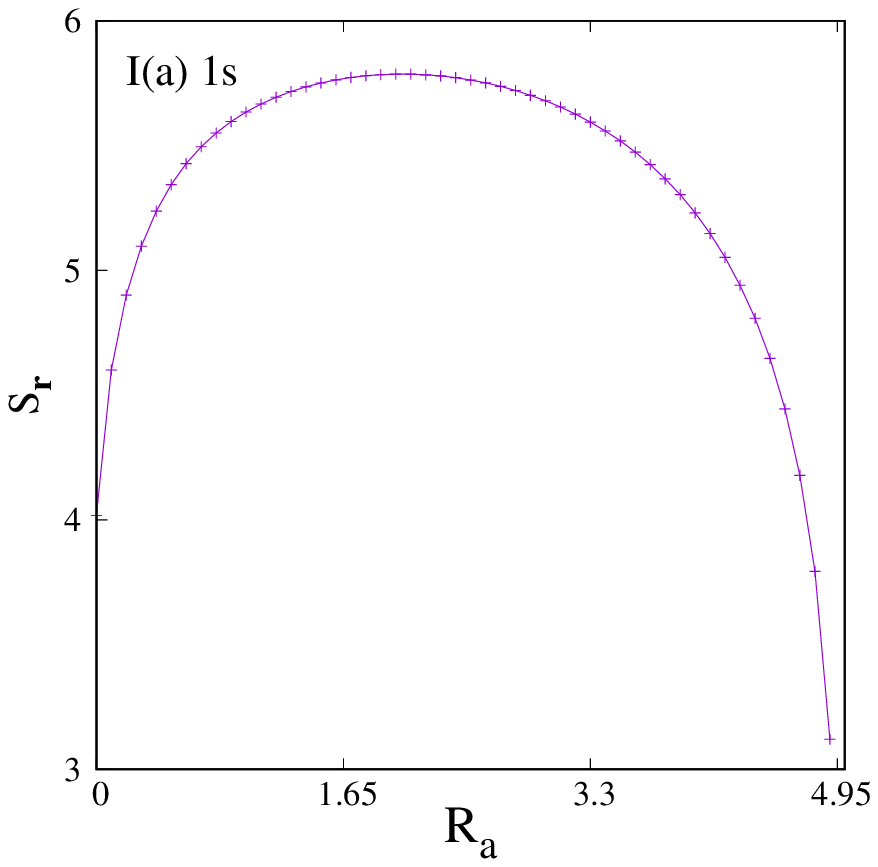}
\end{minipage}%
\begin{minipage}[c]{0.2\textwidth}\centering
\includegraphics[scale=0.33]{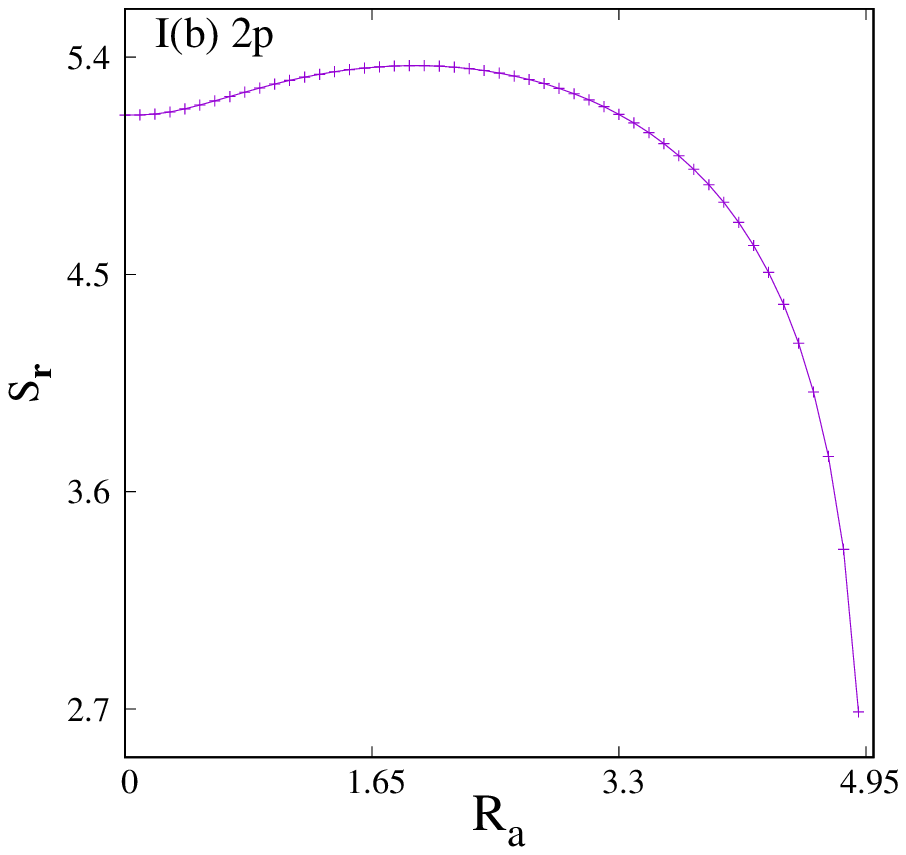}
\end{minipage}%
\begin{minipage}[c]{0.2\textwidth}\centering
\includegraphics[scale=0.33]{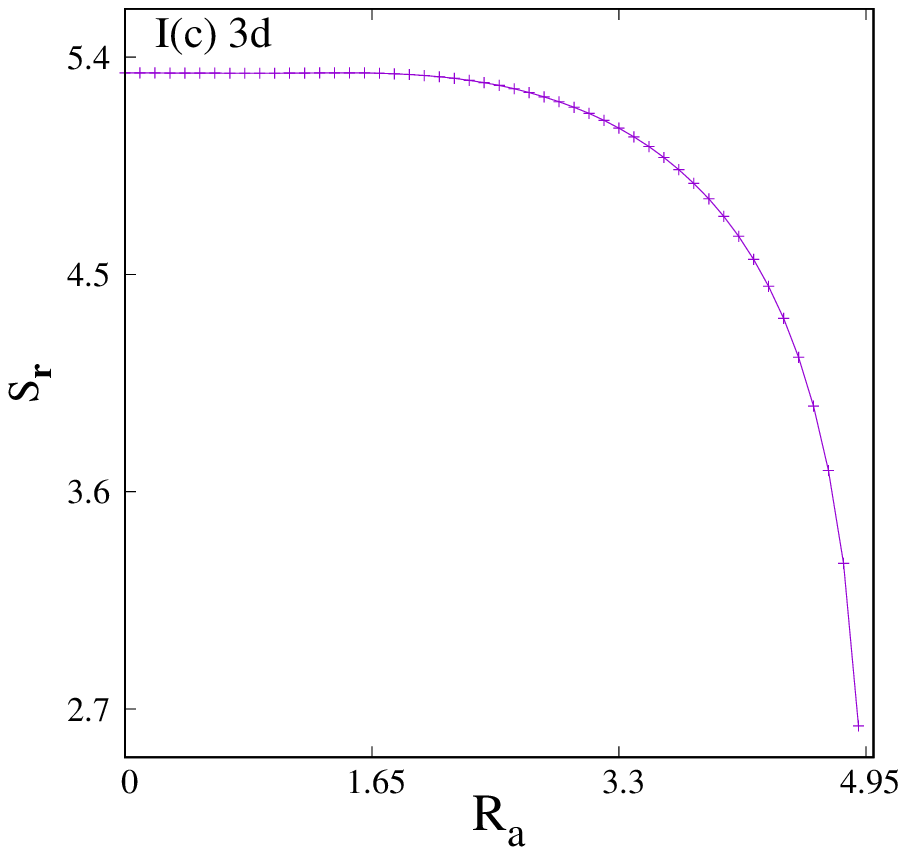}
\end{minipage}%
\begin{minipage}[c]{0.2\textwidth}\centering
\includegraphics[scale=0.33]{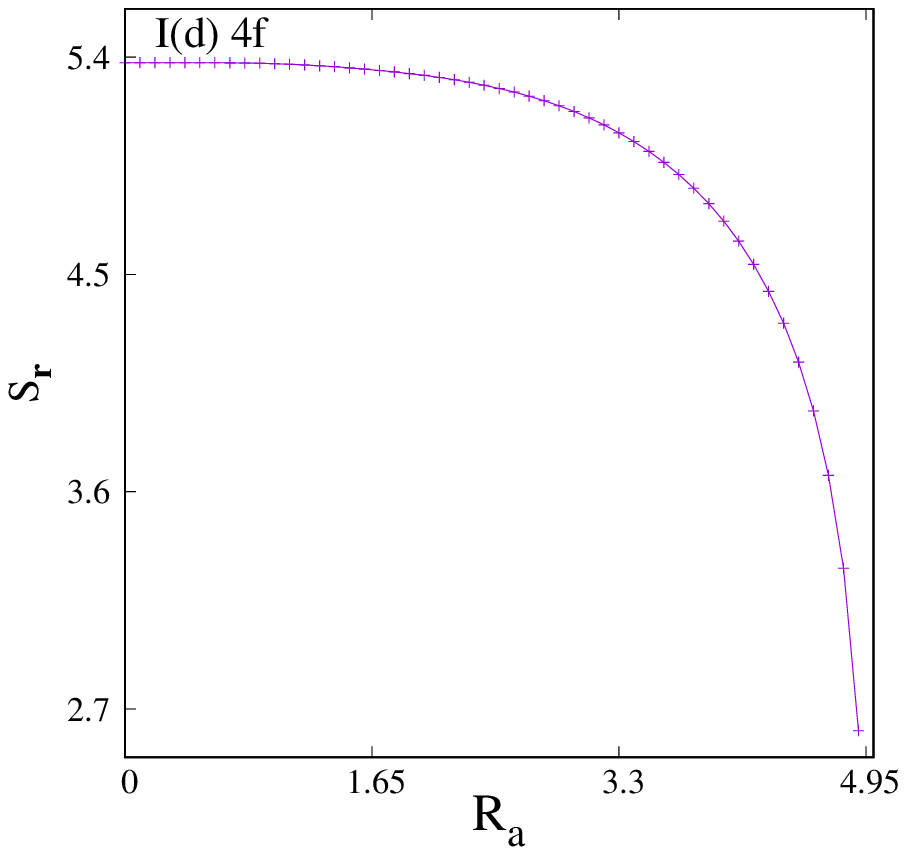}
\end{minipage}%
\begin{minipage}[c]{0.2\textwidth}\centering
\includegraphics[scale=0.33]{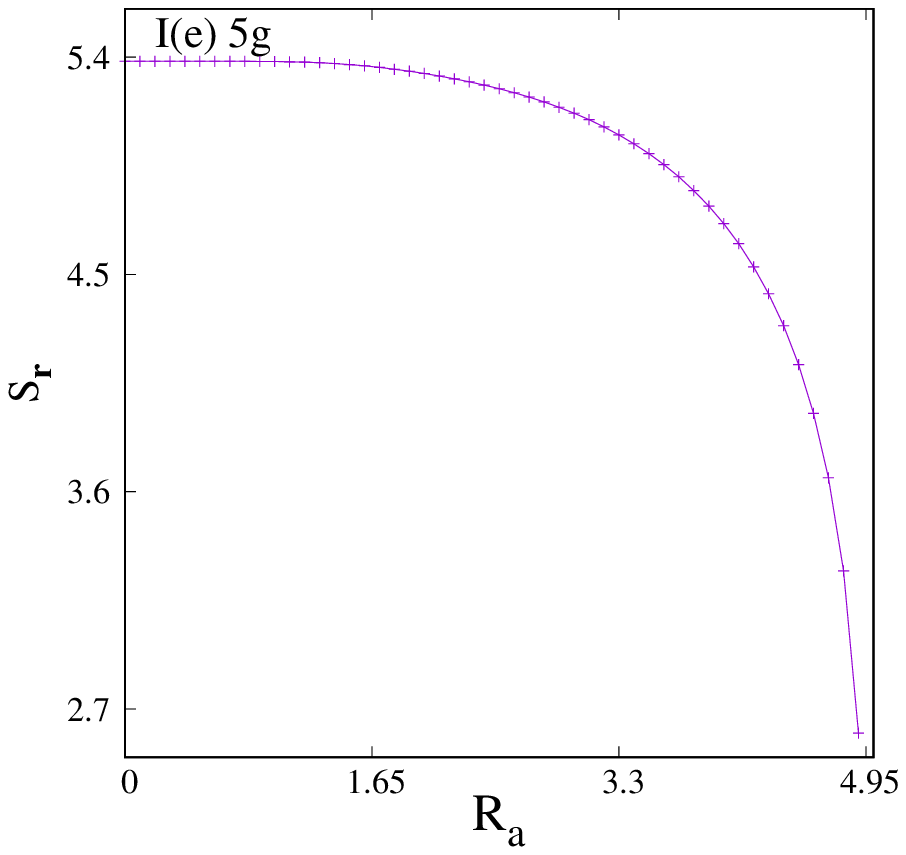}
\end{minipage}%
\caption{Plot of $S_{\rvec}$ as function of $R_{a}$ (in a.u) for $1s,2p,3d,4f,5g$ states in (I) SCHA ($R_b=5$) (II) SCHA, 
keeping $\Delta R$ fixed at $1$ a.u. (III) LCHA. Consult text for details.}
\end{figure} 

\section{future and outlook}
Incidental degeneracy, multipole oscillator strength, multipole polarizability, Shannon entropy and Onicescu 
{\color{red}information} energy have been investigated for 
GCHA model, with special emphasis on SCHA, which has not been done before. The proposed model can explain both \emph{free} and \emph{confined} 
systems effectively. An in-depth analysis reveals several fascinating and hitherto unreported features in such systems. 
The possibility of this degeneracy in Debye and exponential-cosine screened plasma environment has been established. In GCHA with increase in 
$n$ number of these degenerate states increases, while at a fixed $n$, with growth in $\ell$, their count declines. In confined 
condition, \emph{negative} polarizability is encountered in ground and several excited states, which in accordance with Herzfeld criterion, 
suggests metallic character. Simplified analytical expressions of $f^{(k)}, \alpha^{(k)}$ in FHA are reported. The impact 
of $R_{a}, R_{b}$ and $\Delta R$ on spectroscopic and density-dependent properties are examined. Similar calculations in other central 
potentials is highly desirable. Particularly, it is necessary to verify the existence of such degeneracy by imposing \emph{shell confinement} on other
quantum chemical systems. Investigation of Hellmann-Feynman theorem in the context of SCHA is desirable. Further, exploration of two-photon 
transition amplitude, photo-ionization cross-section, relative information in GCHA model would provide critical insight. Apart from that, it would be 
interesting to extend the shell confinement model to many-electron atoms.              

\section{Acknowledgement}
Financial support from BRNS, India (sanction order: 58/14/03/2019-BRNS/10255) is gratefully acknowledged. Partial financial support from 
SERB, India (CRG/2019/000293) is also appreciated. NM thanks CSIR, New Delhi, India, for a Senior Research Associate-ship (Pool No. 9033A).
The authors acknowledge valuable discussion with Prof. K. D. Sen. {\color{red}The authors are thankful to two anonymous referees for 
their constructive comments.} 

\appendix
{\color{red}
\section{Multipole oscillator strength and polarizability in FHA}
Here, we present the \emph{first} transition corresponding to the respective selection rule for $k=1,2,3,4$ respectively. 
The remaining results are provided in Supporting Material.
\subsection{Dipole oscillator strength and polarizability}

The selection rule is $\Delta \ell=\pm 1$. In $s$ states, it changes to $\Delta \ell= 1$. However, 
in $\ell \ne 0$ states, $\alpha^{(1)}_{n \ell} =\alpha^{(1)}_{n \ell}(\ell-1) + \alpha^{(1)}_{n \ell}(\ell+1)$. 

The expressions of $f^{(1)}_{(1s \rightarrow np)}(Z)$ and $f^{(1)}_{(2p \rightarrow ns)}(Z)$ are found as,
\begin{equation}\label{eq:22}
\begin{aligned}
f^{(1)}_{(1s \rightarrow np)}(Z) & = \frac{2^{8}}{3Z^{7}}  \ n^{5} \ \frac{(n-1)^{(2n-4)}}{(n+1)^{(2n+4)}}, \\
f^{(1)}_{(2p \rightarrow ns)}(Z) & = \frac{2^{13}}{27Z^{7}}  \ n^{7} \ \frac{(n-2)^{(2n-5)}}{(n+2)^{(2n+5)}}, \  \ (n \ne 2)\\
\end{aligned}
\end{equation}
Now, using Eq.~(\ref{eq:22}) in Eq.~(\ref{eq:4}), one easily gets $\alpha^{(1)}_{1s}(p)(Z),\alpha^{(1)}_{2p}(s)(Z)$ of FHA. They are obtained as,  
\begin{equation}\label{eq:23}
\begin{aligned}
\alpha^{(1)}_{1s}(p)(Z) & = \sum^{n}_{j=2} \frac{2^{10}}{3Z^{9}}  \ j^{9} \ \frac{(j-1)^{(2j-6)}}{(j+1)^{(2j+6)}}, \\
\alpha^{(1)}_{2p}(s)(Z) & = \sum^{n}_{\substack{j=1 \\ j \ne 2}} \frac{2^{19}}{27Z^{9}}  \ j^{11} \ 
\frac{(j-2)^{(2j-7)}}{(j+2)^{(2j+7)}}, \\
\end{aligned}
\end{equation}

\subsection{Quadrupole oscillator strength and polarizability}

In this case, the selection rule is $\Delta \ell=0, \pm 2$. In $s$ states it is $\Delta \ell= 2$. Similarly, in $p$ states $\Delta = 0, 2$. Hence,
$\alpha^{(2)}_{n \ell=1}=\alpha^{(2)}_{n \ell}(\ell)+ \alpha^{(2)}_{n \ell}(\ell+2)$. Moreover, for $\ell \ge 2$, 
$\alpha^{(2)}_{n \ell}=\alpha^{(2)}_{n \ell}(\ell-2)+ \alpha^{(2)}_{n \ell}(\ell)+ \alpha^{(2)}_{n \ell}(\ell+2)$.  

The closed-form expression of $f^{(2)}_{(1s \rightarrow nd)}(Z)$, $f^{(2)}_{(2p \rightarrow np)}$ and $f^{(2)}_{(3d \rightarrow ns)}(Z)$ are obtained as,
\begin{equation}\label{eq:32}
\begin{aligned}
f^{(2)}_{(1s \rightarrow nd)}(Z) & = \frac{2^{12}}{5Z^{9}}  \ n^{7} \ (n^{2}-4) \frac{(n-1)^{(2n-6)}}{(n+1)^{(2n+6)}}, \\
f^{(2)}_{(2p \rightarrow np)}(Z) & = \frac{2^{22}}{75Z^{9}}  \ n^{9} \ (n^{2}-1) \frac{(n-2)^{(2n-7)}}{(n+2)^{(2n+7)}}, \\
{(2)}_{(3d \rightarrow ns)}(Z) & = \frac{2^{17}3^{7}}{125Z^{9}}  \ n^{13} \ (n^{2}-6)^{2} \frac{(n-3)^{(2n-9)}}{(n+3)^{(2n+9)}}.
\end{aligned}
\end{equation}
Now, applying Eq.~(\ref{eq:32}) in Eq.~(\ref{eq:4}), one easily gets $\alpha^{(2)}_{1s}(d)(Z), \alpha^{(2)}_{2p}(p)(Z)$ and 
$\alpha^{(2)}_{3d}(s)(Z)$ of FHA. They take the following forms, 
\begin{equation}\label{eq:33}
\begin{aligned}
\alpha^{(2)}_{1s}(d)(Z) & = \sum^{n}_{j=3} \frac{2^{12}}{5Z^{11}}  \ j^{11} \ (j^{2}-4) \frac{(j-1)^{(2j-8)}}{(j+1)^{(2j+8)}}, \\	
\alpha^{(2)}_{2p}(p)(Z) & = \sum^{n}_{j=3} \frac{2^{28}}{75Z^{11}}  \ j^{13} (j^{2}-1) \frac{(j-2)^{(2j-8)}}{(j+2)^{(2j+8)}}, \\
\alpha^{(2)}_{3d}(s)(Z) & = \sum^{n}_{\substack{j=1 \\ j \ne 3}} \frac{2^{19}3^{11}}{5^{3}Z^{11}}  \ j^{17} (j^{2}-6)^{2} \
\frac{(j-3)^{(2j-11)}}{(j+3)^{(2j+11)}}.
\end{aligned}
\end{equation}

\subsection{Octupole oscillator strength and polarizability}
The selection rule is $\Delta \ell=\pm 1, \pm 3$. For $\ell =0$, $\Delta \ell =3$. For $\ell = 1$, $\alpha^{(3)}_{n \ell}=\alpha^{(3)}_{n \ell}(\ell+1)+
\alpha^{(3)}_{n \ell}(\ell+3)$. Next, $\ell=2$, the relation is, $\alpha^{(3)}_{n \ell}=\alpha^{(3)}_{n \ell}(\ell-1)+\alpha^{(3)}_{n \ell}(\ell+1)+
\alpha^{(3)}_{n \ell}(\ell+3)$. Further, for $\ell \ge 3$, $\alpha^{(3)}_{n \ell}=\alpha^{(3)}_{n \ell}(\ell-1)+\alpha^{(3)}_{n \ell}(\ell-3)+\alpha^{(3)}_{n \ell}(\ell+1)+ \alpha^{(3)}_{n \ell}(\ell+3)$.

Now, $f^{(3)}_{(1s \rightarrow nf)}(Z)$, $f^{(3)}_{(2p \rightarrow nd)}(Z)$, $f^{(3)}_{(3d \rightarrow np)}(Z)$ and 
$f^{(3)}_{(4f \rightarrow ns)}(Z)$ are written as, 
\begin{equation}\label{eq:46}
\begin{aligned}
f^{(3)}_{(1s \rightarrow nf)}(Z) & = \frac{9 \ 2^{12}}{7Z^{11}} \ n^{9} \ (n^{2}-4)(n^{2}-9) \frac{(n-1)^{(2n-8)}}{(n+1)^{(2n+8)}}, \\
f^{(3)}_{(2p \rightarrow nd)}(Z) & = \frac{2^{27}}{49Z^{11}}  \ n^{13} \ (n^{2}-1)(n^{2}-16)^{2} \frac{(n-2)^{(2n-10)}}{(n+2)^{(2n+10)}}, \\
f^{(3)}_{(3d \rightarrow np)}(Z) & = \frac{2^{18}3^{13}}{5^{2} \ 7^{2} \ Z^{11}}  \ n^{13} \ (n^{2}-1)(4n^{2}-9)^{2} \frac{(n-3)^{(2n-12)}}{(n+3)^{(2n+12)}},\\
f^{(3)}_{(4f \rightarrow ns)}(Z) & = \frac{2^{34}}{245Z^{11}}  \ n^{15} \ (141n^{4}-3008n^{2}+18176)^{2} \frac{(n-4)^{(2n-13)}}{(n+4)^{(2n+13)}}.
\end{aligned}
\end{equation}
Putting Eq.~(\ref{eq:46}) in Eq.~(\ref{eq:4}), one easily gets $\alpha^{(3)}_{4f}(s)(Z)$,
$\alpha^{(3)}_{1s}(f)(Z)$, $\alpha^{(3)}_{2p}(d)(Z)$, $\alpha^{(3)}_{3d}(p)(Z)$ and $\alpha^{(3)}_{4f}(s)(Z)$. They have following forms, 
\begin{equation}
\begin{aligned}
\alpha^{(3)}_{1s}(f)(Z) & = \sum^{n}_{j=4} \frac{9 \ 2^{14}}{7Z^{13}}  \ j^{13} \ (j^{2}-4)(j^{2}-9) \frac{(j-1)^{(2j-10)}}{(j+1)^{(2j+10)}}, \\
\alpha^{(3)}_{2p}(d)(Z) & = \sum^{n}_{j=3} \frac{2^{33}}{49Z^{13}}  \ j^{17} (j^{2}-1)(j^{2}-16)^{2} \frac{(j-2)^{(2j-12)}}{(j+2)^{(2j+12)}}, \\
\alpha^{(3)}_{3d}(p)(Z) & = \sum^{n}_{\substack{j=1 \\ j \ne 3}} \frac{2^{20}3^{17}}{5^{2} \ 7^{2} Z^{13}}  \ j^{17} (j^{2}-1)(4j^{2}-9)^{2} \
\frac{(j-3)^{(2j-14)}}{(j+3)^{(2j+14)}}, \\	
\alpha^{(3)}_{4f}(s)(Z) & = \sum^{n}_{\substack{j=2 \\ j \ne 4}} \frac{2^{44}}{245 \ Z^{13}}  \ j^{19} (141j^{4}-3008j^{2}+18176)^{2} \
\frac{(j-4)^{(2j-15)}}{(j+4)^{(2j+15)}}.
\end{aligned}
\end{equation}

the expression of $\alpha^{(4)}_{n\ell}$ changes with alteration of $\ell$ values. They are, 
\begin{equation}
\begin{aligned}
\ell & =1, \ \ \ \alpha^{(4)}_{n \ell} & = & \ \alpha^{(4)}_{n \ell}(\ell+2)+\alpha^{(4)}_{n \ell}(\ell+4), \\
\ell & =2, \ \ \ \alpha^{(4)}_{n \ell} & = & \ \alpha^{(4)}_{n \ell}(\ell)+\alpha^{(4)}_{n \ell}(\ell+2)+\alpha^{(4)}_{n \ell}(\ell+4), \\
\ell & =3, \ \ \ \alpha^{(4)}_{n \ell} & = & \ \alpha^{(4)}_{n \ell}(\ell-2)+\alpha^{(4)}_{n \ell}(\ell)+\alpha^{(4)}_{n \ell}(\ell+2)+\alpha^{(4)}_{n \ell}(\ell+4), \\
\ell & =4, \ \ \ \alpha^{(4)}_{n \ell} & = & \ \alpha^{(4)}_{n \ell}(\ell-4)+\alpha^{(4)}_{n \ell}(\ell-2)+\alpha^{(4)}_{n \ell}(\ell)+\alpha^{(4)}_{n \ell}(\ell+2)+\alpha^{(4)}_{n \ell}(\ell+4).  
\end{aligned}
\end{equation}

$f^{(4)}_{(1s \rightarrow ng)}(Z), f^{(4)}_{(2p \rightarrow nf)}(Z), f^{(4)}_{(3d \rightarrow nd)}(Z), f^{(4)}_{(4f \rightarrow np)}(Z) \  
\mathrm{and} \ f^{(4)}_{(5g \rightarrow ns)}(Z)$ as,
\begin{equation}\label{eq:59}
\begin{aligned}
f^{(4)}_{(1s \rightarrow ng)}(Z) & = \frac{2^{18}}{9Z^{13}}  \ n^{11} \ (n^{2}-16)(n^{2}-9)(n^{2}-4) \ \frac{(n-1)^{(2n-10)}}{(n+1)^{(2n+10)}}, \\
f^{(4)}_{(2p \rightarrow nf)}(Z) & = \frac{2^{33}}{3^{5}Z^{13}}  \ n^{13} \ (n^{2}-1)(n^{2}-9)(7n^{2}+68)^{2} \frac{(n-2)^{(2n-12)}}{(n+2)^{(2n+12)}}, \\
f^{(4)}_{(3d \rightarrow nd)}(Z) & = \frac{2^{20}3^{15}}{35Z^{13}}  \ n^{17} \ (n^{2}-1)(n^{2}-4)(n^{2}-21)^{2} \frac{(n-3)^{(2n-13)}}{(n+3)^{(2n+13)}}, \\
f^{(4)}_{(4f \rightarrow np)}(Z) & = \frac{2^{43}}{8505Z^{13}}  \ n^{17} \ (n^{2}-1)(31n^{4}-4768n^{2}+43776)^{2} \frac{(n-4)^{(2n-15)}}{(n+4)^{(2n+15)}}, \\
f^{(4)}_{(5g \rightarrow ns)}(Z)  & = \frac{2^{21}5^{11}}{7 \ 3^{6}Z^{13}}  \ n^{19} \ (187n^{6}-9350n^{4}+204625n^{2}+1743750)^{2} 
\\ & \, \, \, \, \, \frac{(n-5)^{(2n-17)}}{(n+5)^{(2n+17)}}.
\end{aligned}
\end{equation}

Now, applying Eqs.~(\ref{eq:59}) in Eq.~(\ref{eq:4}), one easily obtains $\alpha^{(4)}_{1s}(g)(Z)$,
$\alpha^{(4)}_{2p}(f)(Z)$, $\alpha^{(4)}_{3d}(d)(Z)$, $\alpha^{(4)}_{4f}(p)(Z)$ and $\alpha^{(4)}_{5g}(s)(Z)$. They 
take following forms, 
\begin{equation}\label{eq:60}
\begin{aligned}
\alpha^{(4)}_{1s}(g)(Z) & = \sum^{n}_{i=5}\frac{2^{20}}{9Z^{15}}  \ i^{15} \ (i^{2}-16)(i^{2}-9)(i^{2}-4) \ \frac{(i-1)^{(2i-12)}}{(i+1)^{(2i+12)}}, \\
\alpha^{(4)}_{2p}(f)(Z) & = \sum^{n}_{j=4} \frac{2^{39}}{3^{5}Z^{15}}  \ j^{17} (j^{2}-1)(j^{2}-9)(7j^{2}+68)^{2}
\frac{(j-2)^{(2j-14)}}{(j+2)^{(2j+14)}}, \\
\alpha^{(4)}_{3d}(d)(Z) & = \sum^{n}_{j=3} \frac{2^{22}3^{19}}{35Z^{15}}  \ j^{21} (j^{2}-1)(j^{2}-4)(j^{2}-21)^{2} \
\frac{(j-3)^{(2j-13)}}{(j+3)^{(2j+13)}}, \\	
\alpha^{(4)}_{4f}(p)(Z) & = \sum^{n}_{\substack{j=2 \\ j \ne 4}} \frac{2^{53}}{8505\ Z^{15}}  \ j^{13} (31j^{4}-4768j^{2}+43776)^{2} \
\frac{(j-4)^{(2j-13)}}{(j+4)^{(2j+13)}}, \\	
\alpha^{(4)}_{5g}(s)(Z) & = \sum^{n}_{\substack{j=1 \\ j \ne 5}} \frac{2^{27}5^{19}}{7 \ 3^{6} \ Z^{15}}  \ j^{23}(187j^{6}-9350j^{4}+204625j^{2}-1743750)^{2}\ \frac{(j-5)^{(2j-19)}}{(j+5)^{(2j+19)}}.
\end{aligned}
\end{equation} 
}

\bibliography{ref}

\begin{thebibliography}{10}

\bibitem{grochala07}
W.~Grochala, R.~Hoffmann, J.~Feng, and N.~W. Ashcroft.
\newblock {\em Angew.~Chem.~Int.~Ed.}, 46:3620, 2007.

\bibitem{snider20}
E.~Snider, N.~Dasenbrock-Gammon, R.~McBride, M.~Debessai, H.~Vindana,
  K.~Vencatasamy, K.~V. Lawler, A.~Salammat, and R.~P. Dias.
\newblock {\em Nature}, 586:373, 2020.

\bibitem{sen02}
K.~D. Sen, J.~Garza, R.~Vargas, and N.~Aquino.
\newblock {\em Phys. Lett. A}, 295:299, 2002.

\bibitem{sen14}
{K.~D.~Sen (Ed.)}.
\newblock {\em Electronic {S}tructure of {Q}uantum {C}onfined {A}toms and
  {M}olecules}.
\newblock Springer International Publishing, Switzerland, 2014\color{red}.

\bibitem{aquil21}
\color{red}V. Aquilanti, H.~E.~Montgomery Jr., C.~N. Ramachandran, and
  N.~Sathyamurthy.
\newblock {\em Eur. Phys. J. D}, 75:187, 2021.

\bibitem{jean20}
J.-P. Connerade.
\newblock {\em Eur. Phys. J. D}, 74:211, 2020\color{black}.

\bibitem{raggi14}
G.~Raggi, A.~J. Stace, and E.~Bichoutskaia.
\newblock {\em Phys.~Chem.~Chem.~Phys.}, 16:23869, 2014.

\bibitem{raggi16}
G.~Raggi, E.~Besley, and A.~J. Stace.
\newblock {\em Phil.~Trans.~R.~Soc.~A}, 374:20150319, 2016.

\bibitem{carbon18}
F.~J. Dominguez-Gutierrez, P.~S. Krstic, S.~Irle, and R.~Cabrera-Trujillo.
\newblock {\em Carbon}, 134:189, 2018.

\bibitem{mitroy10}
J.~Mitroy, M.~S. Safronova, and C.~W. Clark.
\newblock {\em J.~Phys.~B}, 43:202001, 2010.

\bibitem{tiihonen17}
J.~Tiihonen, I.~Kylänp\"a\"a, and T.~T. Rantala.
\newblock {\em J.~Chem.~Phys.}, 147:204101, 2017.

\bibitem{michel37}
A.~Michels, J.~de~Boer, and A.~Bijl.
\newblock {\em Physica}, 4:981, 1937.

\bibitem{sommerfeld38}
A.~Sommerfeld and H.~Welker.
\newblock {\em Ann.~Phys.}, 32:56, 1938.

\bibitem{guillot99}
T.~Guillot.
\newblock {\em Planet.~Space~Sci.}, 47:1183, 1999.

\bibitem{garza98}
J.~Garza, R.~Vargas, and A.~Vela.
\newblock {\em Phys.~Rev.~E}, 58:3949, 1998.

\bibitem{loughlin02}
M.~Cohen C.~Laughlin, B.~L.~Burrows.
\newblock {\em J. Phys. B}, 35:701, 2002.

\bibitem{burrows06}
B.~L. Burrows and M.~Cohen.
\newblock {\em Int.~J.~Quantum~Chem.}, 106:478, 2006.

\bibitem{aquino07}
N.~Aquino, G.~Campoy, and H.~E.~Montgomery Jr.
\newblock {\em Int.~J.~Quantum~Chem.}, 107:1548, 2007.

\bibitem{baye08}
D.~Baye and K.~D. Sen.
\newblock {\em Phys.~Rev.~E}, 78:026701, 2008.

\bibitem{roy15}
A.~K. Roy.
\newblock {\em Int.~J.~Quantum~Chem.}, 115:937, 2015.

\bibitem{centeno17}
D.~Puertas-Centeno, N.~M. Temme, I.~V. Toranzo, and J.~S. Dehesa.
\newblock {\em J.~Math.~Phys.}, 58:103302, 2017.

\bibitem{coll17}
N.~Sobrino-Coll, D.~Puertas-Centeno, I.~V. Toranzo, and J.~S. Dehesa.
\newblock {\em J. Stat. Mech.}, 8:083102, 2017.

\bibitem{montgomery07}
H.~E.~Montgomery Jr., N.~A. Aquino, and K.~D. Sen.
\newblock {\em Int.~J.~Quantum~Chem.}, 107:798, 2007.

\bibitem{mukherjee19}
N.~Mukherjee and A.~K. Roy.
\newblock {\em Phys.~Rev.~A}, 99:022123, 2019.

\bibitem{mukherjee18}
N.~Mukherjee and A.~K. Roy.
\newblock {\em Int.~J.~Quant.~Chem.}, 118:e25596, 2018.

\bibitem{mukherjee18a}
N.~Mukherjee, S.~Majumdar, and A.~K. Roy.
\newblock {\em Chem.~Phys.~Lett.}, 691:449, 2018\color{red}.

\bibitem{jiao17}
\color{red}L. G.~Jiao, L.~R. Zan, Y.~Z. Zhang, and Y.~K. Ho.
\newblock {\em Int.~J.~Quant.~Chem.}, 117:e25375, 2017\color{black}.

\bibitem{mukherjee17}
S.~Majumdar, N.~Mukherjee, and A.~K. Roy.
\newblock {\em Chem.~Phys.~Lett.}, 687:322, 2017.

\bibitem{mukherjee20}
N.~Mukherjee and A.~K. Roy.
\newblock {\em J.~Phys.~B}, 53:235002, 2020.

\bibitem{sabin09}
{J.~R.~Sabin and E.~Brändas and S.~A.~Cruz (Eds)}.
\newblock {\em The {T}heory of {C}onfined {Q}uantum {S}ystems, {P}arts {I} and
  {II}}.
\newblock Advances in Quantum Chemistry, Vols. 57 and 58, Academic Press,
  Cambridge, Massachusetts, 2009.

\bibitem{sen12}
{K.~D.~Sen (Ed.)}.
\newblock {\em Statistical {C}omplexity: {A}pplications in {E}lectronic
  {S}tructure}.
\newblock Springer International Publishing, Berlin, 2012\color{red}.

\bibitem{zan17}
\color{red}L. R.~Zan, L.~G. Jiao, J.~Ma, and Y.~K. Ho.
\newblock {\em Phys. Plasmas}, 24:122101, 2017.

\bibitem{salazar21}
\color{red}S. J.~C.~Salazar, H.~G. Laguna, B.~Dahiya, V.~Prasad, and R.~P.
  Sagar.
\newblock {\em Eur. Phys. J. D}, 75:127, 2021.

\bibitem{ou19}
\color{red}J. H.~Ou and Y.~K. Ho.
\newblock {\em Int.~J.~Quant.~Chem.}, 119:e25928, 2019.

\bibitem{he21}
\color{red}Y. Y.~He, L.~G. Jiao, A.~Liu, Y.~Z. Zhang, and Y.~K. Ho.
\newblock {\em Eur. Phys. J. D}, 75:126, 2021.

\bibitem{yadav21}
\color{red}C. Yadav, S.~Lumb, and V.~Prasad.
\newblock {\em Eur. Phys. J. D}, 75:21, 2021.

\bibitem{zhu20}
L.~Zhu, Y.~Y. He, L.~G. Jiao, Y.~C. Wang, and Y.~K. Ho.
\newblock {\em Int.~J.~Quant.~Chem.}, 120:e26245, 2020.

\bibitem{lin13}
\color{red}C. Y.~Lin and Y.~K. Ho.
\newblock {\em Few.~Body.~Syst.}, 54:425, 2013\color{black}.

\bibitem{sen05}
K.~D. Sen.
\newblock {\em J. Chem. Phys.}, 122:194324, 2005.

\bibitem{burrows08}
M.~Cohen B.~L.~Burrows.
\newblock {\em Mol.~Phys.}, 106:267, 2008.

\bibitem{sen81}
K.~D. Sen and P.~C. Schmidt.
\newblock {\em Phys.~Rev.~A}, 23:1026, 1981.

\bibitem{wood32}
J.~G. Kirkwood.
\newblock {\em Phys. Z}, 33:57, 1932.

\bibitem{buck37}
R.~A. Buckingham.
\newblock {\em Proc.~R.~Soc.~London, Ser. A}, 160:94, 1937.

\bibitem{stern54}
R.~M. Sternheimer.
\newblock {\em Phys.~Rev.}, 96:951, 1954.

\bibitem{pupyshev19}
V.~I. Pupyshev and H.~E.~Montgomery Jr.
\newblock {\em Int.~J.~Quant.~Chem.}, 119:e25887, 2019.

\bibitem{nascimento11}
E.~M. Nascimento, F.~V. Prudente, M.~N.~Guimar\ aes, and A.~M. Maniero.
\newblock {\em J.~Phys.~B}, 44:015003, 2011.

\bibitem{efros16}
A.~L. Efros and D.~J. Nesbitt.
\newblock {\em Nature Nanotechnology}, 11:661, 2016.

\bibitem{fei21}
Z.~Fei, Z.~Wang, D.~Li, F.~Xue, C.~Cheng, Q.~Liu, X.~Chen, M.~Cui, and X.~Qiao.
\newblock {\em Nanoscale}, 13:10765, 2021.

\bibitem{peng18}
H.~Peng, C.~Rao, N.~Zhang, X.~Wang, W.~Liu, W.~Mao, L.~Han, P.~Zhang, and
  S.~Dai.
\newblock {\em Angew.~Chem.~Int.~Ed.}, 57:8953, 2018.

\bibitem{rao18}
C.~Rao, C.~Peng, H.~Peng, L.~Zhang, W.~Liu, X.~Wang, N.~Zhang, and P.~Wu.
\newblock {\em ACS Appl. Mater. Interfaces}, 10:9220, 2018.

\bibitem{kumar19}
T.~Raj kumar, G.~Gnana kumar, and Arumugam Manthiram.
\newblock {\em Adv.~Energy~Mater.}, 9:1803238, 2019.

\bibitem{lai21}
Y.~Lai, W.~Xia, J.~Li, J.~Pan, C.~Jiang, Z.~Cai, C.~Wu, X.~Huang, T.~Wang, and
  J.~He.
\newblock {\em Electrochimica Acta}, 375:137966, 2021.

\bibitem{fan20}
M.~Fan, D.~Liao, M.~F.~Aly Aboud, I.~Shakir, and Y.~Xu.
\newblock {\em Angew.~Chem.~Int.~Ed.}, 59:8247, 2020.

\bibitem{shi13}
A.-C. Shi and B.~Li.
\newblock {\em Soft Matter}, 9:1398, 2013.

\bibitem{khadilkar18}
M.~R. Khadilkar and A.~Nikoubashman.
\newblock {\em Soft Matter}, 14:6903, 2018.

\bibitem{qin20}
L.~Qin, C.~Li, X.~Li, X.~Zhang, C.~Shen, Q.~Meng, L.~Shen, Y.~Lu, and G.~Zhang.
\newblock {\em J.~Mater.~Chem.~A}, 8:1929, 2020.

\bibitem{zhang20}
M.~Zhang, C.~Xiao, X.~Yan, S.~Chen, C.~Wang, R.~Luo, J.~Qi, X.~Sun, L.~Wang,
  and J.~Li.
\newblock {\em Environ. Sci. Technol.}, 54:10289, 2020.

\bibitem{hastings19}
D.~E. Hastings and H.~D.~H. St\"over.
\newblock {\em ACS Appl. Polym. Mater.}, 1:2055, 2019.

\bibitem{kumar18}
G.~Gnana kumar, S.-H. Chung, T.~Raj kumar, and Arumugam Manthiram.
\newblock {\em ACS Appl. Mater. Interfaces}, 10:20627, 2018.

\bibitem{wei19}
W.~Shuang, H.~Huang, L.~Kong, M.~Zhong, A.~Li, D.~Wang, Y.~Xu, and X.-H. Bu.
\newblock {\em Nano Energy}, 62:154, 2019.

\bibitem{wang20}
J.~Wang, L.~Zhu, F.~Li, T.~Yao, T.~Liu, Y.~Cheng, Z.~Yin, and H.~Wang.
\newblock {\em Small}, 16:2002487, 2020.

\bibitem{roy2014}
A.~K. Roy.
\newblock {\em Mod.~Phys.~Lett.~A}, 29:1450104, 2014.

\bibitem{roy2014a}
A.~K. Roy.
\newblock {\em J.~Math.~Chem.}, 52:1405, 2014.

\bibitem{roy2014b}
A.~K. Roy.
\newblock {\em Mod.~Phys.~Lett.~A}, 29:1450042, 2014.

\bibitem{majumdar2020}
S.~Majumdar and A.~K. Roy.
\newblock {\em Quant.~Rep.}, 2:189, 2020.

\bibitem{majumdar2021}
S.~Majumdar and A.~K. Roy.
\newblock {\em Int.~J.~Quant.~Chem.}, 121:e26630, 2021.

\bibitem{das12}
M.~Das.
\newblock {\em Phys. Plasmas}, 19:092707, 2012.

\bibitem{dalgarno62}
A.~Dalgarno.
\newblock {\em Adv.~Phys.}, 11:281, 1962.

\bibitem{zhu20a}
\color{red}L. Zhu, Y.~Y. He, L.~G. Jiao, Y.~C. Wang, and Y.~K. Ho.
\newblock {\em Phys. Plasmas}, 27:072101, 2020\color{black}.

\bibitem{bbi75}
I.~Bialynicki-Birula and J.~Mycielski.
\newblock {\em Commun.~Math.~Phys.}, 44:129, 1975\color{red}.

\bibitem{shiner99}
\color{red}J. S.~Shiner, M.~Davison, and P.~T. Landsberg.
\newblock {\em Phys.~Rev.~E}, 59:1459, 1999\color{black}.

\bibitem{solyu12}
A.~Solyu.
\newblock {\em Phys.~Plasmas}, 19:072701, 2012.

\bibitem{jiao21}
L.~G. Jiao, Y.~Y. He, Y.~Z. Zhang, and Y.~K. Ho.
\newblock {\em J.~Phys.~B}, 2021.

\bibitem{paul09}
S.~Paul and Y.~K. Ho.
\newblock {\em Phys.~Rev.~A}, 79:032714, 2009.

\bibitem{bahar14}
M.~K. Bahar and A.~Solyu.
\newblock {\em Phys.~Plasmas}, 092703:21, 2014.

\bibitem{bahar16}
M.~K. Bahar, A.~Soylu, and A.~Poszwa.
\newblock {\em IEEE~Trans.~Plasma~Sci.}, 44:2297, 2016.

\bibitem{jung95}
Y.-D. Jung.
\newblock {\em Phys.~Plasmas}, 332:2, 1995.

\bibitem{jung96}
Y.-D. Jung and J.-S. Yoon.
\newblock {\em J.~Phys.~B}, 29:3549, 1996.

\bibitem{song03}
M.~Y. Song and Y.-D. Jung.
\newblock {\em Phys.~Plasmas}, 36:2119, 2003.

\bibitem{lin10}
C.~Y.Lin and Y.~K. Ho.
\newblock {\em Eur.~Phys.~J.~D}, 57:21, 2010.

\bibitem{lin11}
C.~Y. Lin and Y.~K. Ho.
\newblock {\em Comp.~Phys.~Commun.}, 182:125, 2011.

\bibitem{baye12}
D.~Baye.
\newblock {\em Phys.~Rev.~A}, 86:06254, 2012.

\bibitem{harzfeld27}
K.~F. Herzfeld.
\newblock {\em Phys.~Rev.}, 29:701, 1927.

\end{thebibliography}
\bibliographystyle{unsrt} 
\end{document}